\documentclass[twocolumn, tighten, twocolappendix, times]{aastex701}
\hypersetup{linkcolor=blue,citecolor=blue,filecolor=cyan,urlcolor=blue}

\usepackage{amsmath}
\usepackage{amsfonts}
\usepackage{amssymb}
\usepackage{graphicx}
\usepackage{txfonts}
\usepackage{xcolor}
\usepackage{comment}
\usepackage{mathrsfs}
\usepackage{caption}
\usepackage{subcaption}
\usepackage{float}
\usepackage{placeins}
\usepackage[ruled,vlined]{algorithm2e}
\usepackage{algorithmic}
\usepackage{tikz}

\newcommand{\norm}[1]{|\!|#1|\!|}
\newcommand{\cs}{c_\mathrm{s}}
\newcommand{\tff}{t_\mathrm{ff}}
\newcommand{\kms}{\mathrm{km~s^{-1}}}
\newcommand{\DS}{\displaystyle}
\newcommand{\HALF}{\frac{1}{2}}
\newcommand{\del}{\partial}

\newcommand{\rphalf}{r_{i+\HALF}}
\newcommand{\rmhalf}{r_{i-\HALF}}
\newcommand{\tphalf}{\theta_{j+\HALF}}
\newcommand{\tmhalf}{\theta_{j-\HALF}}
\newcommand{\Rphalf}{R_{i+\HALF}}
\newcommand{\Rmhalf}{R_{i-\HALF}}
\newcommand{\zphalf}{z_{k+\HALF}}
\newcommand{\zmhalf}{z_{k-\HALF}}
\newcommand{\pjphalf}{\phi_{j+\HALF}}
\newcommand{\pjmhalf}{\phi_{j-\HALF}}
\newcommand{\pkphalf}{\phi_{k+\HALF}}
\newcommand{\pkmhalf}{\phi_{k-\HALF}}

\begin{document}

\title{A fast spectral-multigrid Poisson solver in non-Cartesian geometries}

\correspondingauthor{Ankush Mandal}

\author[orcid=0000-0002-0912-8081, gname=Ankush, sname='Mandal']{Ankush Mandal}
\affiliation{Leibniz-Institut f\"ur Astrophyisk Potsdam (AIP), An der  Sternwarte 16, 14482 Potsdam, Germany}
\email[show]{amandal@aip.de}  

\author[orcid=0000-0002-5398-9225, gname=Oliver, sname='Gressel']{Oliver Gressel} 
\affiliation{Leibniz-Institut f\"ur Astrophyisk Potsdam (AIP), An der  Sternwarte 16, 14482 Potsdam, Germany}
\email{ogressel@aip.de}

\author[gname=Udo, sname='Ziegler']{Udo Ziegler}
\affiliation{Leibniz-Institut f\"ur Astrophyisk Potsdam (AIP), An der  Sternwarte 16, 14482 Potsdam, Germany}
\email{uziegler@aip.de}

\author[orcid=0000-0002-8352-6635, gname=Andrea, sname='Mignone']{Andrea Mignone}
\affiliation{Dipartimento di Fisica, Universit\`a degli Studi di Torino, Via Pietro Giuria 1, I-10125 Torino, Italy}
\email{andrea.mignone@unito.it}

\shorttitle{Spectral–Multigrid Poisson Solver}
\shortauthors{Mandal et al.}

\begin{abstract}
Accurate and efficient computation of self-gravity is essential in astrophysical fluid dynamics, particularly in spherical and cylindrical geometries where large radial dynamic ranges and non-axisymmetric structures frequently arise. Poisson solvers in such settings must simultaneously achieve high accuracy, scalability, and flexibility across a wide range of grid configurations and physical regimes.
We present a robust and scalable Poisson solver for three-dimensional non-Cartesian geometries, supporting both spherical and cylindrical coordinates with either uniform or logarithmic radial discretizations. The method employs azimuthal Fourier decomposition to transform the 3D Poisson equation into a set of independent 2D Helmholtz equations. These are solved using a geometrically consistent multigrid algorithm that preserves second-order accuracy on both uniform and non-uniform grids.
Vacuum boundary conditions are implemented through a screening-mass approach, enabling accurate solutions in domains with open boundaries, inner cavities, and strongly non-axisymmetric mass distributions. Owing to the differing convergence rates of Fourier modes -- where higher-order modes converge more rapidly -- the solver allows efficient mode-by-mode treatment. The combination of spectral decomposition and multigrid acceleration provides a highly efficient and flexible computational framework.
The solver is implemented in the \textsc{Pluto} code and validated against both analytical solutions and dynamical test problems in spherical and cylindrical geometries. Results demonstrate second-order convergence and excellent agreement with reference solutions. Weak-scaling tests up to 4096 cores show strong parallel performance, with the Poisson solve remaining subdominant to magnetohydrodynamic update costs. This makes the method well suited for large-scale simulations of star formation, accretion disks, and gravitational instabilities.

\end{abstract}



\section{Introduction}
\label{sec:introduction}
Self-gravity governs the evolution of a wide range of astrophysical systems, including molecular cloud collapse and formation of stars \citep{Shu_1987, Ostriker_2001, McKee2007}, protoplanetary disk evolution and planet formation \citep{Rice_2005, Armitage_2011}, galactic dynamics \citep{Mo_2010, Binney_2011}, and large scale structure formation \citep{Bertschinge_1998}. 
In numerical simulations, the gravitational potential is obtained by solving the Poisson equation at every timestep for the evolving mass distribution. 
As this elliptic solver is typically one of the most computationally expensive components of a self-gravitational calculation, the efficiency, accuracy, and scalability of the Poisson solver are central to the overall performance and reliability of such simulations.

Many problems in astrophysics possess intrinsic geometric symmetries that are more naturally represented in non-Cartesian coordinates. Gravitational collapse is fundamentally radial and is most accurately described in spherical geometry, whereas rotating systems such as accretion disks and collapsing cores are better modeled in spherical or cylindrical coordinates \cite[e.g.,][]{Hawley_2000,Suzuki_2014}. These geometries enable efficient coverage of large spatial dynamic ranges, especially when combined with logarithmically stretched radial grids that concentrate resolution near the center while retaining computational tractability at large radii. 
Aligning the computational grid with the physical symmetry improves accuracy, reduces numerical artifacts, and facilitates angular-momentum conservation. In contrast, Cartesian grids often require substantially higher resolution to achieve comparable fidelity and may introduce preferred directions that can influence symmetry and long-term evolution.

Despite these advantages, solving the Poisson equation in spherical and cylindrical coordinates poses significant numerical challenges. 
In curvilinear geometries the Laplacian contains spatially varying metric coefficients, leading to anisotropic and variable-coefficient operators. 
These effects become particularly pronounced on logarithmic grids, where strong radial stretching introduces additional stiffness. 
Methods developed for uniform Cartesian grids do not directly retain their efficiency or robustness in this setting \cite[e.g.,][]{Briggs_2000, Trottenberg_2000}.

Several classes of numerical methods have been developed to solve the Poisson equation in astrophysical contexts. In Cartesian geometry, fast Fourier transform (FFT)–based solvers are widely used in periodic domains due to their computational efficiency and spectral accuracy \cite[e.g.,][]{Hockney_1988, Mayani_2024, Restrepo_2026}. 
However, these methods are not directly applicable to isolated boundary conditions or curvilinear geometries without significant modification. 
Tree-based methods \citep{Barnes_1986, Hernquist_1987} and fast multipole methods \cite[e.g.,][]{Dehnen_2000} provide flexible alternatives that naturally handle open boundary conditions and adaptive spatial distributions, and are widely used in particle-based and hybrid simulations \cite[e.g.,][]{Hernquist_1989,Springel_2005,Price_2018,Springel_2021,Schaller_2024}. 
Nonetheless, these approaches introduce approximation errors and are generally less efficient for grid-based hydrodynamics simulations, where the density field is already defined on a mesh and direct solution of the discretized Poisson equation is preferable \cite[see however][]{Wunsch_2018}.

For grid-based methods, multigrid algorithms \citep{Brandt_1977,Briggs_2000,Trottenberg_2000} provide one of the most efficient approaches for solving elliptic equations, offering optimal computational complexity and excellent convergence properties. 
Multigrid solvers have been successfully implemented in many astrophysical fluid dynamics codes \cite[e.g.,][]{Ziegler_2005, Ricker_2008, Almgren_2010, Guillet_2011, Wang_2020, Mandal_2023}. 
However, their performance can degrade in curvilinear coordinate systems, particularly on non-uniform grids. 
Metric terms introduce variable coefficients and anisotropies that reduce the effectiveness of standard smoothers, while geometric consistency must be maintained across grid levels to preserve accuracy and convergence \cite[see][for discussion]{Trottenberg_2000}. 
These challenges make the construction of efficient and robust multigrid solvers in spherical and cylindrical geometries significantly more complex than in Cartesian coordinates.

For problems defined on the full azimuthal domain, spherical and cylindrical coordinates naturally admit periodicity in the azimuthal direction.
This symmetry permits decomposition of the Poisson equation into azimuthal Fourier modes, reducing the original three-dimensional problem to a sequence of independent two-dimensional Helmholtz equations. 
The effectiveness of this strategy has been demonstrated in recent work. \citet{Moon_2019} constructed a fast Poisson solver in cylindrical coordinates using FFTs in the azimuthal and vertical directions, reducing the system to a tridiagonal radial solve. 
In spherical geometry, where full separability is not available, \citet{Ahn_2025} employed azimuthal Fourier decomposition and solved the resulting two-dimensional Helmholtz equations using a divide-and-conquer strategy combined with the \citet{James_1977} method, achieving an efficient solver on logarithmic grids.

Beyond dimensional reduction, the Fourier decomposition has important implications for iterative solution strategies. 
The Helmholtz equations arising from azimuthal decomposition of the Poisson equation remain positive definite. 
The mode-dependent term enhances diagonal dominance as the azimuthal wavenumber increases, leading to distinct convergence behavior for different modes. 
Higher-order modes typically converge rapidly under iterative schemes \cite[see, e.g.,][]{Trottenberg_2000}, while lower-order modes dominate the overall computational cost. 
Therefore, a mode-by-mode multigrid strategy can achieve superior convergence and computational efficiency compared to solving the full three-dimensional system directly.

Building on these insights, we present an efficient hybrid Poisson solver for self-gravitating fluid dynamics simulations in spherical and cylindrical coordinates. 
The method combines azimuthal Fourier decomposition with a geometrically consistent multigrid algorithm to efficiently solve the resulting Helmholtz equations.
Vacuum boundary conditions are imposed using the \citet{James_1977} method, based on discrete Green’s functions, which provides an accurate and efficient formulation well suited to strongly non-axisymmetric density distribution as well as domain with inner cavity. 
The solver is designed to operate efficiently on both uniform and logarithmically stretched grids, allowing accurate simulations across large radial dynamic ranges characteristic of gravitational collapse, while also supporting more uniform grids appropriate for disk evolution problems.

We implement the Poisson solver as a self-gravity module in the \textsc{Pluto} code \citep{Mignone_2007, Mignone_2012}. 
Through an extensive suite of static and dynamical tests, we demonstrate that the solver achieves second-order accuracy, exhibits robustness in high–dynamic-range and strongly time-dependent regimes, and exhibits excellent computational efficiency, with runtimes substantially lower than those of the magnetohydrodynamics (MHD) update. 
By combining geometric flexibility, accuracy, and computational efficiency, the presented method provides a robust and scalable tool for self-gravitating astrophysical simulations in non-Cartesian geometries.

This paper is organized as follows. Sec.~\ref{sec:poisson_equation} presents the discretization of the Poisson equation in curvilinear coordinates, along with the azimuthal Fourier decomposition and the resulting Helmholtz formulation. The multigrid algorithm is described in Sec.~\ref{sec:multigrid_solver}, with its individual components detailed in the corresponding subsections. The implementation of vacuum boundary conditions is presented in Sec.~\ref{sec:boundary_calculation}. Sec.~\ref{sec:algo_par} outlines the overall algorithm, workflow, and parallelization strategy. Test results and parallel scaling results are given in Sec.~\ref{sec:test_result} and \ref{sec:scaling_test}, respectively. Finally, Sec.~\ref{sec:summary} summarizes our findings.

\section{Poisson equation}
\label{sec:poisson_equation}
The gravitational potential $\Phi$ generated by a mass distribution $\rho$ satisfies
\begin{equation}\label{eq:poisson}
    \nabla^{2}\Phi = 4\pi G\rho.
\end{equation}
The explicit form of the Laplacian ($\nabla^2$) depends on the underlying coordinate geometry.  
In spherical coordinates $(r,\theta,\phi)$, it reads
\begin{equation}\label{eq:poisson_sph}
    \nabla^2\Phi = \frac{1}{r^{2}}\frac{\partial}{\partial r}
    \left(r^{2}\frac{\partial\Phi}{\partial r}\right)
    +
    \frac{1}{r^{2}\sin\theta}\frac{\partial}{\partial\theta}
    \left(\sin\theta\frac{\partial\Phi}{\partial\theta}\right)
    +
    \frac{1}{r^{2}\sin^{2}\theta}
    \frac{\partial^{2}\Phi}{\partial\phi^{2}}
\end{equation}
In cylindrical coordinates $(R,\phi,z)$, it becomes
\begin{equation}\label{eq:poisson_cyl}
    \nabla^2\Phi = \frac{1}{R}\frac{\partial}{\partial R}
    \left(R\frac{\partial\Phi}{\partial R}\right)
    +
    \frac{1}{R^{2}}\frac{\partial^{2}\Phi}{\partial\phi^{2}}
    +
    \frac{\partial^{2}\Phi}{\partial z^{2}}.
\end{equation}
These expressions form the starting point for constructing the numerical discretization presented in the following section.

\subsection{Finite-volume discretization}
\label{sec:discretization}
The Poisson equation in Eq.~\eqref{eq:poisson} can be written in conservative (divergence) form as,
\begin{equation}\label{eq:poisson_div}
\boldsymbol{\nabla}\cdot(\boldsymbol{\nabla}\Phi) = 4\pi G\rho.
\end{equation}
which makes it naturally suited for a finite-volume discretization.

Let the computational domain be partitioned into non-overlapping ``control volumes'' $\{V_i\}$, each with boundary $\partial V_i$ composed of faces $f\in\{\mathcal{F}_i\}$. Each ``control volume'' has volume
\begin{equation}
    \Delta V_i = \int_{V_i} dV,
\end{equation}
and each face has area
\begin{equation}
    A_f = \int_f dA,
\end{equation}
with outward unit normal vector $\boldsymbol{n}_f$.

Integrating Eq.~\eqref{eq:poisson_div} over the control volume $V_i$ gives
\begin{equation}\label{eq:volume_int_poisson}
    \int_{V_i} \boldsymbol{\nabla}\cdot(\boldsymbol{\nabla}\Phi) dV = 4\pi G\int_{V_i} \rho dV.
\end{equation}
Applying the divergence theorem to the left-hand side yields
\begin{equation}\label{eq:poisson_finite_vol}
    \oint_{\del V_i} (\boldsymbol{\nabla}\Phi)\cdot\boldsymbol{n} dA = 
    \sum_{f\in \{\mathcal{F}_i\}} \int_{f} (\boldsymbol{\nabla}\Phi)\cdot \boldsymbol{n}_f dA = 4\pi G\int_{V_i} \rho dV.
\end{equation}

On each face $f$, we approximate the normal derivative using a two-point difference between neighboring cell centers:
\begin{equation}\label{eq:gradient}
    (\boldsymbol{\nabla}\Phi)\cdot\boldsymbol{n}_f \approx \frac{\Phi_{N_f} - \Phi_i}{d_{iN_f}},
\end{equation}
where $N_f$ is the cell sharing face $f$ with the $i^{\rm th}$ cell and 
\begin{equation}
    d_{iN_f} = (\boldsymbol{x}_{N_f} - \boldsymbol{x}_i)\cdot \boldsymbol{n}_f  
\end{equation}
is the projected distance between the cell centers along the face normal. On non-uniform grids where the face does not coincide with the midpoint between cell centers, this approximation is formally first-order accurate at the face. Nevertheless, the finite-volume discretization of the Poisson operator remains second-order accurate on smoothly varying orthogonal grids due to conservative flux cancellation and smooth metric variation \cite[e.g.,][]{Domelevo_2005,Barth_2017, Bonaventura_2018}, as we demonstrate in Sec.~\ref{sec:convergence_test}.

The flux through face $f$ is then approximated as
\begin{equation}\label{eq:face_contribution}
    \int_{f} (\boldsymbol{\nabla}\Phi)\cdot \boldsymbol{n}_f dA  = A_f\frac{\Phi_{N_f}-\Phi_i}{d_{iN_f}}.
\end{equation}

Substituting Eq.~\eqref{eq:face_contribution} into Eq.~\eqref{eq:poisson_finite_vol} gives
\begin{equation}\label{eq:poisson_discrete}
    \frac{1}{\Delta V_i}\sum_{f\in\{\mathcal{F}_i\}} A_f \frac{\Phi_{N_f} - \Phi_i}{d_{iN_f}} = 4\pi G \rho_i,
\end{equation}
where we approximate the volume integral of the density as 
\begin{equation}
    \int_{V_i}\rho dV \approx \rho_i\times \Delta V_i
\end{equation}

Eq.~\eqref{eq:poisson_discrete} is the general finite-volume discretization of the Poisson equation, independent of the underlying coordinate system. By inserting the appropriate expressions for $\Delta V_i$, the face areas $A_f$, and the projected distances $d_{iN_f}$ for a chosen geometry (Cartesian, cylindrical, spherical, etc.), the corresponding discretized form is obtained.
The remainder of this section specializes this general finite-volume form to spherical and cylindrical coordinates.

\subsubsection{Spherical coordinates}
\label{sec:discrete_sph}
We adopt a finite-volume discretization on a spherical $(r,\theta,\phi)$ grid.
The radial, polar, and azimuthal directions are divided into faces $\{r_{i\pm\HALF}\}$, $\{\theta_{j\pm\HALF}\}$, and $\{\phi_{k\pm\HALF}\}$, which define non-overlapping control volumes $V_{i,j,k}$.
The corresponding cell centers are defined as the volume-weighted centroids of each coordinate interval:
\begin{align}\label{eq:cell_centers_sph}
    r_i =& \frac{\int_{\rmhalf}^{\rphalf} r^3 dr}{\int_{\rmhalf}^{\rphalf} r^2 dr} = \frac{3}{4}\frac{\rphalf^4 - \rmhalf^4}{\rphalf^3 - \rmhalf^3}, \\
    \begin{split}
    \theta_{j} =&\frac{\int_{\tmhalf}^{\tphalf}\theta\sin\theta d\theta}{\int_{\tmhalf}^{\tphalf}\sin\theta d\theta} \\
    =& \frac{\sin{\tphalf}-\sin{\tmhalf} + \tmhalf\cos{\tmhalf} - \tphalf\cos{\tphalf}}{\cos{\tmhalf} - \cos{\tphalf}}
    \end{split}\\
    \phi_k = & \frac{\pkphalf + \pkmhalf}{2}.
\end{align}

Taking into account the cell widths 
$\Delta r_i = \rphalf - \rmhalf$, 
$\Delta\theta_j = \tphalf - \tmhalf$, and 
$\Delta\phi_k = \phi_{k+\HALF} - \phi_{k-\HALF}$, 
the ``control-volume`` geometry in spherical coordinates is fully determined by the corresponding volume
\begin{equation}\label{eq:vol_sph}
    \Delta V_{i,j,k} = \frac{1}{3}\left(\rphalf^3 -\rmhalf^3\right)\left(\cos{\tmhalf} - \cos{\tphalf}\right)\Delta\phi_k
\end{equation}
and face areas
\begin{equation}\label{eq:surface_sph}
    \begin{aligned}
        A_{i\pm\HALF,j,k}  = & r_{i\pm\HALF}^2\left(\cos{\tmhalf} - \cos{\tphalf}\right)\Delta \phi_k\\
        A_{i,j\pm\HALF,k}  = & \frac{1}{2}\left(\rphalf^2 - \rmhalf^2\right)\sin{\theta_{j\pm\HALF}}\Delta\phi_k\\
        A_{i,j,k\pm\HALF}  = & \frac{1}{2}\left(\rphalf^2 - \rmhalf^2\right)\Delta\theta_j
    \end{aligned}
\end{equation}
The gravitational potential $\Phi_{i,j,k}$ and density $\rho_{i,j,k}$ are stored at these cell centers, whereas fluxes are evaluated on the cell faces.

Using Eq.~\eqref{eq:poisson_discrete}, \eqref{eq:vol_sph} and \eqref{eq:surface_sph}, the discrete Poisson equation in spherical coordinates is derived as
\begin{equation}\label{eq:discrete_poisson_sph}
    \nabla^2_r\Phi_{i,j,k} + \nabla^2_\theta\Phi_{i,j,k} + \nabla^2_\phi\Phi_{i,j,k} = 4\pi G\rho_{i,j,k},
\end{equation}
where
\begin{align}
&
\begin{aligned}
 \nabla^2_r \Phi_{i,j,k}  = \frac{3}{\rphalf^3 - \rmhalf^3}
  & \Biggl(
    \rphalf^2
    \frac{\Phi_{i+1,j,k}-\Phi_{i,j,k}}{r_{i+1}-r_i}
\\
    &
    -\, \rmhalf^2
    \frac{\Phi_{i,j,k}-\Phi_{i-1,j,k}}{r_i-r_{i-1}}
  \Biggr),
\end{aligned}
\\
&
\begin{aligned}
\nabla^2_\theta \Phi_{i,j,k} = \frac{\mathcal{R}_i}{\cos\tmhalf-\cos\tphalf}
  &\Biggl(
    \sin{\tphalf}
    \frac{\Phi_{i,j+1,k}-\Phi_{i,j,k}}{\theta_{j+1}-\theta_j}
\\
    -\,\sin&{\tmhalf}
    \frac{\Phi_{i,j,k}-\Phi_{i,j-1,k}}{\theta_j-\theta_{j-1}}
  \Biggr),
\end{aligned}
\\
&
\nabla^2_\phi \Phi_{i,j,k} = \frac{\mathcal{R}_i\mathcal{T}_j}{\Delta\phi_k}
  \Biggl(
    \frac{\Phi_{i,j,k+1}-\Phi_{i,j,k}}{\phi_{k+1}-\phi_k} - \frac{\Phi_{i,j,k}-\Phi_{i,j,k-1}}{\phi_k-\phi_{k-1}}
  \Biggr),
\end{align}
with
\begin{equation}
    \mathcal{R}_i = \frac{3}{2r_i}\frac{\rphalf^2 - \rmhalf^2}{\rphalf^3 - \rmhalf^3},
    \quad {\rm and} \quad
    \mathcal{T}_j = \frac{\Delta\theta_j}{\sin{\theta}_j(\cos{\tmhalf} - \cos{\tphalf})}.
\end{equation}

\subsubsection{Cylindrical coordinates}
\label{sec:discrete_cyl}
Analogous to the spherical discretization in Sec.~\ref{sec:discrete_sph}, we define the cell centers in cylindrical coordinates as
\begin{align}
    R_i =& \frac{\int_{\Rphalf}^{\Rmhalf} R^2dR}{\int_{\Rphalf}^{\Rmhalf} RdR} = \frac{2}{3}\frac{\Rphalf^3 - \Rmhalf^3}{\Rphalf^2 - \Rmhalf^2}, \\
    \phi_{j} =& \frac{\pjphalf + \pjmhalf}{2},\\
    z_k =& \frac{\zphalf + \zmhalf}{2},
\end{align}
The volume element is given by
\begin{equation}\label{eq:vol_cyl}
    \Delta V_{i,j,k} = \frac{1}{2}\left(\Rphalf^2 - \Rmhalf^2\right)\Delta\phi_j \Delta z_k.
\end{equation}
The corresponding face surface areas are
\begin{equation}\label{eq:surface_cyl}
    \begin{aligned}
        A_{i\pm\HALF,j,k} = &\; R_{i\pm\HALF}\Delta\phi_j\Delta z_k, \\
        A_{i,j\pm\HALF,k} = &\; \Delta R_i\Delta z_k, \\
        A_{i,j,k\pm\HALF} = &\; \frac{1}{2}\left(\Rphalf^2 - \Rmhalf^2\right)\Delta\phi_j,
    \end{aligned}
\end{equation}

Therefore, the finite-volume discretization of the Poisson equation in cylindrical coordinates yields
\begin{equation}\label{eq:discrete_poisson_cyl}
    \nabla^2_R\Phi_{i,j,k} + \nabla^2_\phi\Phi_{i,j,k} + \nabla^2_z\Phi_{i,j,k} = 4\pi G\rho_{i,j,k},
\end{equation}
where the individual directional contributions are given by
\begin{align}
&
\begin{aligned}
\nabla^2_R \Phi_{i,j,k} = \frac{2}{\Rphalf^2 - \Rmhalf^2}
  &\Biggl(
    \Rphalf
    \frac{\Phi_{i+1,j,k}-\Phi_{i,j,k}}{R_{i+1}-R_i}\\
    &
    -\, \Rmhalf
    \frac{\Phi_{i,j,k}-\Phi_{i-1,j,k}}{R_i-R_{i-1}}
  \Biggr),
\end{aligned}
\\
&
\nabla^2_\phi \Phi_{i,j,k} = \frac{\mathcal{W}_i}{\Delta\phi_j}
  \Biggl(
    \frac{\Phi_{i,j+1,k}-\Phi_{i,j,k}}{\phi_{j+1}-\phi_j} - \frac{\Phi_{i,j,k}-\Phi_{i,j-1,k}}{\phi_j-\phi_{j-1}}
  \Biggr),
  \\&
\nabla^2_z \Phi_{i,j,k} = \frac{1}{\Delta z_k}
    \Biggl(
    \frac{\Phi_{i,j,k+1} - \Phi_{i,j,k}}{z_{k+1}-z_k} - \frac{\Phi_{i,j,k} - \Phi_{i,j,k-1}}{z_{k}-z_{k-1}}
    \Biggr),
\end{align}
with 
\begin{equation}
    \mathcal{W}_i = \frac{2}{R_i(\Rphalf + \Rmhalf)}
\end{equation}

\subsection{Azimuthal Fourier decomposition}
\label{sec:fourier_decompostion}
As the azimuthal ($\phi$) direction is periodic in both spherical and cylindrical coordinates, it is natural to apply a Fourier decomposition along $\phi$. 

Let
\begin{equation}\label{eq:eigenfunction}
    \mathcal{P}^m_k = \exp\left(\frac{2\pi\sqrt{-1} m k}{N_{\phi}}\right)
\end{equation}
denote the eigenfunctions of the discrete $\nabla^2_\phi$ operators. 
The real-space potential $\Phi_{i,j,k}$ and density $\rho_{i,j,k}$ can then be decomposed as
\begin{equation}\label{eq:fourier_decomposition}
    \Phi_{i,j,k} = \sum_{m=1}^{N_\phi}\Phi_{i,j}^m \mathcal{P}^m_k
    \qquad \text{and}\qquad
    \rho_{i,j,k} = \sum_{m=1}^{N_\phi}\rho_{i,j}^m \mathcal{P}^m_k,    
\end{equation}
where $\Phi^m_{i,j}$ and $\rho^m_{ij}$ are the Fourier coefficients of $\Phi_{i,j,k}$ and $\rho_{i,j,k}$, respectively. 
For the cylindrical case, the same decomposition applies with the indices $j$ and $k$ interchanged to account for the radial-azimuthal-vertical ordering.

Application of Eq.~\eqref{eq:fourier_decomposition} to the three-dimensional (3D) discrete Poisson equations in spherical (Eq.~\ref{eq:discrete_poisson_sph}) and cylindrical (Eq.~\ref{eq:discrete_poisson_cyl}) geometries reduces the problem to a set of independent two-dimensional (2D) Helmholtz equations, one for each Fourier mode, given by
\begin{align}
    \nabla^2_m\Phi^m = &\; \left(\nabla^2_r + \nabla^2_\theta + \lambda^{m}_{i,j}\right)\Phi^m_{i,j} = 4\pi G\rho^m_{i,j}, \label{eq:helmholz_sph}\\
    \nabla^2_m\Phi^m =&\; \left(\nabla^2_R + \nabla^2_z + \gamma^{m}_{i,k}\right)\Phi^m_{i,k} = 4\pi G\rho^m_{i,k}, \label{eq:helmholz_cyl}
\end{align}
for the spherical and cylindrical cases, respectively. The corresponding eigenvalues of the discrete $\nabla^2_\phi$ operator are given by
\begin{align}
    \lambda_{i,j}^m =& -\mathcal{R}_i\mathcal{T}_j \left[\sin{\left(\frac{m\pi}{N_\phi}\right)}/\left(\frac{\pi}{N_\phi}\right)\right]^2, \label{eq:eigenv_sph}\\
    \gamma_{i,k}^m =& -\mathcal{W}_i \left[\sin{\left(\frac{m\pi}{N_\phi}\right)}/\left(\frac{\pi}{N_\phi}\right)\right]^2, \label{eq:eigenv_cyl}
\end{align}
for the spherical and cylindrical cases, respectively.

Eq.~\eqref{eq:helmholz_sph} and \eqref{eq:helmholz_cyl} lead to a set of $N_\phi$ Fourier modes, each of which corresponds to a Helmholtz equation defined on a two-dimensional grid.
In spherical geometry, the grid spans the $(r,\theta)$ plane, while in cylindrical geometry it spans the $(R,z)$ plane.

For a real-valued gravitational potential, however, the Fourier coefficients satisfy the Hermitian symmetry
\begin{equation}\label{eq:Hermitian_symmetry}
    \Phi^{\,N_\phi - m}_{i,j} = \left(\Phi^{\,m}_{i,j}\right)^{*},
    \qquad m = 1,2,\ldots,\frac{N_\phi}{2}-1,
\end{equation}
which implies that the upper half of the spectrum is uniquely determined by the lower half.
Therefore, the solver only computes the independent modes $m = 0,1,\ldots,N_\phi/2$, while the remaining coefficients are reconstructed using the Hermitian symmetry during the inverse Fourier transform.
This reduces the computational cost without any loss of physical information.

\section{Multigrid solver 2D Helmholz equation}
\label{sec:multigrid_solver}
The multigrid method \citep{Brandt_1977, Trottenberg_2000} provides an efficient framework for solving discretized elliptic equations by combining relaxation schemes with a hierarchy of grids of increasing mesh spacing. 
For a linear elliptic operator $\mathcal{L}$, the discrete system on a grid with spacing $h$ is written as
\begin{equation}
    \mathcal{L}_h \Phi_h = f_h,
\end{equation}
where $f_h$ denotes the discretized source term.
Given an approximate solution $\Psi_h$, the defect or residual
\begin{equation}
    \mathscr{D}_h = \mathcal{L}_h \Psi_h - f_h
\end{equation}
quantifies the local error of the discrete solution. Owing to the linearity of $\mathcal{L}$ , the associated error $\mathscr{E}_h = \Phi_h-\Psi_h$ satisfies
\begin{equation}\label{eq:mg_error_equation}
    \mathcal{L}_h \mathscr{E}_h = -\mathscr{D}_h ,
\end{equation}
which forms the foundation of the correction–scheme multigrid approach.

The efficiency of multigrid stems from its scale-aware treatment of errors. 
Relaxation schemes efficiently reduce short-wavelength error components but leave long-wavelength modes largely unaffected. Multigrid explicitly addresses this limitation by transferring residuals to coarser grids through restriction operators, where long-wavelength errors on the fine grid are represented by fewer degrees of freedom and can therefore be reduced more efficiently. 
The resulting coarse-grid error estimate is interpolated back to the fine grid using prolongation operators and applied as a correction,
\begin{equation}
    \Psi_h^{\mathrm{new}} = \Psi_h + \tilde{\mathscr{E}}_h .
\end{equation}
Therefore, the construction and accuracy of these grid-transfer operators play a central role in multigrid performance and are discussed in detail later.

At each grid level, the choice of smoother governs how efficiently oscillatory errors are damped. 
While point-wise relaxation methods can be effective on isotropic Cartesian grids, their performance deteriorates in non-Cartesian geometries or in the presence of strong directional couplings \cite[e.g.,][]{Briggs_2000,Trottenberg_2000}. 
In such situations, line-based or block-based smoothers provide substantially improved convergence by coupling updates along preferred grid directions, as described in detail in Sec.~\ref{sec:line_smoother}.

These components -- smoothing, restriction, coarse-grid correction, and prolongation -- are organized into a multigrid cycle. 
In the widely used V-cycle, a small number of smoothing steps is applied while descending to successively coarser grids, followed by correction and additional smoothing as the solution is propagated back to the finest level. 
On the coarsest grid, the error equation is solved using a fast iterative or direct method, which forms the computational kernel of the multigrid algorithm; which is described in Sec.~\ref{sec:coarse_solver}. 
Repeated application of multigrid cycles yields rapid convergence with a computational cost that scales nearly linearly with the number of unknowns \citep{Briggs_2000,Trottenberg_2000}.

\begin{figure*}
    \centering
     \begin{subfigure}{0.32\textwidth}
         \centering
         \includegraphics[width=\textwidth]{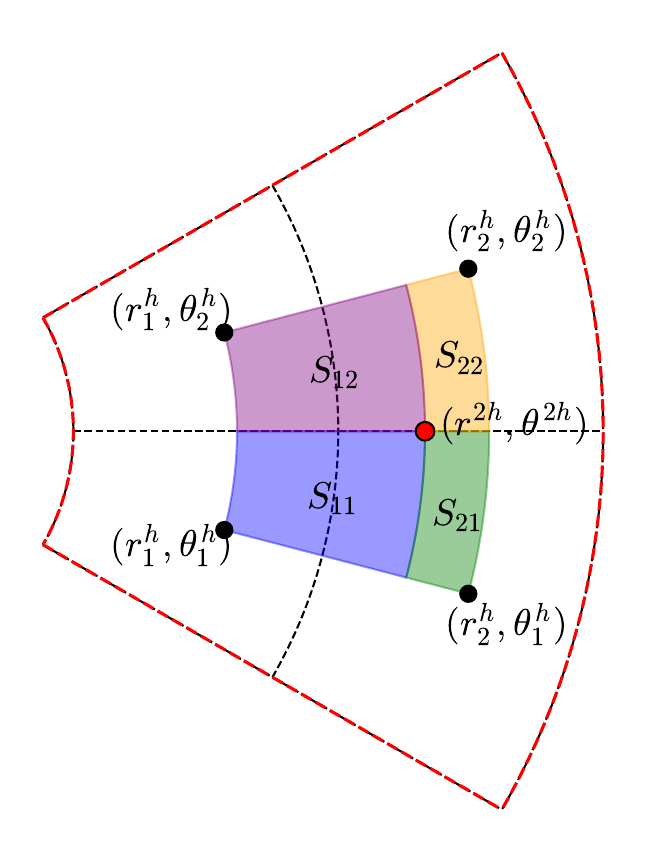}
         \caption{}
         \label{fig:restriction_sph}
     \end{subfigure}
     \hspace{1cm}
     \begin{subfigure}{0.45\textwidth}
         \centering
         \includegraphics[width=\textwidth]{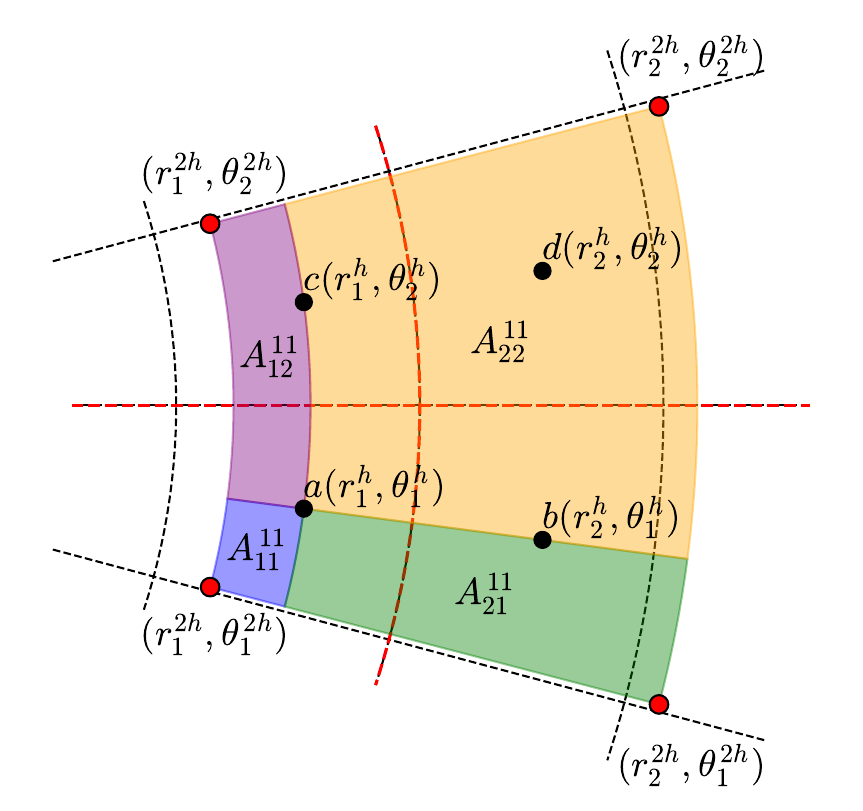}
         \caption{}
         \label{fig:prolongation_sph}
     \end{subfigure}
    \caption{Schematic diagrams of the area-weighted grid transfer operators in spherical coordinates. (a) Restriction operator $\mathscr{R}$: A coarse-grid point ($r^{2h},\theta^{2h}$) (red circle) receives contributions from four surrounding fine-grid points ($r^h_i,\theta^h_j$) (black circles), weighted by the overlapping shaded areas $S_{ij}$. The coarse-grid residual is computed as an area-weighted average of the fine-grid residuals. 
    (b) Prolongation operator $\mathscr{P}$: The interpolation of a fine-grid value from its four surrounding coarse-grid points $(r^{2h}_i, \theta^{2h}_j)$ (red circles) is shown for the example of fine point $a(r^h_1, \theta^h_1)$. The interpolation weight $A^{11}_{ij}$ for this point is proportional to the overlapping shaded area between its dual cell (the region associated with point $a$) and each coarse cell. The same area-weighting principle applies analogously to the other fine points $b$, $c$, and $d$, with weights $A^{mn}_{ij}$ determined by their respective geometric overlaps.}
    \label{fig:stencil}
\end{figure*}

\subsection{Restriction and Prolongation operators}
\label{sec:transfer_operator}
Having outlined the multigrid V-cycle algorithm, we now detail the grid transfer operators essential for communication between different grid levels. 
The effectiveness of a multigrid solver depends critically on these operators, which must preserve the mathematical properties of the underlying problem while efficiently transferring information between grids. 
In our implementation, we employ area-weighted restriction and bilinear interpolation for prolongation, adapted for curvilinear coordinates. 
These operators are constructed to respect the geometry of our spherical (or cylindrical) coordinate system, ensuring that coarse-grid corrections accurately represent the fine-grid error.

The restriction operator $\mathscr{R}$ transfers residuals from a fine grid to a coarse grid. A proper restriction scheme should maintain the integral consistency of the residual field. 
We implement an area-weighted restriction scheme, where the coarse-grid residual at a point is computed as the weighted average of the four surrounding fine-grid residuals, with weights proportional to the overlapping areas in the $(r,\theta)-$plane.

Fig.~\ref{fig:restriction_sph} schematically illustrates this in spherical coordinates.
The coarse-grid point ($r^{2h},\theta^{2h}$) (red circle) lies at the center of a coarse cell, surrounded by four fine-grid points ($r_1^h,\theta_1^h$), ($r_2^h,\theta_1^h$), ($r_1^h,\theta_2^h$), and ($r_2^h,\theta_2^h$) (black circles). 
The contribution of each fine-grid residual to the coarse-grid value is weighted by the shaded area $S_{ij}$ in the figure, representing the geometric overlap between the fine and coarse cells.

Mathematically, the restriction operation is defined as:
\begin{equation}\label{eq:restriction_sph}
    \begin{aligned}
        \mathscr{D}_{2h}(r^{2h},\theta^{2h})
        = & \mathscr{R}_{2h}^h \, \mathscr{D}_{h}(r^{h},\theta^{h}) \\
        = & \frac{1}{S_T} \Big[
        S_{11}\mathscr{D}_h(r^h_1,\theta^h_1)
        + S_{21}\mathscr{D}_h(r^h_2,\theta^h_1) \\
        &
        \qquad + S_{12}\mathscr{D}_h(r^h_1,\theta^h_2)
        + S_{22}\mathscr{D}_h(r^h_2,\theta^h_2)
        \Big],
    \end{aligned}
\end{equation}
where the area weights in spherical coordinates are
\begin{displaymath}\label{eq:restriction_weight_sph}
    S_{ij} = \left|(r^h_i)^2 - (r^{2h})^2\right|\times\left|\theta^h_j - \theta^{2h}\right|,
\end{displaymath}
and the normalization factor is
\begin{equation}\label{eq:restriction_norm_sph}
    S_T = S_{11} + S_{21} + S_{12} + S_{22}.
\end{equation}
This ensures $\sum S_{ij}/S_T = 1$, preserving the total defect during restriction.

For cylindrical grid ($R, z$), the same area-weighting scheme applies with the weights given by
\begin{equation}\label{eq:restriction_weight_cyl}
    S_{ik} = \left|R^h_i - R^{2h}\right|\times\left|z^h_k-z^{2h}\right|.
\end{equation}

The prolongation operator $\mathscr{P}$ interpolates corrections from a coarse grid back to a fine grid. 
We employ a bilinear interpolation scheme that uses the four nearest coarse-grid points to determine the value at a fine-grid location. 
This approach ensures smoothness and continuity of the corrected solution across grid boundaries.

Fig.~\ref{fig:prolongation_sph} depicts the prolongation process in spherical coordinates for a representative fine-grid point. 
The figure illustrates the case for point $a(r_1^h, \theta_1^h)$, which receives contributions from the four surrounding coarse-grid points $(r_1^{2h}, \theta_1^{2h})$, $(r_2^{2h}, \theta_1^{2h})$, $(r_1^{2h}, \theta_2^{2h})$, and $(r_2^{2h}, \theta_2^{2h})$. 
The same principle applies to any fine point $(r^h_m, \theta^h_n)$. The interpolation weights $A^{mn}_{ij}$ are area-based, derived from the overlapping regions between the fine point's dual cell and the coarse-grid cells. 
In the figure, the shaded regions correspond specifically to the weights for evaluating the interpolation at point $a$.

The prolongation is computed as:
\begin{equation}\label{eq:prolongation_sph}
    \begin{aligned}
    \mathscr{E}_h(r^h_m,\theta^h_n) = & \mathscr{P}^{2h}_{h, mn}\mathscr{E}_{2h}(r^{2h}, \theta^{2h})\\
     =& \frac{1}{A_T}\Big[A^{mn}_{22}\mathscr{E}_{2h}(r_1^{2h},\theta_1^{2h}) + A^{mn}_{12}\mathscr{E}_{2h}(r_2^{2h},\theta_1^{2h}) \\
    & \hspace{0.5cm} + A^{mn}_{21}\mathscr{E}_{2h}(r_1^{2h},\theta^{2h}_2) + A^{mn}_{11}\mathscr{E}_{2h}(r_2^{2h},\theta^{2h}_2)\Big],
    \end{aligned}
\end{equation}
with the area weights in spherical coordinates given by
\begin{equation}\label{eq:prolongation_weight_sph}
    A_{ij}^{mn} = \left|(r^{2h}_i)^2-(r^h_m)^2\right|\times\left|\theta_{j}^{2h} - \theta^h_n\right|,
\end{equation}
and the normalization factor
\begin{equation}\label{eq:prolongation_norm_sph}
\begin{aligned}
        A_T &= A_{11}^{mn} + A_{21}^{mn} + A_{12}^{mn} + A_{22}^{mn} \\
        &= \left|(r_2^{2h})^2 - (r_1^{2h})^2\right|\times\left|\theta_2^{2h}-\theta_1^{2h}\right|.
\end{aligned}
\end{equation}
In cylindrical coordinates ($R, z$), the corresponding weights are
\begin{equation}\label{eq:prolongation_weight_cyl}
    A^{mn}_{ik} = \left|R^{2h}_i - R^h_m\right|\times\left|z^{2h}_k-z^h_n\right|,
\end{equation}
with
\begin{equation}\label{eq:prolongation_norm_cyl}
    A_T = \left|R^{2h}_2-R^{2h}_1\right|\times\left|z^{2h}_2-z^{2h}_1\right|.
\end{equation}

\subsection{Directional smoothing for anisotropy}
\label{sec:line_smoother}
In curvilinear coordinate systems, such as spherical or cylindrical geometries, the discretization of elliptic operators often introduces strong anisotropic coupling along one coordinate direction. 
This anisotropy arises from grid stretching near coordinate singularities (e.g., the origin or polar axis) and from metric terms that scale differently with radius. 
Pointwise relaxation methods, which are highly effective in isotropic Cartesian grids, perform poorly under such conditions because they inefficiently damp error modes aligned with the strongly coupled direction \cite[e.g.,][]{Briggs_2000, Trottenberg_2000}.
To address this limitation, line relaxation is commonly employed, wherein unknowns along the direction of strong coupling are solved simultaneously, typically by inverting a tridiagonal (or banded) system. 
This approach provides effective smoothing of error components that are poorly damped by pointwise methods and is essential for maintaining multigrid convergence in non-Cartesian domains.

To illustrate this, consider the discretized form of the Helhmholtz equation (Eq.~\ref{eq:helmholz_sph}) in spherical coordinates :
\begin{equation}
    a\Phi^m_{i+1,j} + b\Phi^m_{i-1,j} + c\Phi^m_{i,j} + d\Phi^m_{i,j+1} + e\Phi^m_{i,j-1} = 4\pi G\rho^m_{i,j}.
\end{equation}
In such grids, the radial coupling often dominates, particularly near the origin or in regions with large grid aspect ratios. In principle, line relaxation would require solving all points along a radial line (constant $\theta$) simultaneously via a tridiagonal solver (e.g., Thomas Algorithm). However, performing such direct solvers in parallel environments can be computationally expensive and may introduce significant communication bottlenecks between processors.

Instead, we adopt an approximate line smoother in which line relaxation is performed using a limited number of Successive Over-Relaxation (SOR) iterations. 
For each grid line, these iterations produce a solution that is sufficiently close to the exact line-relaxation result for effecient smoothing, while avoiding the computational overhead associated with direct tridiagonal solver. 
Therefore, in this work, line relaxation constitutes the multigrid smoother, while SOR is used only as an inner iterative method to approximately solve the resulting one-dimensional systems, replacing a direct tridiagonal solver.

The performance of this method depends critically on the choice of the over-relaxation parameter, $\omega$. 
The classical theory for determining the optimal SOR parameter assumes a uniform grid with constant coefficients, under which the eigenvectors of the iteration matrix correspond to discrete Fourier modes \citep{Young_1954, Press_1992}. 
However, this assumption breaks down on a non-uniform curvilinear grid, where metric terms and grid stretching introduce strong spatial variations and the eigenvectors are not known analytically. 
Consequently, determining $\omega_{\mathrm{opt}}$ requires approximating both the eigenvectors and the corresponding spectral radius. 
In Appendix~\ref{sec:SOR_omega}, we carry out this analysis and derive an approximate expression for the optimal over-relaxation parameter, $\omega_{\mathrm{opt}}$, using a Wentzel–Kramers–Brillouin (WKB) approach adapted to the discrete system. 
In practice, using this value of $\omega$ together with only a few SOR iterations per line allows the smoother to efficiently damp the dominant anisotropic error modes.

For approximate $r-$line relaxation, the update along each radial line is given by
\begin{equation}
    \begin{aligned}
     \Phi^{\rm tmp}_{i,j}  =\;& \frac{1}{c}\left[f^m_{i,j} - a\Phi^m_{i+1,j} - b\Phi^m_{i-1,j}\right], \\
     \Phi^{m,\rm new}_{i,j} =\; & (1-\omega_j)\Phi^{m}_{i,j} + \omega_j \Phi^{\rm tmp}_{i,j},
    \end{aligned}
\end{equation}
where $\omega_j$ is the over-relaxation parameter for the $j^{\rm th}$ line. The right hand-side containing the source term and the contribution from the other direction, $f^m_{i,j} = 4\pi G\rho^m_{i,j} - d\Phi^m_{i,j+1} - e\Phi^m_{i,j-1}$, is kept fixed during the relaxation process. The sweeps are performed radially from $i=1$ to $i=N_r$.

Similarly, for approximate $\theta-$line smoothing,
\begin{equation}
    \begin{aligned}
         \Phi^{\rm tmp}_{i,j}   =&\; \frac{1}{c}\left[g^m_{i,j} - d\Phi^m_{i,j+1} - e\Phi^m_{i,j-1}\right], \\
         \Phi^{m,\rm new}_{i,j} =&\; (1-\omega_i)\Phi^{m}_{i,j} + \omega_i \Phi^{\rm tmp}_{i,j},
    \end{aligned}
\end{equation}
with $g^m_{i,j} = 4\pi G\rho^m_{i,j} - a\Phi^m_{i+1,j} - b\Phi^m_{i-1,j}$, sweeping along $\theta$ from $j=1$ to $j=N_\theta$.

These approximate line relaxations are embedded within an alternating line smoothing strategy. 
One smoothing cycle consists of a complete relaxation along all radial ($r$) lines, followed by relaxation along all angular ($\theta$) lines. 
At each multigrid level, a small number of such smoothing cycles (typically 2–3) is applied before transferring the residual/error solution to the next coarser/finer grid. 
This hierarchical approach retains much of the smoothing efficiency of true line relaxation while preserving the computational efficiency and scalability required for multigrid solvers on curvilinear grids.

\begin{algorithm}
    \caption{Alternating line smoother}
    \begin{algorithmic}[1]
    \STATE Given solution $u$ and right-hand side $f$ at a multigrid level
    \FOR{$n = 1 \text{ to } N_{\mathrm{smooth}}$}
        \FOR{each angular grid line $r = \mathrm{const}$}
            \STATE Perform $N_{\mathrm{SOR}}$ SOR iterations along the line
        \ENDFOR
        \FOR{each radial grid line $\theta = \mathrm{const}$}
            \STATE Perform $N_{\mathrm{SOR}}$ SOR iterations along the line
        \ENDFOR
    \ENDFOR
    \STATE Return updated solution $u$
    \end{algorithmic}
\end{algorithm}

This similar strategy is also used for Cylindrical grid using Eq.~\eqref{eq:helmholz_cyl}.

\subsection{Coarsest grid solver}
\label{sec:coarse_solver}
At the coarsest level of the multigrid hierarchy, the error equation (Eq.~\ref{eq:mg_error_equation}) should be solved efficiently to higher accuracy while preserving overall scalability. Since the system size is small, the solver only needs to rapidly eliminate the remaining smooth (low-frequency) error components, forming the \emph{kernel} of the algorithm \cite[e.g.,][]{Briggs_2000}.

While direct solvers could be used, their global communication overhead and reduced parallel efficiency often outweigh their advantage in distributed environments. Therefore, we employ a pointwise SOR scheme, which is simple, scalable, and converges rapidly for small systems \citep[e.g.,][]{Varga_2009}.

The iteration for the $m^{\rm th}$ mode is given by
\begin{equation}
    \begin{aligned}
     \Phi_{i,j}^{\rm tmp} = & \frac{1}{c_{i,j}}\big[4\pi G\rho_{i,j}^m - a_{i,j}\Phi_{i+1, j}^m - b_{i,j}\Phi_{i-1, j}^m \\ 
     & \hspace{2cm} - d_{i,j}\Phi_{i, j+1}^m - e_{i,j}\Phi_{i, j-1}^m\big],\\
    \Phi^{m,\rm new}_{i,j} = & (1-\omega)\Phi^{m}_{i,j} + \omega\Phi_{i,j}^{\rm tmp},
    \end{aligned}
\end{equation}
where the relaxation parameter $\omega$ is chosen from the two-dimensional counterpart of Eq.~\eqref{eq:omega_opt_mu}.

\section{Exact Boundary potential calculation: James method}
\label{sec:boundary_calculation}
To obtain an accurate solution of the Poisson equation, it is crucial to specify the gravitational potential on the domain boundaries with high fidelity. 
In our implementation, we apply Dirichlet boundary conditions (BC) along the radial and polar (or vertical) directions, which naturally align with the physical requirements of isolated mass distributions. Over the years, various strategies have been proposed to determine these boundary values. 
A commonly adopted technique is the multipole expansion, in which the potential is decomposed into spherical harmonics and truncated at an appropriate maximum meridional mode $l_\mathrm{max}$ \cite[e.g., see][]{Katz_2016, Mandal_2021}.

While effective in many contexts, this method becomes unsuitable in spherical or cylindrical coordinate systems where an inner cavity is introduced to avoid the coordinate singularity at the origin. 
The presence of this central hole violates the core assumptions of the multipole expansion, causing the boundary evaluation to lose accuracy and consistency.

To overcome these limitations, \cite{Cohl_1999} introduced the compact cylindrical Green’s function (CCGF) expansion, an approach specifically designed to deliver high accuracy in geometries such as strongly flattened disks. 
The method provides significantly improved convergence properties compared to traditional multipole techniques. 
However, its efficiency deteriorates for strongly non-axisymmetric mass distributions, where the accuracy depends sensitively on the chosen maximum azimuthal mode 
$m_\mathrm{max}$. 
As $m_\mathrm{max}$ increases, both the computational cost and memory requirements grow rapidly, making the approach increasingly impractical for highly asymmetric systems \cite[e.g., see][]{Gressel_2024}.

A particularly powerful and general method for imposing vacuum boundary conditions is the four-step algorithm introduced by \cite{James_1977}. This approach is widely used in surface-potential calculations \cite[e.g.,][]{Magorrian_2007, Moon_2019, Gressel_2024, Ahn_2025} and is grounded in the correspondence between Newtonian gravity and electrostatics. In electrostatics, a charge distribution enclosed within a grounded conducting box induces surface “screening charges’’ that enforce zero potential on the boundaries. The resulting solution is the superposition of the potentials due to the interior charges and the induced surface charges.

Translating this analogy to gravity, we interpret the mass density as the analogue of charge density and denote the interior potential, the induced surface potential, and their sum by $\Phi$, $\Theta$, and $\Psi$ respectively. 
This leads to a clean and intuitive four-step procedure:
\begin{enumerate}
    \item Solve $\nabla^2\Psi(\boldsymbol{x}) = 4\pi G\rho(\boldsymbol{x})$ in the interior of the domain ($\boldsymbol{x}\in\Omega$) setting $\Psi|_{\del\Omega} = 0$ at the boundary, using the method described in Sec.~\ref{sec:multigrid_solver}.
    \item Compute the screening mass density
    \begin{equation}\label{eq:screen_mass}
        \sigma = \frac{1}{4\pi G}\nabla^2\Psi|_{\del\Omega}
    \end{equation}
    assuming $\Psi=0$ exterior to the domain.

    \item Evaluate the surface potential $\Theta$ using a convolution integral with a suitable Green's function (as described in Sec.~\ref{sec:DGF}), using $\sigma$ as the source term, e.g.,
    \begin{equation}\label{eq:convolution}
        \Theta(\boldsymbol{x}) = \oint_{\del\Omega}\mathcal{G}(\boldsymbol{x}-\boldsymbol{x'})\sigma(\boldsymbol{x'}) dS(\boldsymbol{x'})
    \end{equation}

    \item Solve the actual gravitational potential $\nabla^2\Phi = 4\pi G\rho$, with $\Phi|_{\del\Omega} = -\Theta|_{\del\Omega}$, again using the solver described in Sec.~\ref{sec:multigrid_solver}.
\end{enumerate}
This procedure requires two Poisson solves: one to determine $\Psi$ and thereby infer the boundary potential contribution, and a second to compute the final gravitational potential $\Phi$. 
In earlier applications \cite[e.g.,][]{Gressel_2024}, the full three-dimensional Poisson equation is solved in real space, and the convolution in step 3 requires Fourier transformation of the screening mass obtained in step 2.

In contrast, because we solve for each Fourier mode independently, the above procedure is applied separately to each mode. This approach naturally produces the Fourier coefficients of the screening mass and thereby eliminates the need for additional FFT operations between intermediate steps.

\subsection{Screening mass computation}
\label{sec:screening_mass}
For each $m$-mode, given that $\Psi^m$ is non-zero only at the interior and $\Psi^m=0$ otherwise, the corresponding screening mass 
\[\sigma^m = \frac{\nabla^2_m\Psi^m}{4\pi G}\] 
from Eq.~\eqref{eq:helmholz_sph} and Eq.~\eqref{eq:helmholz_cyl} for spherical and cylindrical coordinates, respectively. 
Therefore, The $m^{\rm th}$ Fourier mode of the screening mass at the four boundaries at $(i=0, j)$ (left), $(i=N_r+1, j)$ (right), $(i,j=0)$ (bottom) and $(i,j=N_\theta+1)$ (top) in spherical coordinate are calculated as:
\begin{align}
\sigma^{m}_{j,\rm L} =& \frac{1}{4\pi G}\frac{3r_{\HALF}^2}{(r_{\HALF}^3 - r_{-\HALF}^3)(r_{1}-r_{0})}\Psi^m|_{i=1}\nonumber\\
\sigma^{m}_{j,\rm R} =& \frac{1}{4\pi G}\frac{3r_{N_r+\HALF}^2}{(r_{N_r+\frac{3}{2}}^3 - r_{N_r+\HALF}^3)(r_{N_r+1}-r_{N_r})}\Psi^m|_{i=N_r}\nonumber\\
\sigma^{m}_{i,\rm B} =& \frac{1}{4\pi G}\frac{\mathcal{R}_i\sin{\theta_{\HALF}}}{(\cos{\theta_{-\HALF}} - \cos{\theta_{\HALF}})(\theta_{1}-\theta_{0})}\Psi^m|_{j=1} \label{eq:charge_sph}\\
\sigma^{m}_{i,\rm T} =& \frac{1}{4\pi G}\frac{\mathcal{R}_i\sin{\theta_{N_\theta+\HALF}}}{(\cos{\theta_{N_\theta+\HALF}} - \cos{\theta_{N_\theta+\frac{3}{2}}})(\theta_{N_\theta+1}-\theta_{N_\theta})}\Psi^m|_{j=N_\theta}\nonumber
\end{align}
Since $\Psi^m_{i,j}=0$, along the boundaries, the terms involving $\lambda_{i,j}^m$ in the Helmholtz equation (Eq.~\ref{eq:helmholz_sph}) do not contribute to the screening mass in Eq.~\eqref{eq:charge_sph}.

Similarly, For cylindrical grid, the $m^{\rm th}$ Fourier mode of the screening mass at the four boundaries ($i=0, N_r+1$; $k=0, N_z+1$) for $m^{\rm th}$ mode are computed as follows:
\begin{equation}\label{eq:charge_cyl}
    \begin{aligned}
        \sigma^{m}_{j,\rm L} = & \frac{2R_{\HALF}}{(R^2_{\HALF} - R^2_{-\HALF})(R_{1}-R_{0})}\Psi^m|_{i=1} \\
        \sigma^{m}_{j,\rm R} = & \frac{2R_{N_R+\HALF}}{(R^2_{N_R+\frac{3}{2}} - R^2_{N_R+\HALF})(R_{N_R+1}-R_{N_R})}\Psi^m|_{i=N_R}\\
        \sigma^{m}_{i,\rm B} = & \frac{1}{\Delta z_{0}(z_{1} - z_{0})}\Psi^m|_{k=1}\\
        \sigma^{m}_{i,\rm T} = & \frac{1}{\Delta z_{N_z+1}(z_{N_z+1} - z_{N_z})}\Psi^m|_{k=N_z}\\
    \end{aligned}
\end{equation}

\subsection{Discrete Green's function}
\label{sec:DGF}
To compute the boundary potential $\Theta$ generated by the screening mass $\sigma$, an appropriate Green’s function is required to evaluate the convolution (Eq.~\ref{eq:convolution}). As noted by \citet{Moon_2019}, one must use a discrete Green’s function (DGF) associated with the discrete Laplacian operator, rather than the continuous Green’s function, because the computational domain is composed of finite-volume cells. Therefore, the source is represented as a mass distributed over a finite cell volume, rather than as an idealized point mass.

Here, we adopt the approach of \citet{Moon_2019}, which has also been recently used by \citet{Ahn_2025}. The DGF, $\mathcal{G}_{i,i',j,j',k-k'}$, in spherical coordinates represents the gravitational potential at grid point $(i,j,k)$ due to a unit mass located at $(i',j',k')$. It satisfies the discrete Poisson equation
\begin{equation}\label{eq:DGF_full_sph}
    (\nabla^2_r + \nabla^2_\theta + \nabla^2_\phi)\mathcal{G}_{i,i',j,j',k-k'} = 4\pi G \frac{\delta_{ii'}\delta_{jj'}\delta_{kk'}}{\Delta V_{i',j',k'}},
\end{equation}
where $\delta_{ii'}$ is the Kronecker delta and $\Delta V_{i',j',k'}$ is the volume of the source cell. Applying FFT along the azimuthal direction, we obtain
\begin{equation}\label{eq:DGF_fft_sph}
    (\nabla^2_r + \nabla^2_\theta + \lambda_{i,j}^m)\mathcal{G}^m_{i,i',j,j'} = 4\pi G\frac{\delta_{ii'}\delta_{jj'}}{\Delta V_{i',j'}},
\end{equation}
where $\mathcal{G}^m_{i,i',j,j'}$ is the $m^{\rm th}$ Fourier mode of $\mathcal{G}_{i,i',j,j'k-k'}$.

Therefore, the Fourier transform of the potential generated by the screening mass is given by
\begin{equation}\label{eq:Theta_fft}
    \Theta^m_{i,j} =\sum_{i'=1}^{N_r}\sum_{j'=1}^{N_\theta} \mathcal{G}^m_{i,i',j,j'}\sigma^m_{i',j'}\Delta V_{i',j'}.
\end{equation}

To compute $\mathcal{G}^m_{i,i',j,j'}$ from Eq.~\eqref{eq:DGF_fft_sph}, we follow \citet{Moon_2019} and introduce an auxiliary grid of size $(2n_{\rm pad}+1) \times (2n_{\rm pad}+1)$ centered on each source location $(i',j')$ along the boundary of the primary computational domain \cite[see also Fig.~1 of][]{Gressel_2024}, respecting the original grid geometry. 
The density on this auxiliary grid is set to zero everywhere except at the central cell, where it is set to $1/\Delta V_{i',j'}$, corresponding to a unit mass distributed over the source cell volume.

We typically adopt $n_{\rm pad} = 16$, ensuring that the auxiliary grid extends sufficiently far from the source for the continuous Green’s function, $\mathcal{G}(\boldsymbol{x},\boldsymbol{x'}) = -G/|\boldsymbol{x}-\boldsymbol{x'}|$, to provide an accurate  Dirichlet boundary condition. Therefore, the boundary value of $\mathcal{G}^m_{i,i',j,j'}$ is given by
\begin{equation}\label{eq:DGF_bc_sph}
    \mathcal{G}^m_{i,i',j,j'} \approx -G \sum_{k=0}^{N_{\phi}-1}
    \frac{\DS\cos\left(\frac{2\pi mk}{N_\phi}\right)}
    {\DS\sqrt{A - B\cos\left(\frac{2\pi k}{N_\phi}\right)}},
\end{equation}
where
\begin{equation}
    \begin{aligned}
    A &= r_i^2 + r_{i'}^2 - 2r_i r_{i'}\cos\theta_j \cos\theta_{j'}, \\
    B &= 2r_i r_{i'}\sin\theta_j \sin\theta_{j'}.
    \end{aligned}
\end{equation}

With these boundary conditions, each mode of Eq.~\eqref{eq:DGF_fft_sph} is solved at the interior points of the auxiliary grid using a SOR method, as described in Sec.~\ref{sec:coarse_solver}, to obtain $\mathcal{G}^m_{i,i',j,j'}$ throughout the grid. 
The values corresponding to grid points $(i,j)$ on the original domain boundary that lie within the auxiliary grid are then extracted and stored. 
For boundary points outside the auxiliary grid, $\mathcal{G}^m_{i,i',j,j'}$ is directly evaluated using $m^{\rm th}$ Fourier mode of the continuous Green’s function (Eq.~\ref{eq:DGF_bc_sph}).

A similar procedure is applied in cylindrical coordinates, where the $m^{\mathrm{th}}$ Fourier mode of the discrete Green’s function is obtained by solving the cylindrical Helmholtz equation,
\begin{equation}\label{eq:DGF_fft_cyl}
    (\nabla^2_r + \nabla^2_z + \gamma_{i,k}^m)\mathcal{G}^m_{i,i',k,k'} = 4\pi G \frac{\delta_{ii'}\delta_{kk'}}{\Delta V_{i',k'}}.
\end{equation}
The boundary values on the auxiliary grid are specified using,
\begin{equation}\label{eq:DGF_bc_cyl}
    \mathcal{G}^m_{i,i',k,k'} \approx -G \sum_{j=0}^{N_{\phi}-1}
    \frac{\DS\cos\left(\frac{2\pi mj}{N_\phi}\right)}{\DS\sqrt{R_i^2 + R_{i'}^2 - 2R_i R_{i'}\cos\left(\frac{2\pi j}{N_\phi}\right) + (z_k - z_{k'})^2}}.
\end{equation}

For logarithmic grids with large radial extent ratios (i.e., $r_{\rm max}/r_{\rm min}$ or $R_{\rm max}/R_{\rm min} \gtrsim 10^2$), complications can arise, as noted by \citet{Moon_2019}. Near the inner boundary, where the radial cells are extremely narrow, 16 auxiliary cells are insufficient to apply Eqs.~\eqref{eq:DGF_bc_sph} or \eqref{eq:DGF_bc_cyl} as Dirichlet boundary conditions. \citet{Moon_2019} demonstrated that, in this case, using the radial derivative of these equations as a Neumann boundary condition at the inner boundary provides sufficiently accurate DGF values within the original computational domain, a result we have also verified for both spherical and cylindrical grids.

To implement this, instead of constructing a small auxiliary grid at each boundary point, we extend the computational domain. If $N_r$ is the number of radial cells in the original grid, we add $N_r-16$ cells at the inner radial boundary and 16 cells at the outer boundary. In the $\theta$ or $z$ directions, we add $N_\theta/2$ or $N_z/2$ cells at the upper and lower boundaries. For each boundary point in the original domain, a corresponding source is placed in the extended domain, and Eq.~\eqref{eq:DGF_fft_sph} or \eqref{eq:DGF_fft_cyl} is solved using the Multigrid method described in Sec.~\ref{sec:multigrid_solver} for spherical or cylindrical coordinates.

The boundary conditions at the outer, upper, and lower boundaries are Dirichlet, calculated from Eqs.~\eqref{eq:DGF_bc_sph} or \eqref{eq:DGF_bc_cyl}, while the inner radial boundary uses the radial derivative as a Neumann condition. In spherical coordinates, this derivative is given by
\begin{equation}
    \frac{\del \mathcal{G}^m_{i,i', j,j'}}{\del r} = G \sum_{k=0}^{N_\phi}
    \frac{\DS\left[C - D\cos\left(\frac{2\pi k}{N_\phi}\right)\right] \cos\left(\frac{2\pi mk}{N_\phi}\right)}{\DS\left[A - B\cos\left(\frac{2\pi k}{N_\phi}\right)\right]^{3/2}},
\end{equation}
with
\begin{equation}
C = r_i - r_{i'}\cos\theta_j \cos\theta_{j'}, \quad
D = r_{i'} \sin\theta_j \sin\theta_{j'},
\end{equation}
and in cylindrical coordinates,
\begin{equation}
    \frac{\del \mathcal{G}^m_{i,i', k,k'}}{\del R} \approx G \sum_{j=0}^{N_\phi}
    \frac{\DS\left[R_i - R_{i'}\cos\left(\frac{2\pi j}{N_\phi}\right)\right] \cos\left(\frac{2\pi mj}{N_\phi}\right)}{\DS\left[R_i^2 + R_{i'}^2 - 2 R_i R_{i'} \cos\left(\frac{2\pi j}{N_\phi}\right) + (z_k - z_{k'})^2\right]^{3/2}}.
\end{equation}

Once we obtain $\mathcal{G}^m_{i,i',j,j'}$, the potential at the four boundaries for the $m^{\rm th}$ mode can be calculated using Eq.~\eqref{eq:convolution} and is given by,
\begin{align}
    &
    \begin{aligned}\label{eq:theta_left}
        \Theta^{m}_{j, \rm L} = & \sum_{j'=1}^{N_\theta}\left[\tilde{\mathcal{G}}^m_{0,0,j,j'}\sigma^{m}_{j',\rm L}
        + \tilde{\mathcal{G}}^m_{0,N_r+1,j,j'}\sigma^{m}_{j', \rm R}\right] \\
        & + \sum_{i'=1}^{N_r}\left[\tilde{\mathcal{G}}^m_{0,i',j,0}\sigma^{m}_{i',\rm B} 
        + \tilde{\mathcal{G}}^m_{0,i',j,N_\theta+1}\sigma^{m}_{i',\rm T}\right],
    \end{aligned}
    \\
    &
    \begin{aligned}\label{eq:theta_right}
        \Theta^{m}_{j, \rm R} = & \sum_{j'=1}^{N_\theta}\left[\tilde{\mathcal{G}}^m_{N_r+1,0,j,j'}\sigma^{m}_{j',\rm L}
        + \tilde{\mathcal{G}}^m_{N_r+1,N_r+1,j,j'}\sigma^{m}_{j', \rm R}\right] \\
        & + \sum_{i'=1}^{N_r}\left[\tilde{\mathcal{G}}^m_{N_r+1,i',j,0}\sigma^{m}_{i',\rm B}
        + \tilde{\mathcal{G}}^m_{N_r+1,i',j,N_\theta+1}\sigma^{m}_{i',\rm T}\right],
    \end{aligned}
    \\
    &
    \begin{aligned}\label{eq:theta_bot}
        \Theta^{m}_{i, \rm B} = & \sum_{j'=1}^{N_\theta}\left[\tilde{\mathcal{G}}^m_{i,0,0,j'}\sigma^{m}_{j',\rm L}
        + \tilde{\mathcal{G}}^m_{i,N_r+1,0,j'}\sigma^{m}_{j', \rm R}\right] \\
        & + \sum_{i'=1}^{N_r}\left[\tilde{\mathcal{G}}^m_{i,i',0,0}\sigma^{m}_{i',\rm B} 
        + \tilde{\mathcal{G}}^m_{i,i',0,N_\theta+1}\sigma^{m}_{i',\rm T}\right],
    \end{aligned}
    \\
    &
    \begin{aligned}\label{eq:theta_top}
        \Theta^{m}_{i, \rm T} = & \sum_{j'=1}^{N_\theta}\left[\tilde{\mathcal{G}}^m_{i,0,N_\theta+1,j'}\sigma^{m}_{j',\rm L}
        + \tilde{\mathcal{G}}^m_{i,N_r+1,N_\theta+1,j'}\sigma^{m}_{j', \rm R}\right] \\
        & +  \sum_{i'=1}^{N_r}\left[\tilde{\mathcal{G}}^m_{i,i',N_\theta+1,0}\sigma^{m}_{i',\rm B} 
        + \tilde{\mathcal{G}}^m_{i,i',N_\theta+1,N_\theta+1}\sigma^{m}_{i',\rm T}\right],
    \end{aligned}
\end{align}
where we define $\tilde{\mathcal{G}}^m_{i,i',j,j'} = \mathcal{G}^m_{i,i',j,j'}\Delta V_{i',j'}$.

For a spherical grid covering the full polar range $\theta\in[0,\pi]$, the poles correspond to coordinate singularities rather than physical boundaries. Therefore, Dirichlet boundary conditions such as those given by Eqs.~\eqref{eq:theta_bot} and \eqref{eq:theta_top} are not imposed at 
$\theta=0$ and $\theta=\pi$. Instead, the gravitational potential must remain single-valued when approaching the pole from different azimuthal directions. This requirement leads to the symmetry condition
\begin{equation}\label{eq:polar_bc}
    \Phi(r,0,\phi) = \Phi(r, 0, \phi+\pi)
\end{equation}
and an analogous relation at $\theta=\pi$.

Although the active computational cells are located at cell centers within the interval $0<\theta<\pi$, the discrete Laplacian stencil requires values outside this range when evaluating derivatives near the poles. In a finite-volume discretization, this introduces ghost cells in the regions $\theta<0$ and $\theta>\pi$, which must be populated in a manner consistent with the polar symmetry. In spherical coordinates, points across the pole satisfy the geometric relation
\begin{equation}\label{eq:geomtric}
    \begin{aligned}
        (r,-\Delta\theta,\phi) &\equiv (r,\Delta\theta,\phi+\pi), \\
        (r,\pi+\Delta\theta,\phi) &\equiv (r,\pi-\Delta\theta,\phi+\pi)
    \end{aligned}
\end{equation}
meaning that a small displacement across the pole corresponds to a shift of the azimuthal angle by $\pi$.

After applying a Fourier transform in the azimuthal direction, the above symmetry relation in Eq.~\eqref{eq:polar_bc}), together with the geometric relation in Eq.~\eqref{eq:geomtric}, translates into a parity relation for each Fourier mode,
\begin{equation}
    \begin{aligned}
        \Phi^m(r,-\Delta\theta) &= (-1)^m\Phi^m(r,\Delta\theta), \\
        \Phi^m(r,\pi+\Delta\theta) &= (-1)^m\Phi^m(r,\pi-\Delta\theta).
    \end{aligned}
\end{equation}
This relation provides the values required in the ghost cells used by the discrete Laplacian near the poles. Therefore, the ghost-cell values at the $\theta$-boundary in the multigrid solver are set as follows:
\begin{equation}
    \Phi^m_{i,0} = (-1)^m\Phi^m_{i,1}, \quad \text{ and }\quad \Phi^m_{i,N_\theta+1} = (-1)^m\Phi^m_{i, N_\theta}.
\end{equation}
This parity condition ensures regularity of the solution at the poles.
\begin{table*}
\caption{Parameters for mesh-segment test.}
\label{tab:parameter_meshseg}
\centering

\setlength{\tabcolsep}{5pt}
\renewcommand{\arraystretch}{1.2}

\begin{tabular*}{\textwidth}{@{\extracolsep{\fill}}lcccccccccc}
\hline\hline
 & \multicolumn{6}{c}{Source parameters} & \multicolumn{4}{c}{Computational domain} \\
\cline{2-7} \cline{8-11}
Coordinate system 
& $r_1/R_1$ & $r_2/R_2$ & $\theta_1/z_1$ & $\theta_2/z_2$ & $\phi_1$ & $\phi_2$ 
& $r_{\min}/R_{\min}$ & $r_{\max}/R_{\max}$ & $\theta_{\min}/z_{\min}$ & $\theta_{\max}/z_{\max}$ \\
\hline

Spherical logarithmic   
& $\sqrt{2}$ & $2\sqrt{2}$ & $0.38\pi$ & $0.54\pi$ & $0.0$ & $0.25\pi$ 
& $1.0$ & $4.0$ & $0.34\pi$ & $0.66\pi$ \\

Cylindrical logarithmic 
& $\sqrt{2}$ & $2\sqrt{2}$ & $-0.9375$ & $0.9375$ & $0.0$ & $0.25\pi$ 
& $1.0$ & $4.0$ & $-1.5$ & $1.5$ \\

\hline

Spherical uniform       
& $2.1$ & $3.2$ & $0.38\pi$ & $0.54\pi$ & $0.0$ & $0.25\pi$ 
& $1.9$ & $3.5$ & $0.34\pi$ & $0.66\pi$ \\

Cylindrical uniform     
& $2.1$ & $3.2$ & $-0.6$ & $0.6$ & $0.0$ & $0.25\pi$ 
& $1.9$ & $3.5$ & $-1.2$ & $1.2$ \\

\hline
\end{tabular*}
\tablecomments{In all tests, the azimuthal domain $\phi$ spans from $0$ to $2\pi$.}
\end{table*}

\section{Algorithm and parallelization strategy}
\label{sec:algo_par}
\subsection{Algorithm overview}
\label{sec:algorithm_overview}
This section summarizes the complete workflow of the method presented in this paper. 
The procedure consists of a preprocessing stage, performed once at the beginning of the simulation, followed by the main solution phase, which is executed to compute the gravitational potential from a given density distribution. 
This solver is implemented as a self-gravity module in the \textsc{Pluto} code \citep{Mignone_2007, Mignone_2012}, where it is invoked during the simulation to compute the gravitational potential consistently with the evolving density field. 
The individual steps of the algorithm are outlined below:
\begin{enumerate}
    \item At the beginning of each simulation, we compute the DGF using the method described in Sec.~\ref{sec:DGF}. Since the DGFs depend only on the computational grid, they are computed once and stored in memory for reuse throughout the simulation.

    \item Perform a forward FFT of the density field $\rho_{i,j,k}$ to obtain the Fourier components $\rho^m_{i,j}$, which requires a computational cost of $\mathcal{O}[N_rN_\theta N_\phi\log N_\phi]$.

    \item Solve $\nabla^{2}_m\Psi^m = 4\pi\rho^m$ for $m = 0, 1,\cdots, N_{\phi/2}$ using the 2D multigrid method described in Sec.~\ref{sec:multigrid_solver}, with $\Psi^m = 0$ imposed at the domain boundaries. The remaining Fourier modes are reconstructed using the Hermitian symmetry (Eq.~\ref{eq:Hermitian_symmetry}). The overall computational cost scales as $\mathcal{O}[N_{\rm cycle} N_r N_\theta (N_\phi/2)]$, where $N_{\rm cycle}$ denotes the number of V-cycles per mode. This quantity is mode-dependent, with higher-$m$ modes typically converging in fewer cycles (see Appendix~\ref{sec:multigrid_convergence}).

    In the ideal case of constant-coefficient elliptic problems, multigrid convergence is expected to be independent of resolution \citep{Braess_1983, Hackbusch_1985, Bramble_1993}. However, in the presence of directional anisotropies -- as in the present problem, which necessitates the use of line smoothers -- the convergence rate becomes weakly resolution-dependent \citep[e.g., see Chapter~5 of][]{Trottenberg_2000}. Therefore, the number of V-cycles required to reach a prescribed tolerance may vary with resolution, and no general consensus exists on its precise scaling.

    \item Compute the screening mass using Eq.~\eqref{eq:charge_sph} or Eq.~\eqref{eq:charge_cyl}, depending on whether spherical or cylindrical coordinates are used, with a computational cost of $\mathcal{O}[(N_r+N_\theta) (N_\phi/2)]$

    \item Compute the boundary potential ($\Theta^m$) using Eqs.~\eqref{eq:theta_left}--\eqref{eq:theta_top}.For each Fourier mode, the potential at each boundary point is obtained by summing over all boundary source points weighted by the precomputed DGF. With $\mathcal{O}[N_r+N_\theta]$ boundary points, the direct convolution incurs a cost of $\mathcal{O}[(N_r+N_\theta)^2(N_\phi/2)]$ operations.

    \item Use $-\Theta^m$ as the Dirichlet boundary condition and solve $\nabla^{2}_m\Phi^m = 4\pi\rho^m$ using the multigrid method. The computational cost is the same as step 3.

    \item Perform an inverse FFT of $\Phi^m_{i,j}$ to obtain the full three-dimensional gravitational potential $\Phi_{i,j,k}$ in real space, with a computational cost of $\mathcal{O}[N_rN_\theta N_\phi\log N_\phi]$ operations.
\end{enumerate}
To accelerate convergence, the solutions $\Psi^m$ and $\Phi^m$ from the previous timestep are stored and used as initial guesses for the multigrid solver in steps 3 and 6. 
Since the gravitational potential typically evolves smoothly between timesteps, this strategy significantly reduces the number of multigrid iterations required for convergence. Within \textsc{Pluto}, this procedure is executed as part of the gravity update step, ensuring efficient coupling between the Poisson solver and the hydrodynamic evolution.

\subsection{Parallelization details}
\label{sec:parallelization}
To enable efficient computation on distributed-memory systems, the computational domain is decomposed across multiple processors, with inter-processor communication handled via the message passing interface (MPI). 
As described in Sec.~\ref{sec:fourier_decompostion}, we apply Fourier decomposition in the azimuthal direction, reducing the 3D Poisson equation to a set of independent 2D Helmholtz equations (Eqs.~\ref{eq:helmholz_sph} and \ref{eq:helmholz_cyl}). 
To perform this decomposition efficiently, the azimuthal direction ($\phi$) is not partitioned into MPI subdomains, allowing the use of the serial version of the \texttt{FFTW3} library\footnote{\url{https://fftw.org}} \citep{FFTW}. 
The resulting Helmholtz equations are then solved independently for each azimuthal mode $m = 0, \dots, N_\phi/2$ using the 2D multigrid algorithm. 
However, this decomposition increases MPI communication compared to solving a single 3D equation, since boundary exchanges must be performed separately for each mode during the iterative solution. 
Therefore, the total number of MPI communication calls scales with the number of modes, resulting in $(N_\phi/2+1)$ times more communication events.

This increase in communication frequency introduces additional overhead due to MPI latency, which includes the time required to initiate each communication event. This overhead becomes particularly significant in the present formulation because each mode corresponds to a 2D subdomain, and therefore the amount of data exchanged per communication is relatively small. In such cases, the communication latency can be comparable to or even exceed the actual data transfer time, reducing parallel efficiency. Furthermore, the repeated invocation of communication routines introduces additional synchronization costs that limit parallel scalability.

In principle, all azimuthal modes could be iterated simultaneously. However, different modes exhibit different convergence rates, with higher-order modes typically converging much faster than lower-order modes, often within only one or two V-cycles. Iterating all modes simultaneously would therefore result in unnecessary computation for rapidly converging modes, as they would continue to be updated while waiting for the slowest-converging mode to reach the prescribed tolerance. To avoid this inefficiency, we solve the Helmholtz equation sequentially, mode by mode. This approach ensures that each mode is iterated only as long as necessary, minimizing redundant computation and improving overall efficiency, at the expense of increased communication frequency.

To reduce the impact of communication overhead, we employ non-blocking MPI communication routines. Boundary exchanges are initiated using \texttt{MPI\_Isend()} and \texttt{MPI\_Irecv()}, allowing communication to proceed concurrently with computation. After initiating these non-blocking calls, each processor performs relaxation sweeps over interior grid points that do not depend on halo data. Once these independent computations are completed, \texttt{MPI\_Waitall()} is used to ensure that all communications have finished before updating grid points adjacent to subdomain boundaries.

This overlap between communication and computation helps hide communication latency and reduces processor idle time. The effectiveness of this strategy improves with increasing subdomain size, as larger interior regions allow more computation to be performed while communication is in progress. Although this approach does not reduce the total number of MPI calls, it significantly improves parallel efficiency by maximizing resource utilization and minimizing the impact of communication latency on overall solver performance.

Within this framework, the iterative solver itself is structured to maximize parallelism. 
The SOR iterations for each line in the approximate line smoother (Sec.~\ref{sec:line_smoother}) are performed in an even–odd ordering, sweeping first over even-indexed points and then over odd-indexed points. 
Lines are further updated using a zebra pattern, where alternating lines are grouped and processed in two sweeps to complete a single smoothing step, enhancing concurrency and overlap with communication.
On the coarsest grid (Sec.~\ref{sec:coarse_solver}), the 2D SOR solver employs a red–black ordering, updating cell values in two passes according to a checkerboard pattern to enable simultaneous updates across half of the domain in each pass.

Finally, to ensure efficient multigrid operation, the number of grid cells and processors in all directions except the azimuthal direction is required to be a power of two. 
This enables the maximum number of coarsening levels under factor-two coarsening, improving solver efficiency and convergence. 
More importantly, it guarantees that coarse-grid subdomains remain fully aligned with the corresponding fine-grid subdomains on the same processor. 
As a result, restriction and prolongation operations can be performed entirely locally without requiring additional inter-processor communication. 
In principle, arbitrary processor counts could be supported; however, misalignment between fine and coarse subdomains would introduce additional communication during multigrid transfers, increasing overhead and reducing parallel efficiency.

\section{Test results}
\label{sec:test_result}
In this section, we evaluate the accuracy, robustness, and convergence behavior of the proposed Poisson solver. 
We consider a series of test problems, including time-dependent simulations with known analytical or well-established reference solutions, to verify the correctness of the method and quantify numerical errors. 
These tests demonstrate the reliability and efficiency of the algorithm across different grid configurations and coordinate systems.

\begin{figure*}
\centering
\setlength{\tabcolsep}{0pt}
\renewcommand{\arraystretch}{1.0}
\begin{tabular}{cc}
  \includegraphics[width=0.5\linewidth]{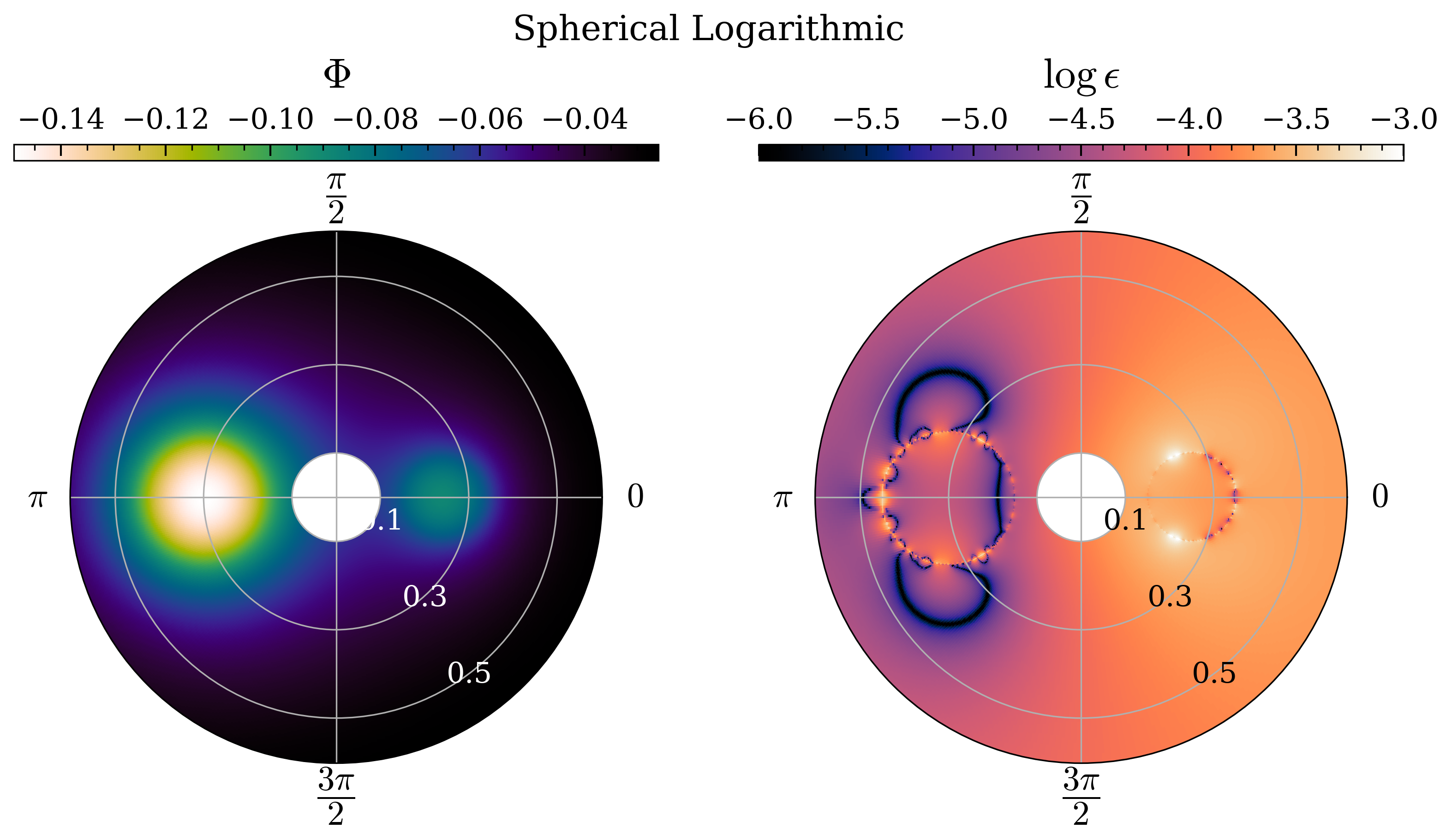} &
  \includegraphics[width=0.5\linewidth]{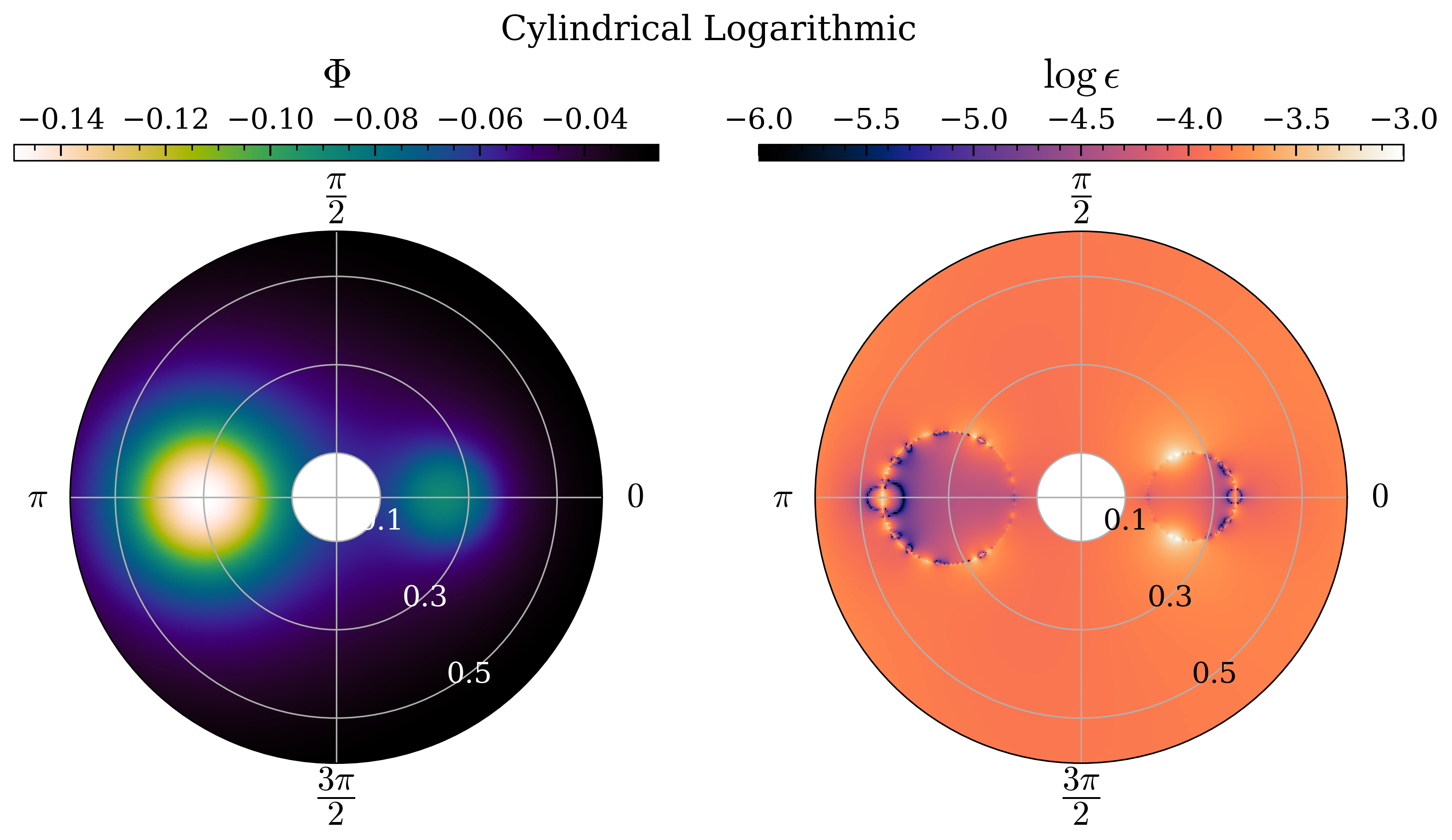}\\
  \includegraphics[width=0.5\linewidth]{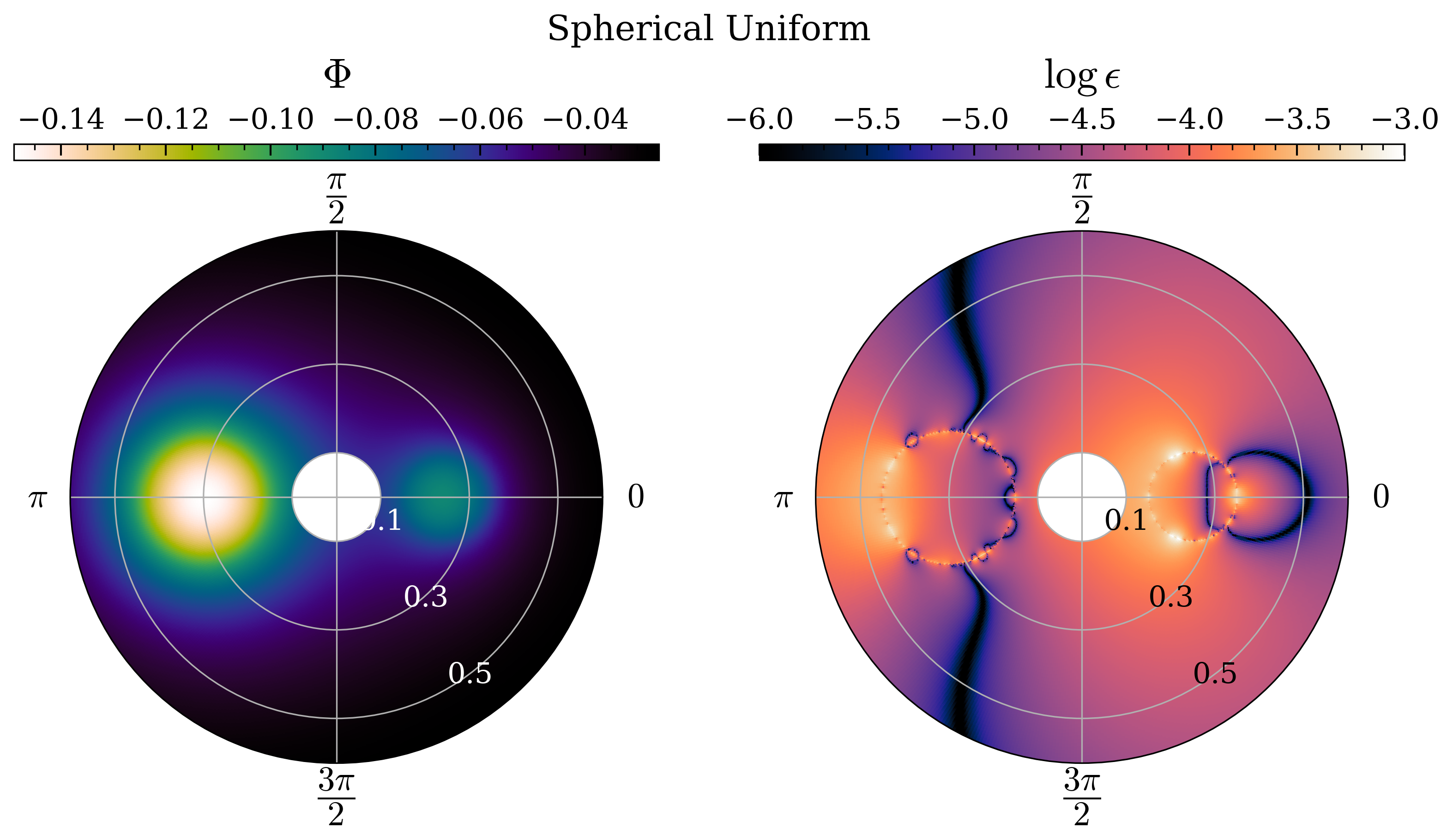} &
  \includegraphics[width=0.5\linewidth]{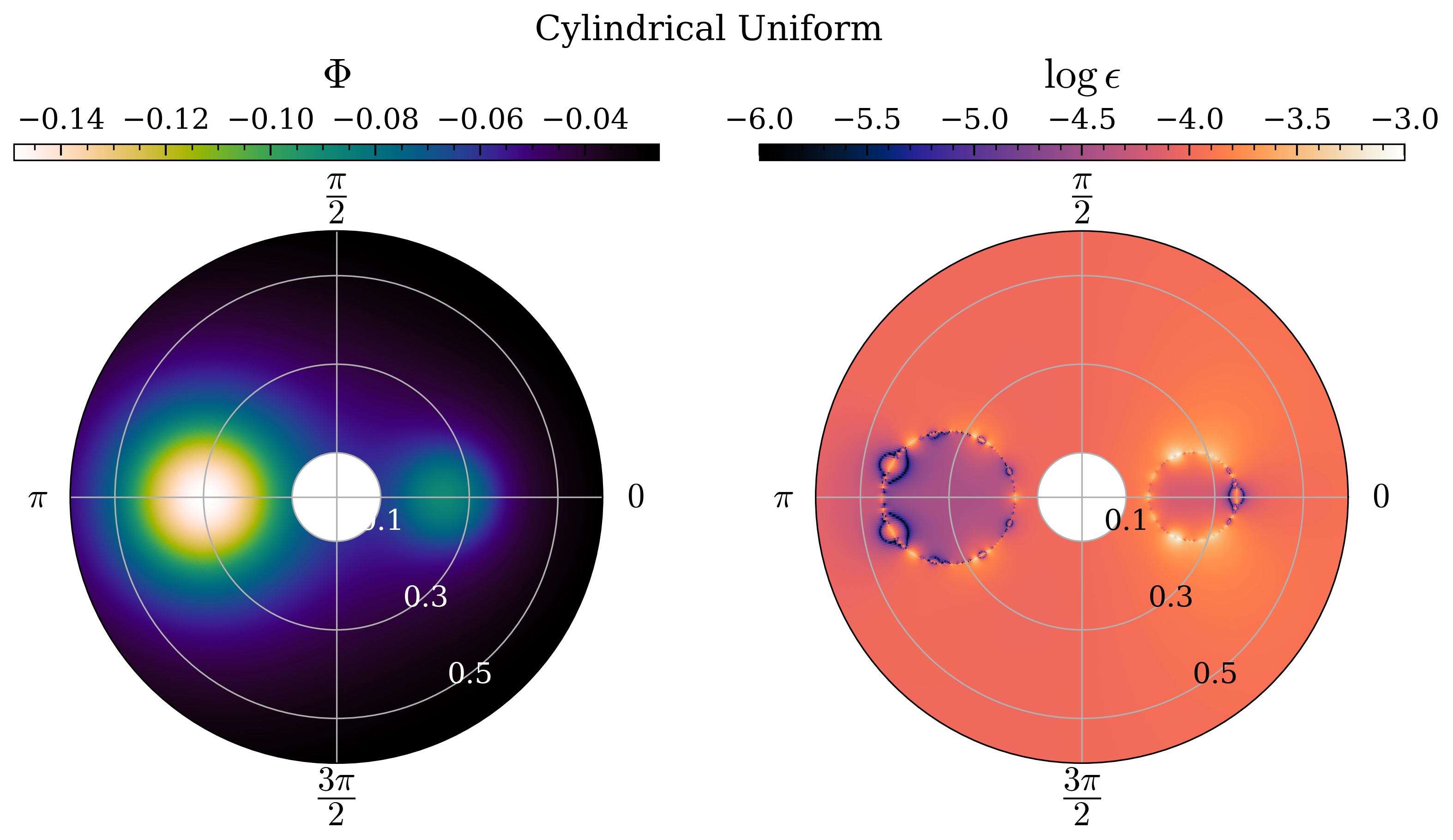}
\end{tabular}
\caption{Equatorial slices ($\theta=\pi/2$ or $z=0$) of the potential and its relative error for the double-sphere test on spherical (left-row) and cylindrical (right-right) grids. The results are shown for both the radially logarithmic (top) and uniform (bottom) grid.}
\label{fig:potential_double_sphere}
\end{figure*}

\subsection{Accuracy Tests}
We assess the accuracy and convergence properties of the Poisson solver using a set of static test problems with known or well-defined reference solutions. 
The tests are designed to probe both grid-aligned and non-grid-aligned density distributions, thereby evaluating the solver performance under idealized as well as more general geometric configurations.

Unless otherwise specified, the computational domain spans the full azimuthal range $[0,2\pi]$. 
We consider both spherical $(r,\theta,\phi)$ and cylindrical $(R,\phi,z)$ coordinate systems, using either radially logarithmic or uniform grids, while the remaining directions are discretized uniformly. 
The domain is discretized using $N \times N \times 2N$ cells in spherical coordinates and $N \times 2N \times N$ cells in cylindrical coordinates. Unless noted otherwise, we set $N=256$.

To quantify the numerical accuracy, we define the pointwise relative error as
\begin{equation}
    \epsilon = \left|\frac{\Phi - \Phi_r}{\Phi_r}\right|,
\end{equation}
where $\Phi$ is the numerical solution and $\Phi_r$ is a reference solution. 
We further compute the global $L_2$ norm of the error as a volume-weighted root-mean-square,
\begin{equation}\label{eq:L2_norm}
    \norm{\epsilon}_2 = \left(\frac{\sum_{i,j,k} \epsilon_{i,j,k}^2 \, \Delta V_{i,j,k}}{\sum_{i,j,k} \Delta V_{i,j,k}}\right)^{1/2}.
\end{equation}

\subsubsection{Double Sphere}
\label{sec:double_sphere}
We begin with a simple yet demanding static test problem to assess the accuracy of the solver. The configuration consists of two uniform-density spheres with different masses placed at arbitrary locations, following \citet{Tomida_2023}. Here, however, the setup is made more challenging by positioning the computational domain boundaries in close proximity to the mass distribution. In such a configuration, a boundary treatment based on multipole expansion would require a large number of moments to achieve high accuracy. Since we instead employ the James method for boundary value evaluation, this setup provides a stringent test of its accuracy in our implementation.

One sphere is centered at $(r_1,\theta_1,\phi_1) = (0.25,\pi/2,0)$ in spherical grid and $(R_1,\phi_1,z_1) = (0.25,0,0)$ in cylindrical grid, with radius $a_1 = 0.1$. The second sphere is located at $(r_2,\theta_2,\phi_2) = (0.3,\pi/2,\pi)$ or $(R_2,\phi_2,z_2) = (0.3,0,\pi)$ in spgerical or cylindrical grid, respectively, with radius $a_2 = 0.15$. Both spheres have unit density, while the density outside them is zero. The computational domains are $[0.1, 0.6] \times [0,\pi] \times [0, 2\pi]$ in spherical coordinates and $[0.1,0.6] \times [0,2\pi] \times [-0.25,0.25]$ in cylindrical coordinates.

\begin{figure*}[t]
\centering
\setlength{\tabcolsep}{0pt}
\renewcommand{\arraystretch}{1.0}
\begin{tabular}{cc}
  \includegraphics[width=0.5\linewidth]{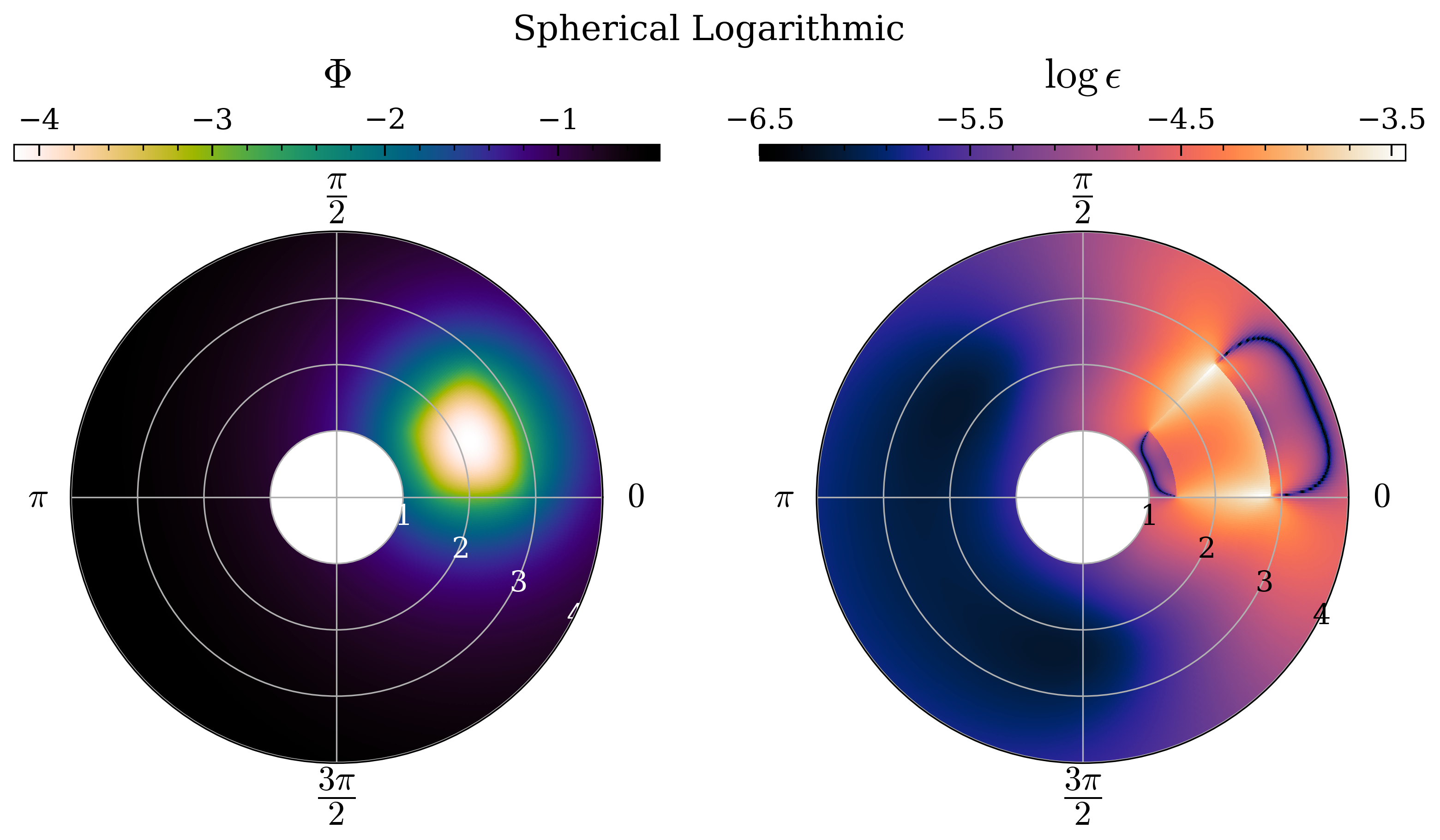} &
  \includegraphics[width=0.5\linewidth]{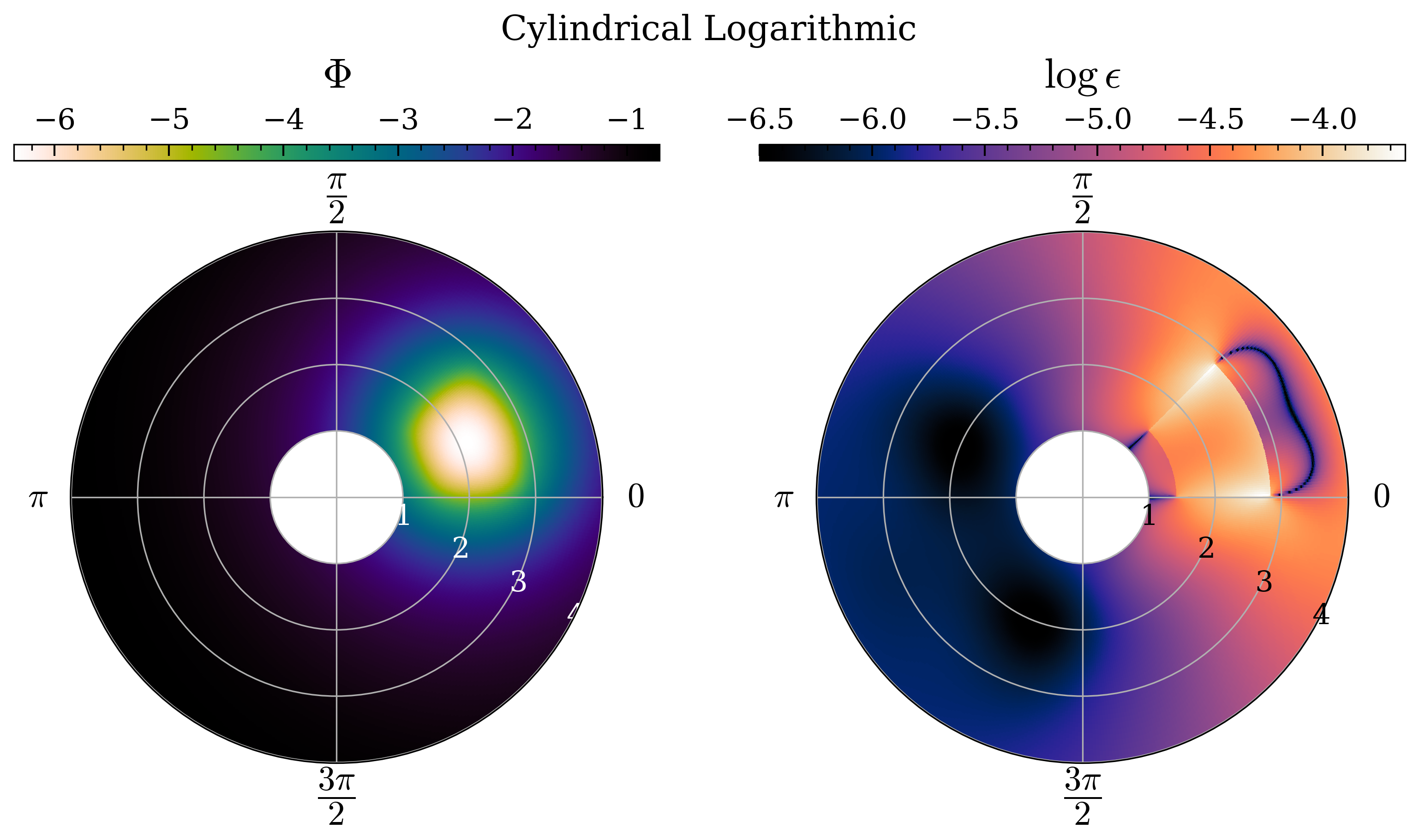}\\
  \includegraphics[width=0.5\linewidth]{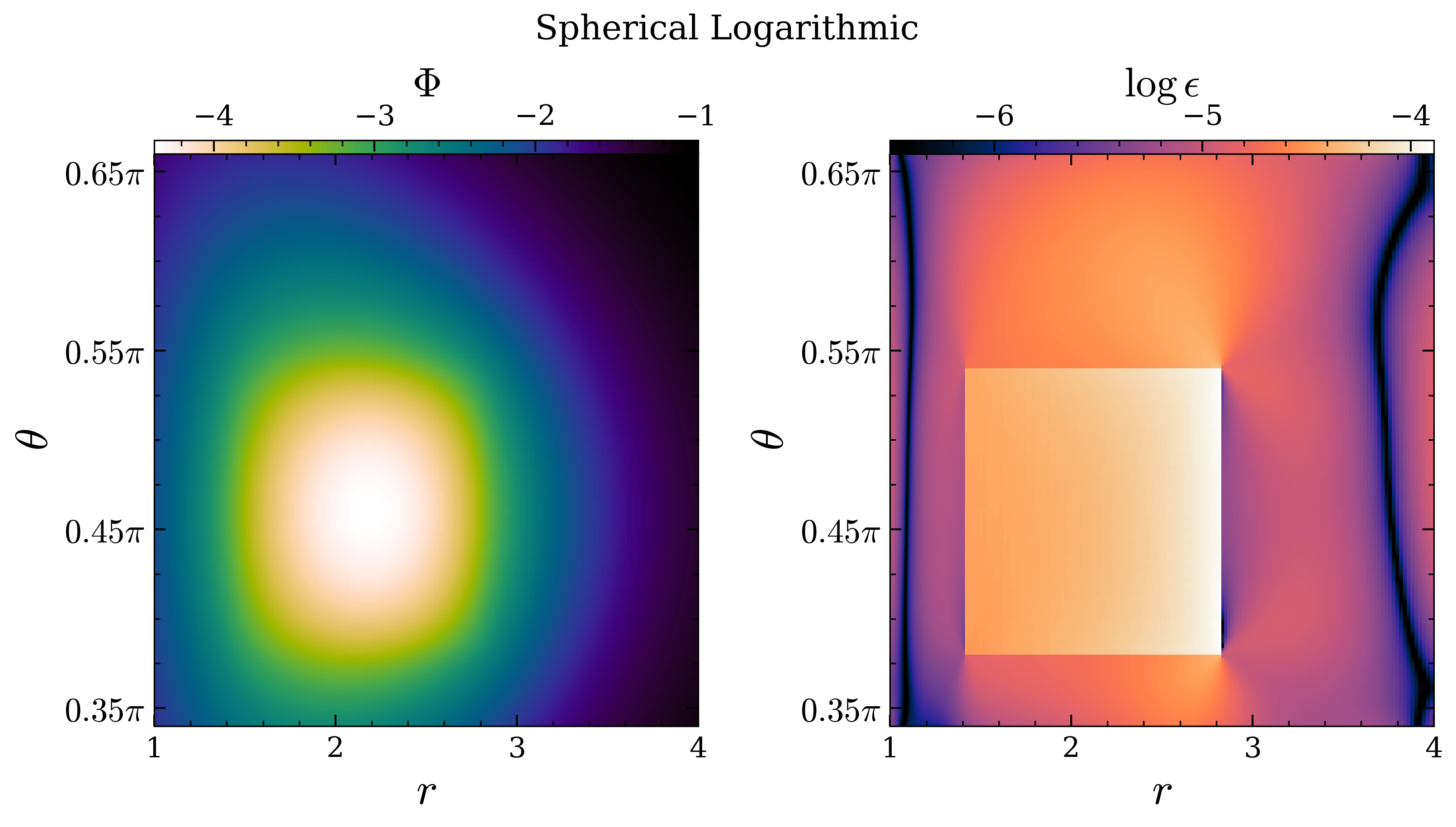} &
  \includegraphics[width=0.5\linewidth]{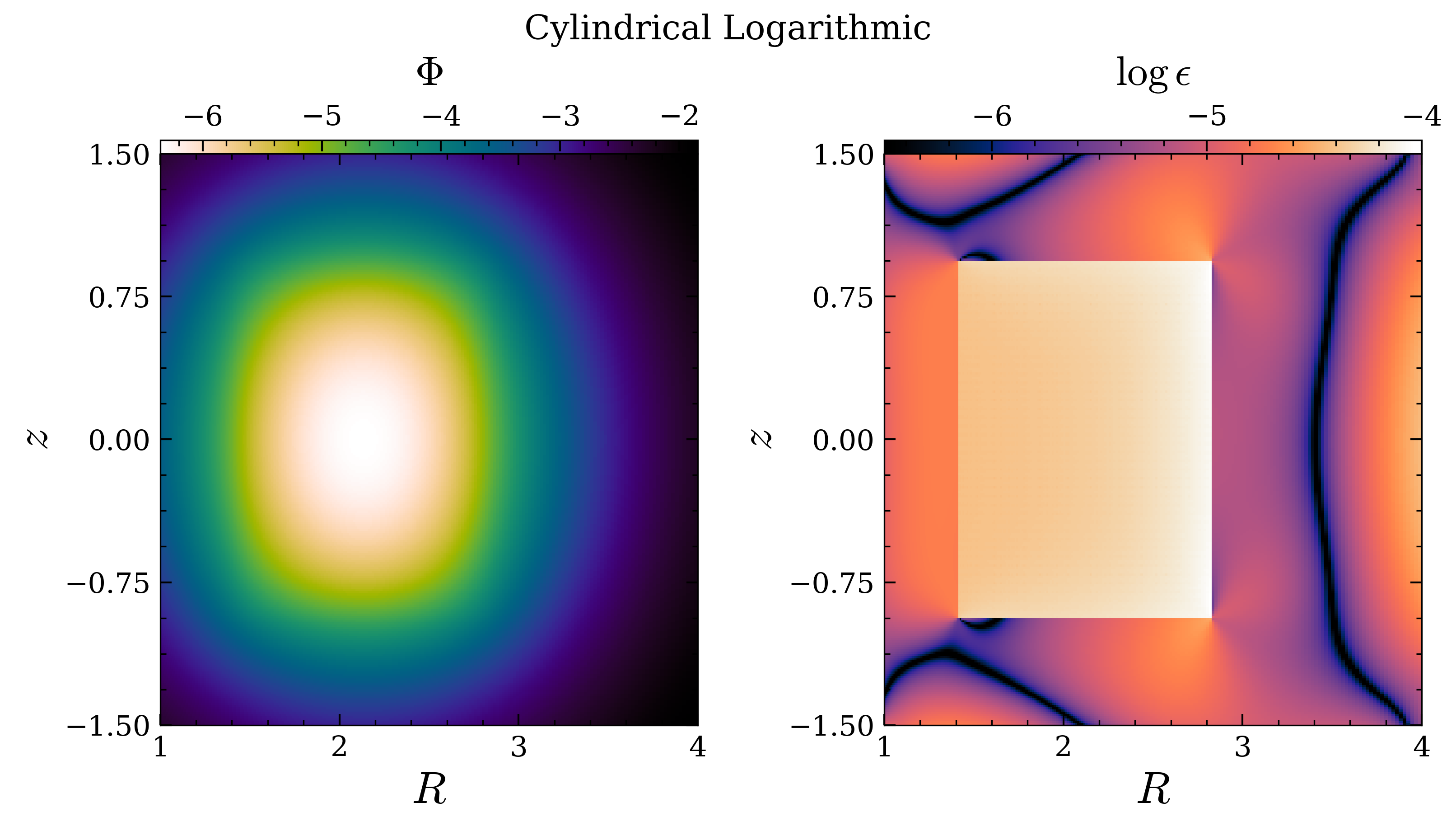}
\end{tabular}
\caption{Comparison of the potential and its relative error for the mesh-segment test on logarithmic spherical (left) and cylindrical (right) grids. The top panels show equatorial slices ($\theta=\pi/2$ or $z=0$, for spherical or cylindrical grid, respectively), while the bottom panels show meridional slices at $\phi=0.125\pi$.}
\label{fig:potential_slice_log}
\end{figure*}

The total gravitational potential is given by the sum of contributions from the two spheres,
\begin{equation}\label{eq:potential_double_sph}
\Phi = \Phi_1 + \Phi_2.
\end{equation}
The potential $\Phi_i$ due to the $i^{\rm th}$ sphere is
\begin{equation}
\Phi_i =
    \begin{cases}
        -\dfrac{G M_i}{2 a_i^3}\left(3a_i^2 - d_i^2\right), & \text{if } d_i < a_i,\\[6pt]
        -\dfrac{G M_i}{d_i}, & \text{if } d_i \ge a_i,
    \end{cases}
\end{equation}
where $d_i$ denotes the distance from a field point to the center of the $i^{\rm th}$ sphere. 
In spherical coordinates $(r,\theta,\phi)$,
\begin{equation}
    d_i = \sqrt{r^2 + r_i^2 - 2 r r_i \left[\cos\theta \cos\theta_i + \sin\theta \sin\theta_i \cos(\phi - \phi_i)\right]},
\end{equation}
and in cylindrical coordinates $(R,\phi,z)$,
\begin{equation}
    d_i = \sqrt{R^2 + R_i^2 - 2 R R_i \cos(\phi - \phi_i) + (z - z_i)^2}.
\end{equation}

Fig.~\ref{fig:potential_double_sphere} shows slices in the $r$–$\phi$ and $R$–$\phi$ planes, taken at $\theta=\pi/2$ and $z=0$, for spherical and cylindrical grid, respectively, of the numerically computed potential and the corresponding point-wise relative error, $\epsilon$, with respect to the analytical solution. Results are presented for both radially logarithmic (top row) and uniform (bottom row) grids, in spherical (left panels) and cylindrical (right panels) geometries.

In all cases, the maximum relative error remains below $0.1\%$, demonstrating the high accuracy of the solver. As expected, the largest point-wise errors are concentrated near the sphere boundaries. Since the spheres are not aligned with the grid, their discrete representation deviates from perfect spherical geometry, leading to resolution-dependent variations in the enclosed mass. Thus, when comparing the numerical solution to the analytical potential for fixed $M_i$, the measured error reflects both the discretization error of the Laplacian and discrepancies in the effective mass distribution. These geometric inaccuracies contribute to the concentration of errors along the sphere boundaries.

\begin{figure*}[!htb]
\centering
\setlength{\tabcolsep}{0pt}
\renewcommand{\arraystretch}{1.0}
\begin{tabular}{cc}
  \includegraphics[width=0.5\linewidth]{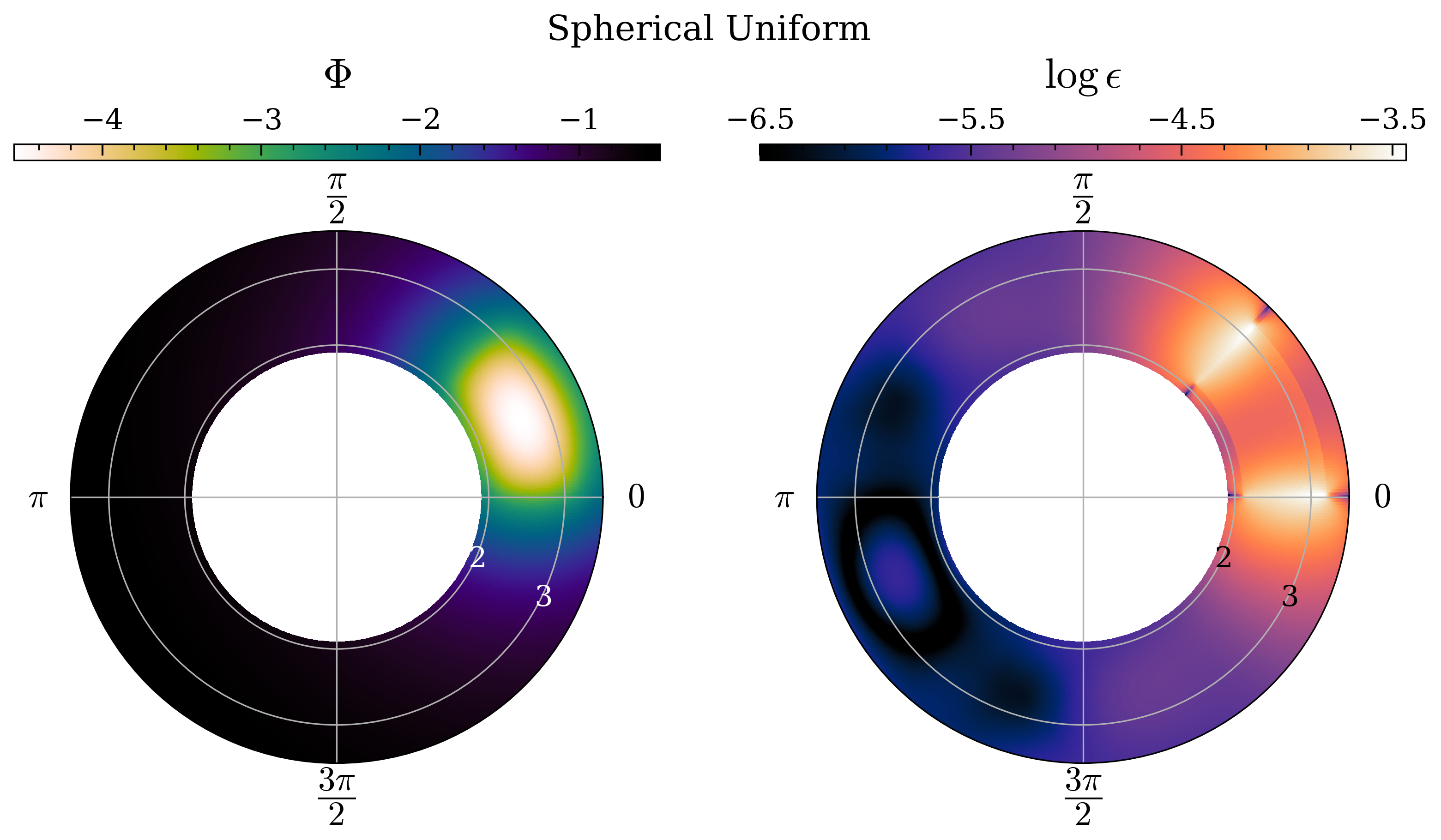} &
  \includegraphics[width=0.5\linewidth]{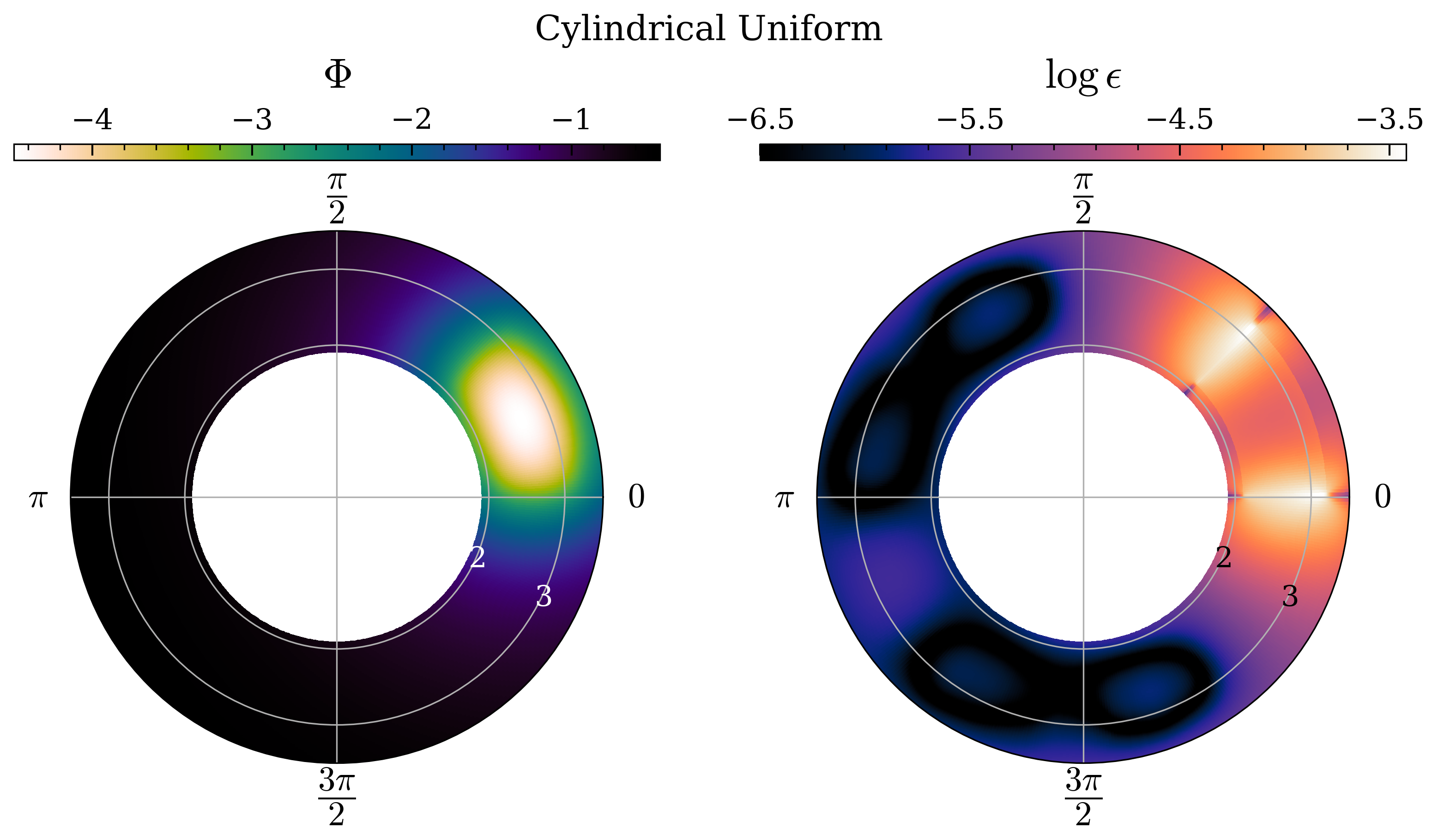}\\
  \includegraphics[width=0.5\linewidth]{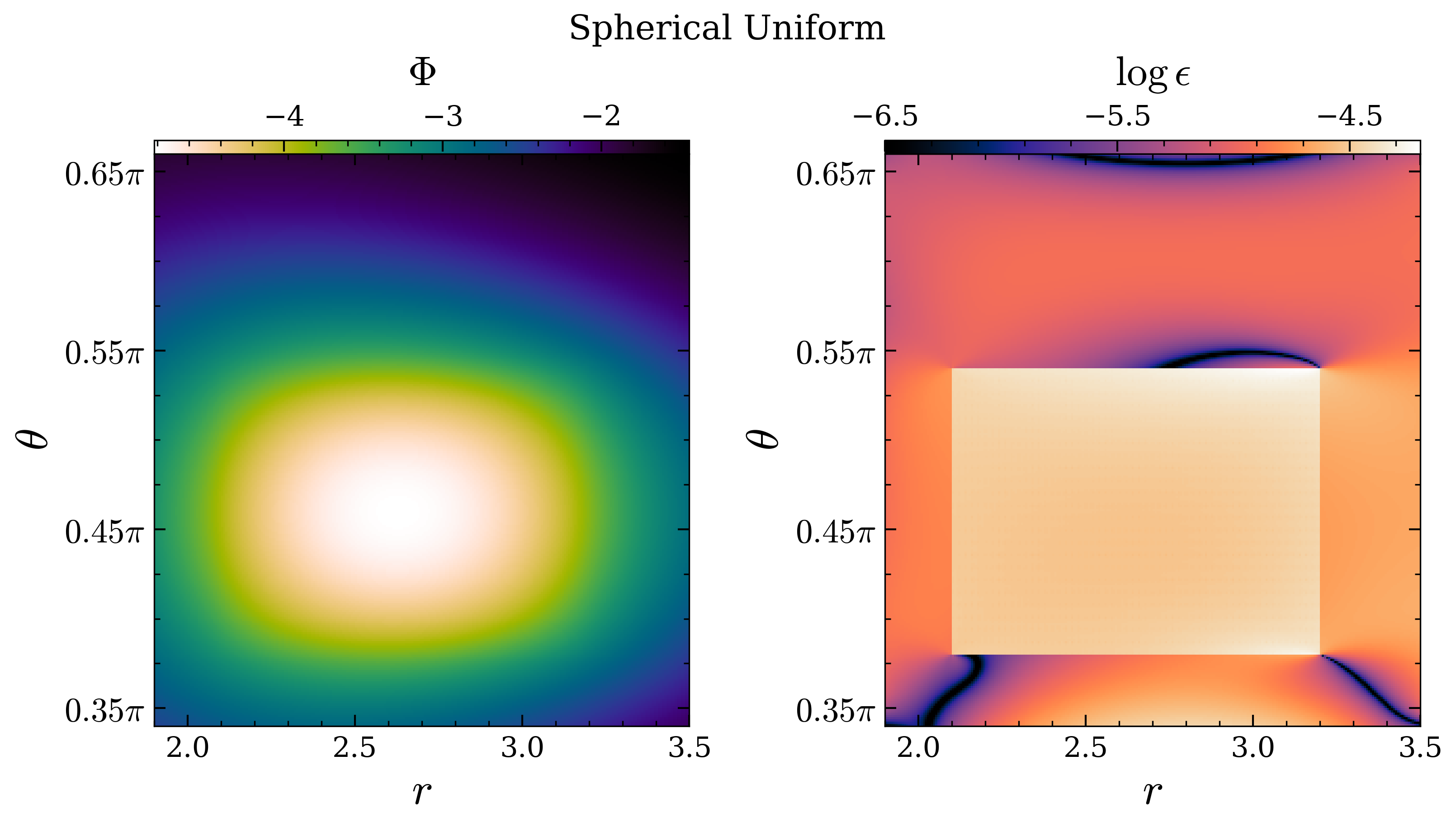} &
  \includegraphics[width=0.5\linewidth]{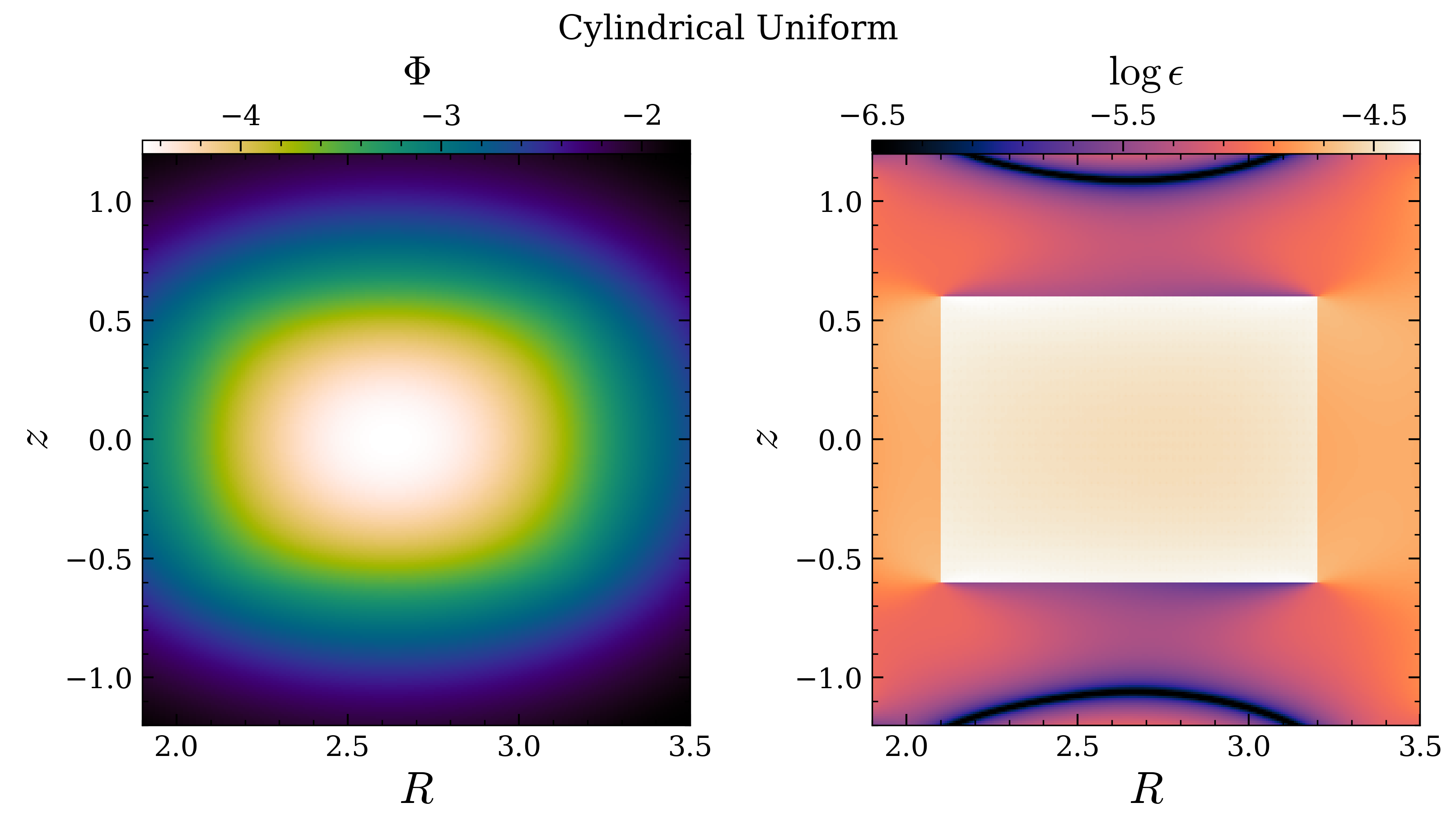}
\end{tabular}
\caption{Same as Fig.~\ref{fig:potential_slice_log}, but for uniform grids in spherical (left) and cylindrical (right) coordinates.}
\label{fig:potential_slice_uni}
\end{figure*}

\subsubsection{Mesh segment}
\label{sec:mesh_segment}
To complement the double-sphere test, where the mass distribution is not aligned with the grid, we now consider a simpler configuration in which the source is fully grid-aligned. This setup removes geometric discretization effects and allows for a more direct assessment of the solver accuracy. Specifically, we examine a problem consisting of a uniform-density source whose shape and spatial extent remain invariant with resolution.

For each coordinate system, we employ both radially logarithmic and uniform grids. The parameters defining the source extent and computational domain are summarized in Table~\ref{tab:parameter_meshseg}.

Although no simple closed-form algebraic expression exists for the analytical potential of these source configurations, \citet{Hure_2014} derived an exact formulation for the gravitational potential of a cylindrical mesh-aligned segment, and \citet{Gressel_2024} presented an analogous expression for the spherical case. 
These formulations involve a combination of line and surface integrals, which we evaluate numerically using Romberg integration with an accuracy of $10^{-8}$. 
This procedure provides a reference solution, $\Phi_r$, allowing for a reliable assessment of the accuracy of the numerical solution.

Fig.~\ref{fig:potential_slice_log} presents the gravitational potential computed by our Poisson solver along with the corresponding pointwise relative error, $\epsilon$, for the radially logarithmic spherical (left) and cylindrical (right) grids. The top panels show equatorial slices ($\theta=\pi/2$, for spherical coordinates and $z=0$, for cylindrical coordinates), the bottom panels show meridional slices at $\phi = 0.125\pi$, corresponding to the $r\mbox{-}\theta$ plane in spherical coordinates and the $R\mbox{-}z$ plane in cylindrical coordinates. The corresponding results for the radially uniform grid are shown in Fig.~\ref{fig:potential_slice_uni}.

These results demonstrate the high accuracy and robustness of the solver, with the maximum relative error remaining below $0.05\%$ across all tested geometries and grid configurations at this resolution. The uniformly low error confirms that the method accurately reproduces the gravitational potential and remains reliable on both logarithmic and uniform grids in spherical and cylindrical coordinates. The slightly elevated error within the inner solid region compared to the rest of the domain reflects the underlying density distribution and is consistent with the behavior reported by \citet{Ahn_2025}.

\begin{figure*}
    \centering
    \includegraphics[width=\linewidth]{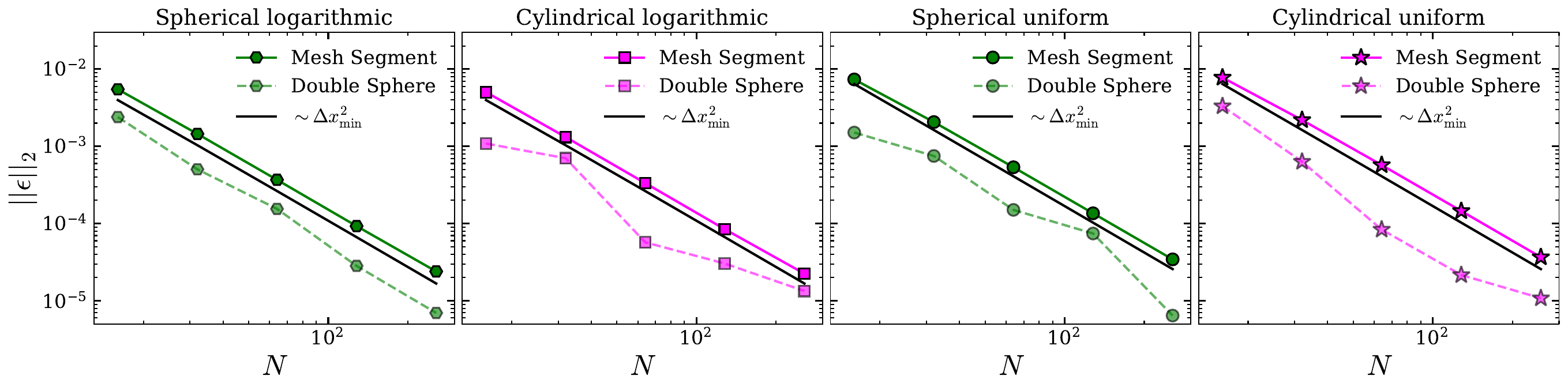}
    \caption{The $L_2$ of the relative error, $\norm{\epsilon}_2$, as a function of resolution $N$, for the double-sphere (daashed lines) and mesh-segment (solid lines) tests. Each panel corresponds to a specific coordinate system and grid type as indicated in the panel title. The solid black line denotes the expected second-order convergence scaling.}
    \label{fig:convergence_scaling}
\end{figure*}

\subsubsection{Convergence test}
\label{sec:convergence_test}
Since the finite-volume discretization of the Laplacian in both spherical and cylindrical coordinates is second-order accurate (as described in Sec.~\ref{sec:discretization}), the numerical error is expected to decrease by approximately a factor of four when grid resolution is doubled, provided that the error is dominated by the truncation error of the discretization rather than by geometric effects associated with the discrete representation of the density field. To verify this behavior, we perform a convergence study for all configurations, considering both the double-sphere (Sec.~\ref{sec:double_sphere}) and mesh-segment (Sec.~\ref{sec:mesh_segment}) test problems. The resolution is varied from $N = 16$ to $N = 256$ for each coordinate system and grid type.

Fig.~\ref{fig:convergence_scaling} summarizes the results of this analysis, with each panel corresponding to a specific grid configuration as indicated in the panel title. Results are shown for both the double-sphere (dashed lines) and mesh-segment (solid lines) tests, and the solid black line indicates the expected second-order convergence rate. In all cases, the error relative to the analytical solution decreases with increasing resolution, consistent with the formal accuracy of the scheme.

For the double-sphere test, however, fluctuations around the ideal scaling are observed. As noted in previous studies \citep[e.g.,][]{Katz_2016, Moon_2019}, such fluctuations do not arise from truncation errors of the finite-difference (or finite-volume) scheme itself, but rather from the inability of a finite grid to exactly represent a spherical mass distribution. In the present case as well, this leads to resolution-dependent variations in the effective density representation, which manifest as deviations from smooth convergence.

By contrast, in the mesh-segment test, where the density distribution is exactly aligned with the grid, the measured error follows the expected second-order scaling across all configurations. This confirms that the solver preserves its formal accuracy for both spherical and cylindrical coordinate systems when geometric discretization effects are absent.

\subsection{Collapse Tests}
\label{sec:collapse_test}
In this section, we test the Poisson solver in fully dynamical collapse simulations for three cases: (i) a non-rotating uniform isothermal sphere, (ii) a rotating uniform sphere, and (iii) a rotating sphere with an $m=2$ perturbation. 
All cases span the full domain ($\theta \in [0,\pi]$ or $z \in [-R,R]$) and use radially logarithmic grids in both spherical and cylindrical coordinate with large $r_{\rm max}/r_{\rm min} \text{ or } R_{\rm max}/R_{\rm min}$ ratios to assess solver accuracy and stability under extreme grid stretching.

A technical complication arises from the coordinate singularity at $r=0$ or $R=0$, which necessitates the introduction of a finite inner radius in the computational domain. As gravitational collapse proceeds, mass flows through this inner radial boundary, leading to a loss of mass from the grid. 
This mass loss would otherwise reduce the self-gravity within the domain and complicate comparisons with available analytical solutions.
To account for this effect, we introduce a sink mass, $m_{\rm sink}$, which is updated at each timestep by accumulating the mass flux through the inner boundary. 
The gravitational potential of this accumulated mass is then included as an external point-mass potential in the hydrodynamic evolution. 
In spherical coordinates, where the excluded region is a small central sphere, this approximation is well justified. 
In cylindrical coordinates, however, the excluded region forms a cylindrical hole of radius $R_{\rm min}$ and height of $\lesssim 2R_{\rm max}$, and representing the removed mass as a point source is formally less accurate. 
Nevertheless, as demonstrated in the following sections, this approximation does not significantly affect the results, and the numerical solutions remain in close agreement with analytical expectations (where available).

\begin{figure}
    \centering
    \includegraphics[width=\linewidth]{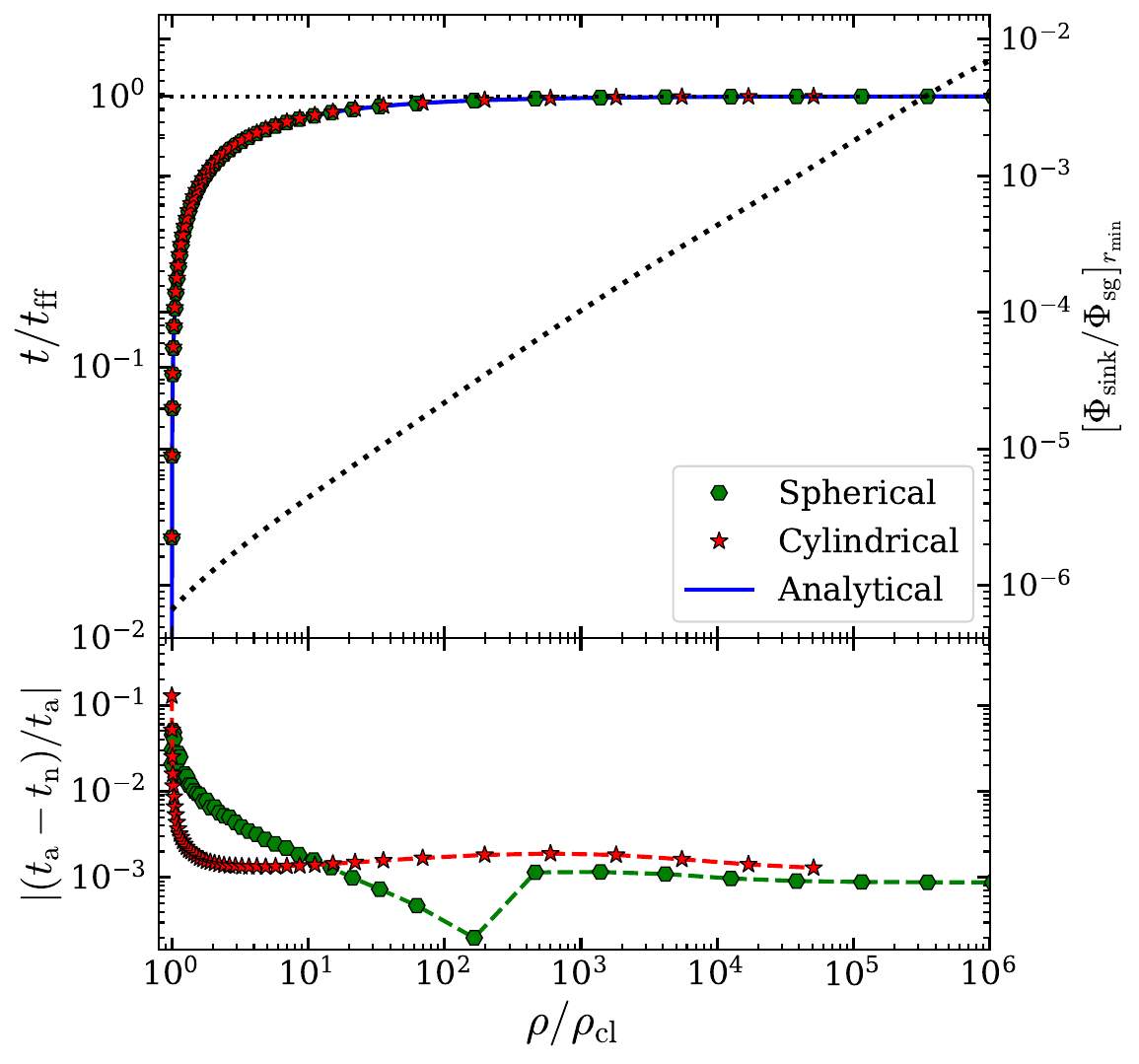}
    \caption{Evolution of the central density during the collapse of a non-rotating uniform sphere.
    The top panel shows the time required for the central density to reach a value $\rho$, comparing simulations performed on a spherical grid (green hexagons) and a cylindrical grid (red stars) with the analytical prediction (solid blue line; Eq.~\eqref{eq:isothermal_rho_t}). The dotted line, referenced to the right-hand axis, shows the relative contribution of the sink potential to the self-gravitational potential at the innermost radius as a function of central density. The bottom panel displays the relative error between the simulation time ($t_{\rm s}$) and the analytical time ($t_{\rm a}$) for both grid geometries.}
    \label{fig:isothermal_time_evolve}
\end{figure}

\subsubsection{Non-rotating uniform isothermal sphere}
\label{sec:isothermal_collapse}
We consider the collapse of a uniform, non-rotating, cold gas sphere, a classical benchmark problem introduced by \citet{Truelove_1998} and widely used to validate grid-based hydrodynamic codes \citep[e.g.,][]{Ziegler_2005,Mandal_2023}. 
The initial configuration consists of a spherical cloud of radius $R = 7.8\times10^{15}\,\mathrm{cm}$ and uniform density $\rho_{\rm cl} = 10^{-15}\,\mathrm{g\,cm^{-3}}$, embedded in a low-density ambient medium with $\rho_{\rm b} = 0.01\,\rho_{\rm cl}$. 
The gas is isothermal with a sound speed $\cs = 0.167\,\mathrm{km\,s^{-1}}$.

This problem is particularly useful because, in the pressureless limit ($T=0$), it admits an analytic self-similar collapse solution. However, in practice, finite pressure leads to the formation of a rarefaction wave that propagates inward from the cloud surface. 
At any given time, the interior region of the rarefaction front remains spatially uniform and follows the analytical self-similar evolution derived by \citet{Truelove_1998}, while the outer layers are progressively affected by the inward-moving rarefaction wave.
As $t$ approaches the free-fall time, $\tff$, the analytic solution predicts a divergence of the central density. 
The temperature is chosen such that the gravitational collapse proceeds more rapidly than the rarefaction wave can reach the center. 
The time required for the density within the rarefaction-free core to reach a value $\rho$ is given analytically by \citep{Truelove_1998}:
\begin{equation}\label{eq:isothermal_rho_t}
    \frac{t_\rho}{\tff} = \frac{2}{\pi}\left[\eta + \frac{1}{2}\sin(2\eta)\right],
     \text{ where  }
    \eta = \arccos\!\left[\left(\frac{\rho}{\rho_{\rm cl}}\right)^{-1/6}\right].
\end{equation}
Here $\tff=\sqrt{3\pi/(32 G\rho_{\rm cl})}$ is the freefall time of the cloud at the initial density.

For the simulations, the radial computational domain spans $[r_{\rm min}, r_{\rm max}]$ or $[R_{\rm min}, R_{\rm max}] = [10^{-3},\,1.1]$. The resolution is set to $128\times64\times64$ for the spherical grid and $128\times64\times128$ for the cylindrical grid.

The top panel of Fig.~\ref{fig:isothermal_time_evolve} shows the time required for the central density, i.e., the density of the region interior to the rarefaction wave, to reach a value of $\rho$, for the spherical (green hexagons) and cylindrical (red stars) grids. For comparison, we show the analytical expectation (Eq.~\ref{eq:isothermal_rho_t}) with the solid blue line. In the simulations, the central density is calculated at the innermost radius ($r_{\rm min}/R_{\rm min}$) of the domain.
The bottom panel presents the corresponding relative error between the numerical and analytical values. 
After an initial deviation of approximately 10\% from the analytical solution, the agreement is excellent throughout the entire time range, with the error remaining below 1\%. 
In particular, during the late phase, when the density changes very rapidly, the numerical error drops below 0.1\%. 
In both coordinate systems, the central density increases by more than six orders of magnitude while closely following the analytical solution, demonstrating the accuracy and stability of the self-gravity implementation during dynamical collapse.

The excellent agreement with the analytical solution naturally raises the question of whether the sink particle, introduced to handle the coordinate singularity, might dominate the gravitational field and bias the results. This is assessed in the top panel (dotted line, right axis), which shows the ratio of the sink potential to the self-gravitational potential at the innermost radius, $\Phi_{\rm sink}/\Phi_{\rm sg}$, as a function of the central density. The ratio remains below $10^{-2}$ (i.e., less than 1\%) throughout the entire collapse and is typically orders of magnitude smaller. Thus, the sink potential is always subdominant, and the dynamics is governed by the Poisson solver. The excellent agreement with the analytical prediction provides a genuine validation of the gravity solver, not merely a test of the sink treatment.

\begin{figure*}
\centering
\setlength{\tabcolsep}{0pt}
\renewcommand{\arraystretch}{1.0}
\begin{tabular}{c}
  \includegraphics[width=0.8\linewidth]{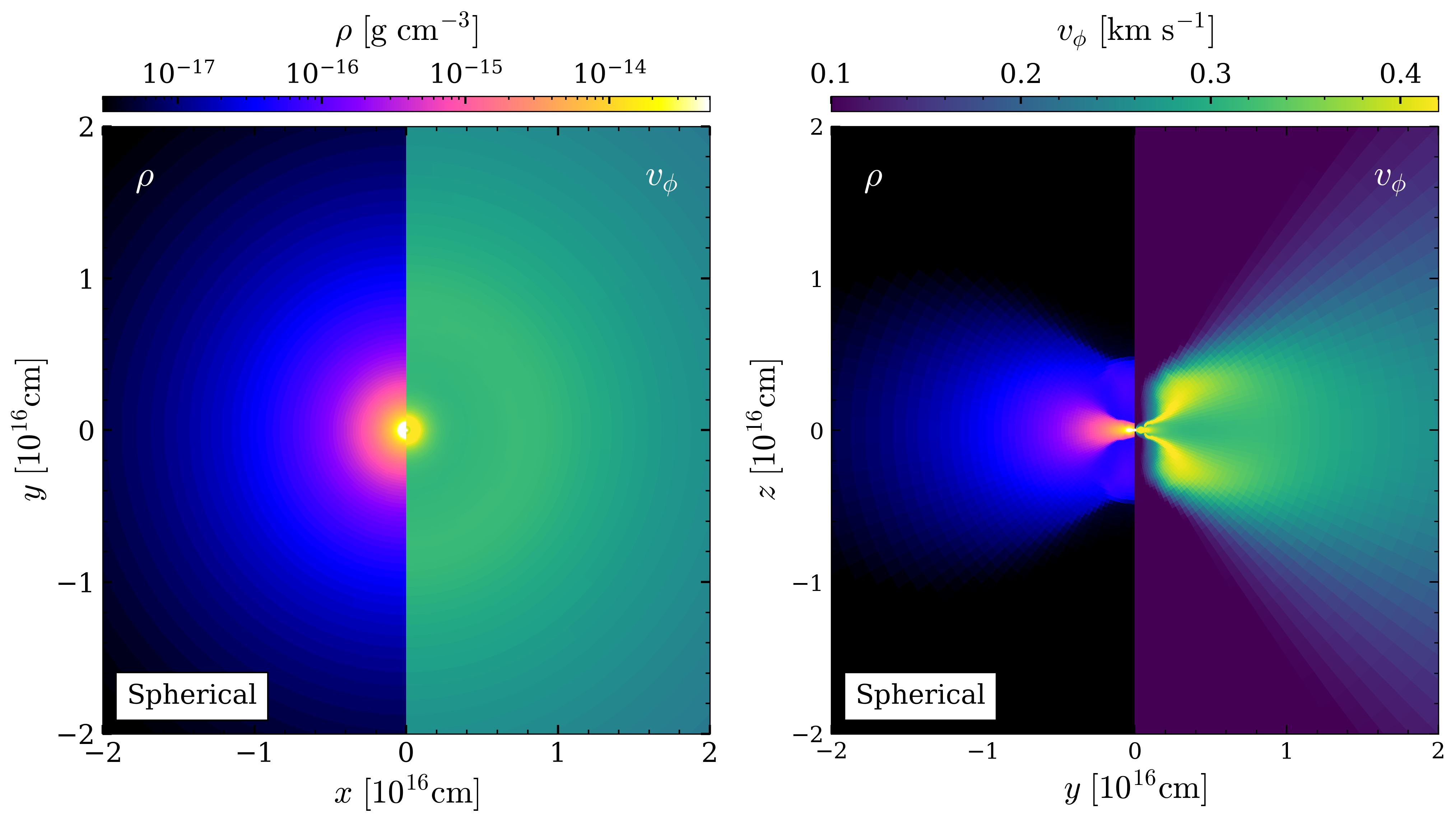} \\
  \includegraphics[width=0.8\linewidth]{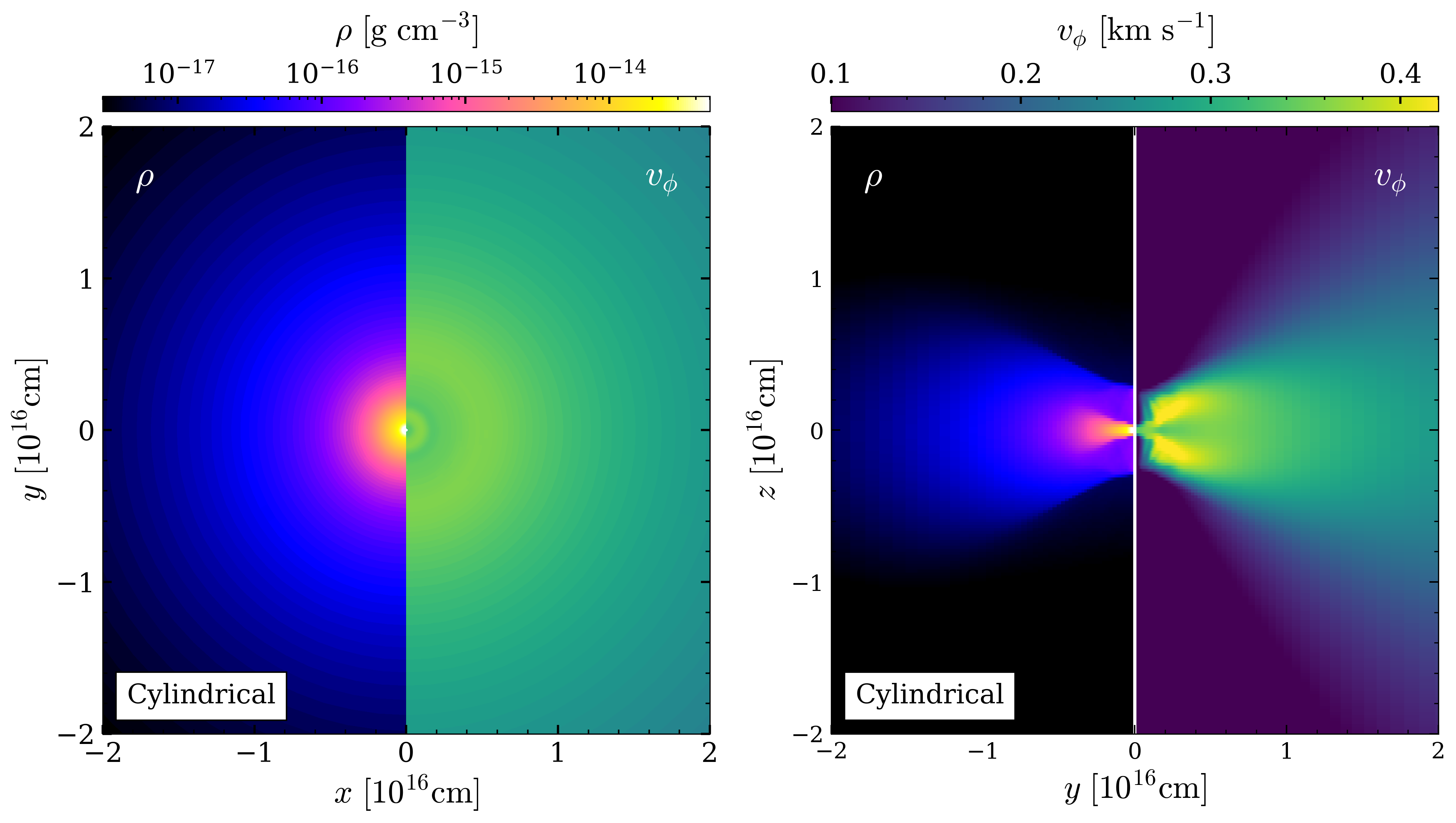}
\end{tabular}
  \caption{Slices of the density (left-half of each panel) and azimuthal velocity, $v_\phi$ (right-half) distribution for the rotating uniform sphere test at $t \sim 1.32~t_{\rm ff}$ are shown in the equatorial plane ($\theta=\pi/2$ in spherical coordinates and $z=0$ in cylindrical coordinates; left panels) and in the meridional plane ($\phi=\pi/2$; right panels). The top panels correspond to results obtained on the spherical grid, while the bottom panels show the corresponding solution computed on the cylindrical grid.}
  \label{fig:rotatingU_slice}
\end{figure*}

\subsubsection{Rotating uniform sphere}
\label{sec:rotating_uniform}
As a second test, we examine the collapse of a isothermal rotating uniform cloud, first introduced by \citet{Norman_1980} and subsequently studied by several authors \citep{Boss_1992, Truelove_1998, Ziegler_2005, Mandal_2023}. 
This setup is particularly valuable for assessing how well a numerical scheme conserves angular momentum. 
The initial conditions consist of a uniform spherical cloud with mass $M = 1~M_\odot$ and radius $R = 7.01\times10^{16}~\mathrm{cm}$, corresponding to a density $\rho_0 = 1.26\times10^{-18}~\mathrm{g~cm^{-3}}$. 
The cloud is isothermal with $T = 5~\mathrm{K}$ and rotates uniformly with angular velocity $\Omega = 3.04\times10^{-13}~\mathrm{rad~s^{-1}}$. 
These parameters yield initial energy ratios of $\alpha = 0.54$ for thermal to gravitational energy and $\beta = 0.08$ for rotational to gravitational energy.
The surrounding medium is initialized with a density of $\rho_\mathrm{b} = 0.01\rho_0$.

The computational domain extends radially over $[r_{\rm min}, r_{\rm max}]$ or $[R_{\rm min}, R_{\rm max}] = [10^{-3},\,1]\times10^{17}~{\rm cm}$. The adopted resolution is $128\times64\times64$ for the spherical grid and $128\times64\times256$ for the cylindrical grid.

\citet{Norman_1980} demonstrated that a poor angular momentum conservation during collapse can result in the formation of an artificial ring rather than a rotationally supported disk. 
In contrast, proper angular momentum conservation leads to the formation of a central disk accompanied by a monotonic increase in the central density. 

Fig.~\ref{fig:rotatingU_slice} shows slices of the density (left-half of each panel) and azimuthal velocity, $v_\phi$ (right half) distribution at $t = 1.32~t_{\rm ff}$ in the equatorial plane ($\theta=\pi/2$ in spherical coordinates and $z=0$ in cylindrical coordinates; left panels) and in the meridional plane ($\phi=\pi/2$; right panels). 
The top panels correspond to results obtained on the spherical grid, while the bottom panels show the corresponding solution computed on the cylindrical grid. 
The resulting morphology is consistent with previous studies \citep[e.g.,][]{Truelove_1998, Ziegler_2005, Mandal_2023}, clearly exhibiting a rotationally supported disk. 
No artificial ring like structures are observed at any stage of the evolution, indicating that angular momentum is accurately conserved in our simulations.

\begin{figure*}
\centering
\setlength{\tabcolsep}{0pt}
\renewcommand{\arraystretch}{1.0}
\begin{tabular}{cc}
  \includegraphics[width=0.4\linewidth]{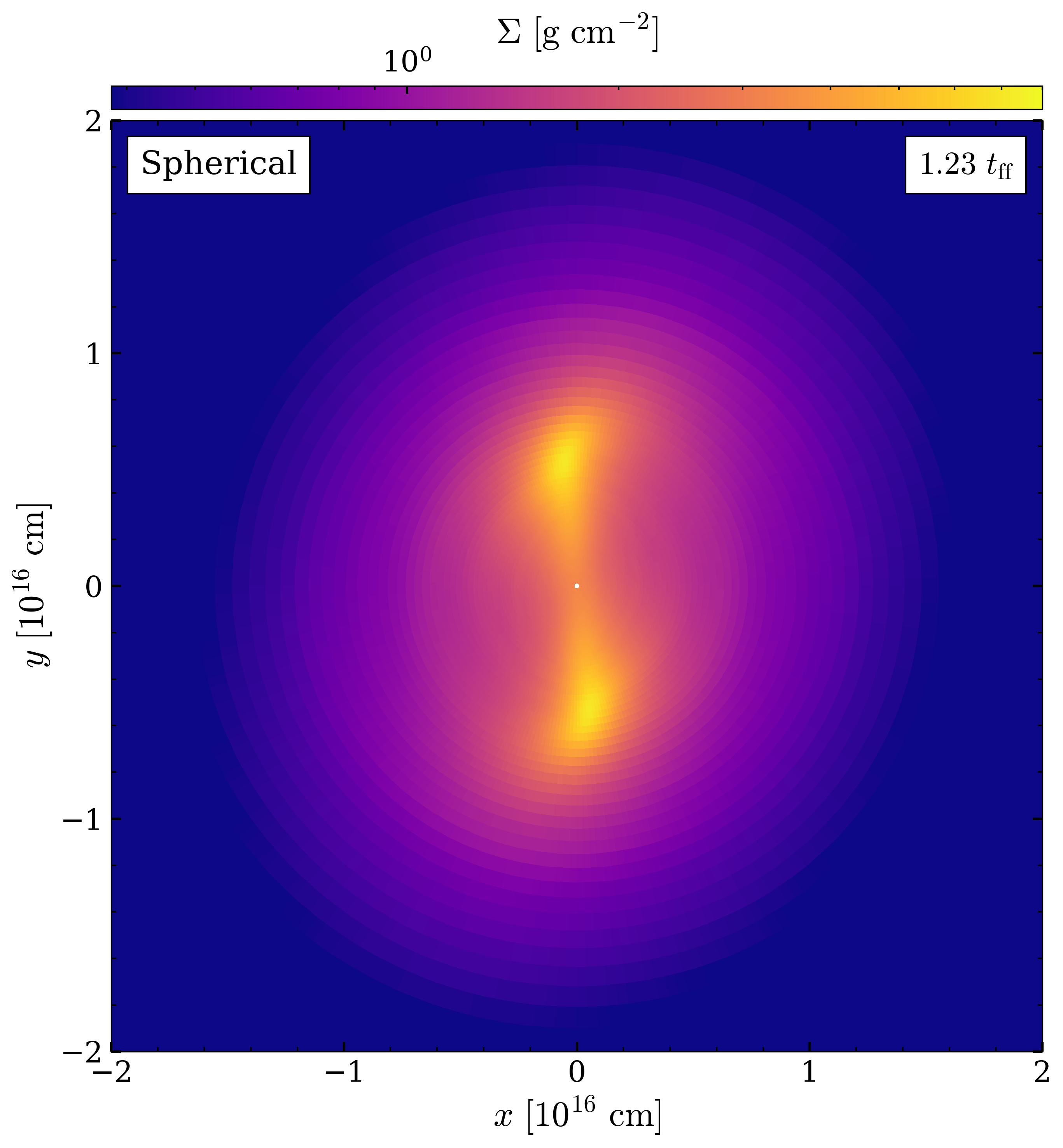} &
  \includegraphics[width=0.4\linewidth]{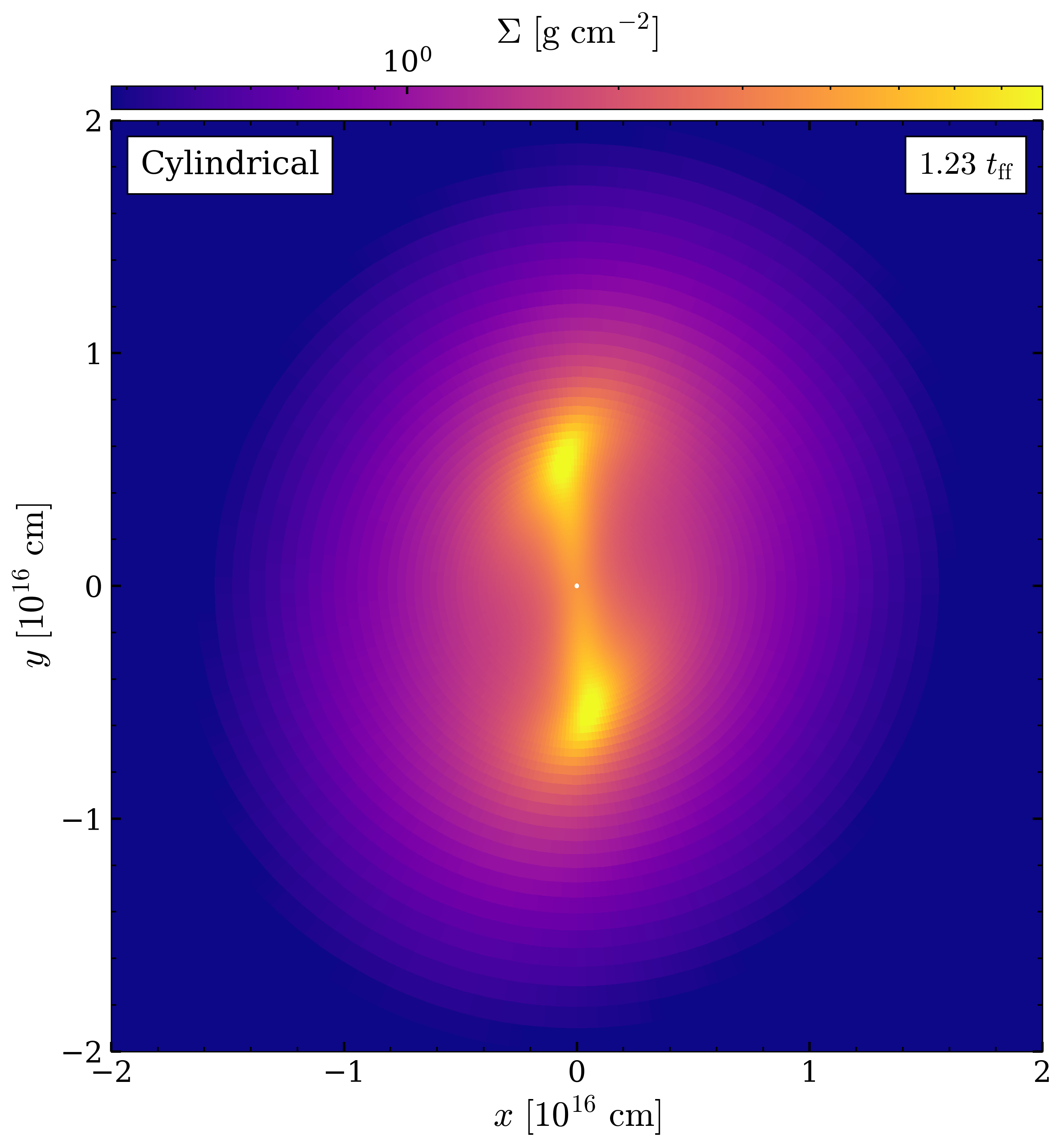} \\
  \includegraphics[width=0.4\linewidth]{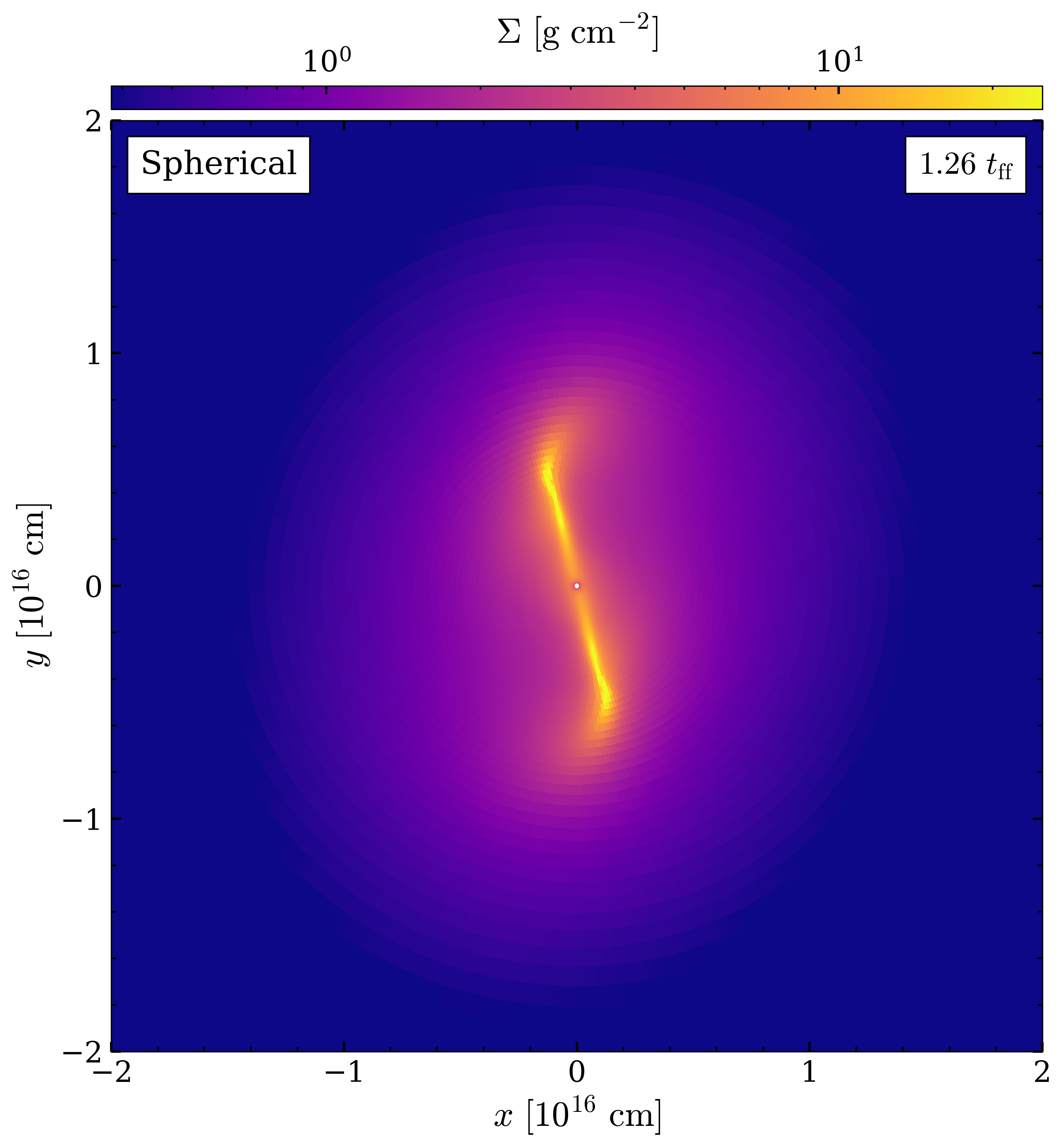} &
  \includegraphics[width=0.4\linewidth]{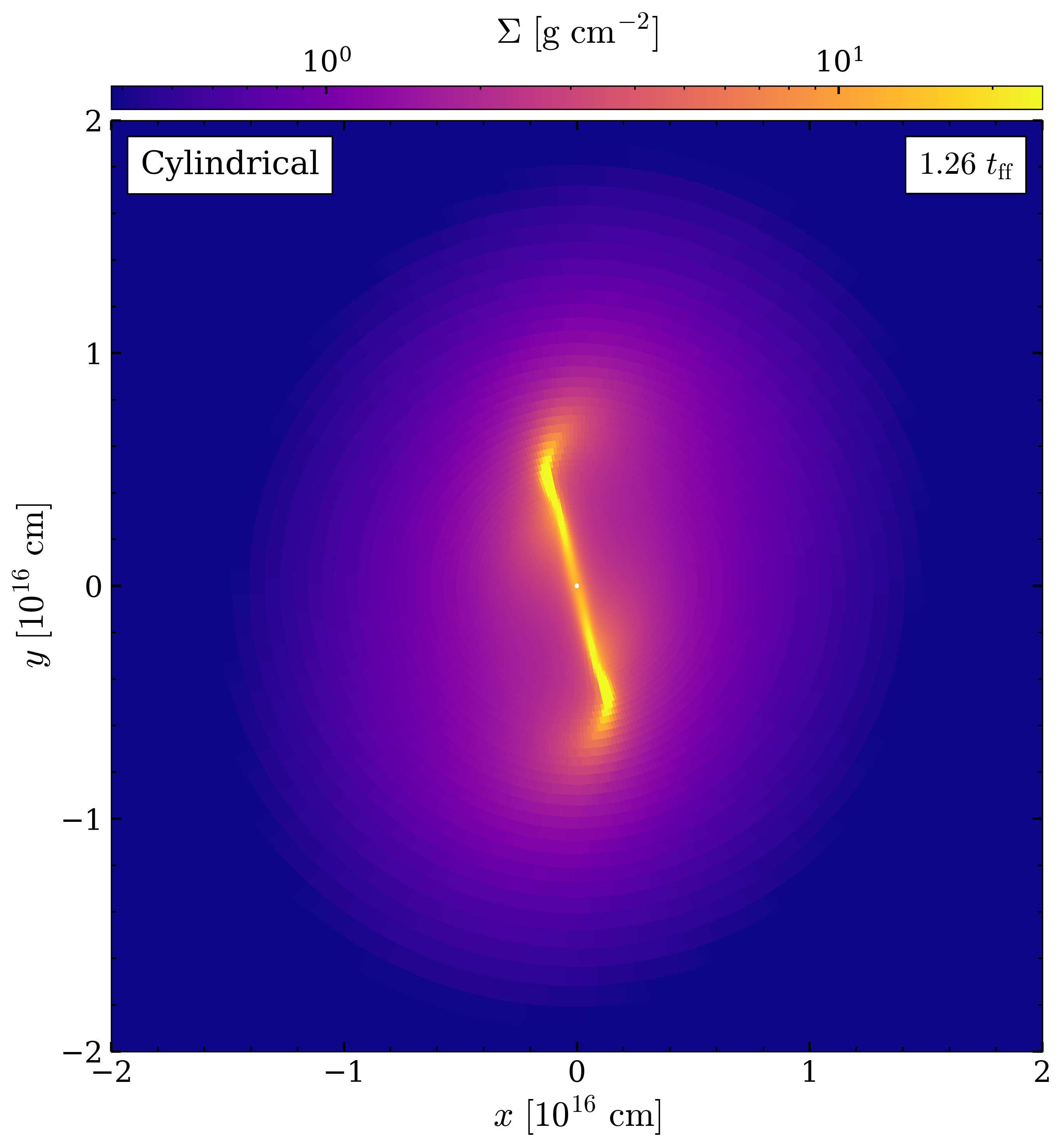}
\end{tabular}
  \caption{The surface density distribution of the rotating cloud with $m=2$ mode perturbation at $t=1.23~\tff$ (top panels) and $1.26~\tff$ (bottom panels), computed on the spherical (left) and cylindrical (right) grids.
  Note that, the top and bottom panels uses different color scale in order to show the structures at the respective time more clearly.}
  \label{fig:rotatingP_proj}
\end{figure*}
\subsubsection{Rotating sphere with \mbox{m=2} mode perturbation}
\label{sec:rotating_perturbed}
Lastly, we consider the highly demanding problem of fragmentation in a rotating, initially perturbed cloud, first introduced by \citet{Boss_1979} and subsequently adopted in various forms as a benchmark for self-gravitating hydrodynamics codes \cite[e.g.,][]{Burkert_1993,Truelove_1997,Truelove_1998,Ziegler_2005,Springel_2005,Mandal_2023}. 
This test probes the nonlinear interplay between gravity, rotation, and non-axisymmetric perturbations, and is particularly sensitive to the accurate treatment of self-gravity.

The original Boss \& Bodenheimer configuration employed an isothermal, uniformly dense cloud seeded with a $50\%$ azimuthal $m=2$ density perturbation. 
\citet{Burkert_1993} demonstrated that a reduced perturbation amplitude of $10\%$ produces a more challenging and discriminating test, as fragmentation then results from the natural amplification of a small non-axisymmetric seed rather than from a strongly imposed asymmetry. 
This weaker perturbation has therefore become a standard benchmark for assessing numerical accuracy and gravitational solvers. 
Here we adopt the \citet{Burkert_1993} version.

The setup consists of an isothermal spherical cloud with sound speed $\cs = 0.167~\kms$, mass $M = 1~M_\odot$, and radius $R = 5\times10^{16}~\mathrm{cm}$. The cloud rotates rigidly with angular velocity $\omega = 7.4\times10^{-13}~\mathrm{rad~s^{-1}}$, corresponding to energy ratios $\alpha = 0.26$ (thermal-to-gravitational) and $\beta = 0.16$ (rotational-to-gravitational). The initial density distribution is
\begin{equation}
    \rho =
    \begin{cases}
        \rho_0 \left[1 + 0.1 \cos(2\phi)\right], & r \le R, \\
        0, & r > R,
    \end{cases}
\end{equation}
where $\rho_0 = 3.82\times10^{-18}~\mathrm{g~cm^{-3}}$ is the uniform background density and $\phi$ is the azimuthal angle about the rotation axis. 
Outside the cloud, the ambient density is set to $0.01~\rho_0$.

The computational domain spans the radial range $[0.001,0.6]\times10^{17}~{\rm cm}$, with a resolution of $128\times128\times256$ in spherical geometry and $128\times256\times256$ in cylindrical geometry. 
This test is particularly demanding for the Poisson solver because the fragmentation process is driven by growth of the $m=2$ mode and subsequent nonlinear gravitational interaction between the forming condensations. 
Accurate mode representation and consistent gravitational coupling are therefore essential.

The top panel of Fig.~\ref{fig:rotatingP_proj} shows the surface density distribution in the equatorial plane at $t = 1.23~\tff$ for both spherical (left) and cylindrical (right) grids. 
The initially small $m=2$ perturbation has grown significantly, leading to the formation of two dense condensations connected by a bar-like structure. 
Each condensation is elongated along the bar axis and is proceeding toward local collapse. 
The morphology and symmetry of the solution closely resemble those reported by \citet{Truelove_1998}, indicating that the non-axisymmetric growth is captured accurately in both coordinate systems.

\begin{figure*}[t]
\centering
\setlength{\tabcolsep}{0pt}
\renewcommand{\arraystretch}{1.0}
\begin{tabular}{cc}
  \includegraphics[width=0.5\linewidth]{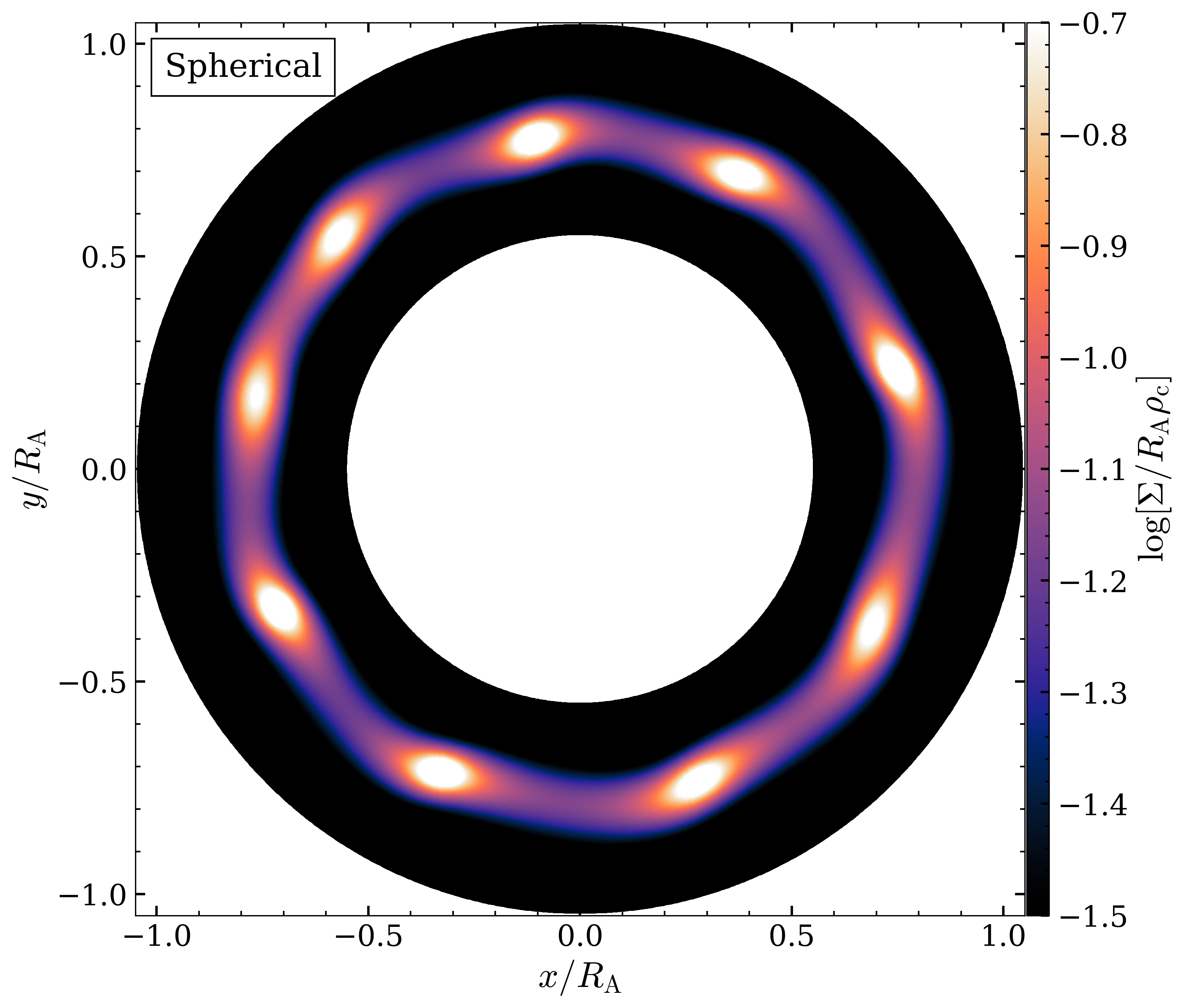} &
  \includegraphics[width=0.5\linewidth]{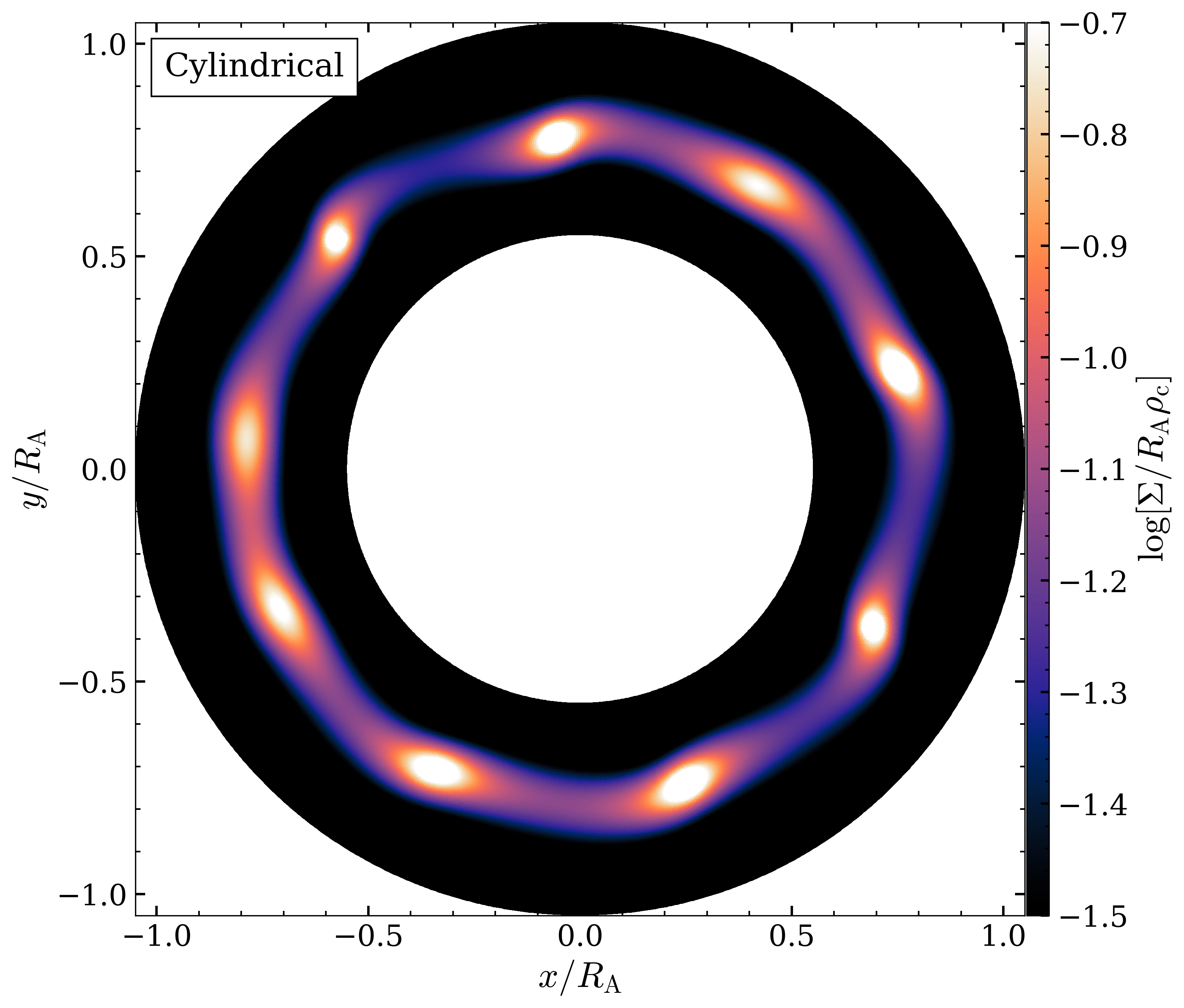} \\
    \includegraphics[width=0.5\linewidth]{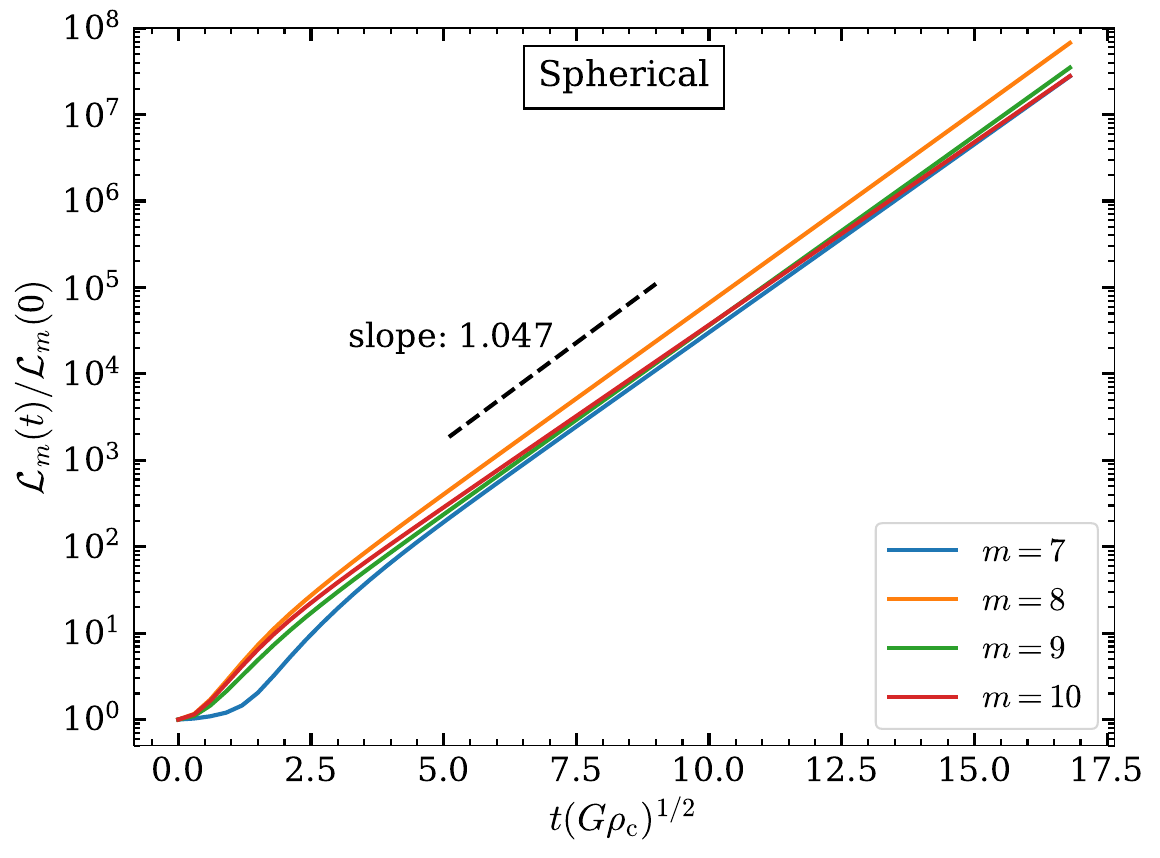} &
  \includegraphics[width=0.5\linewidth]{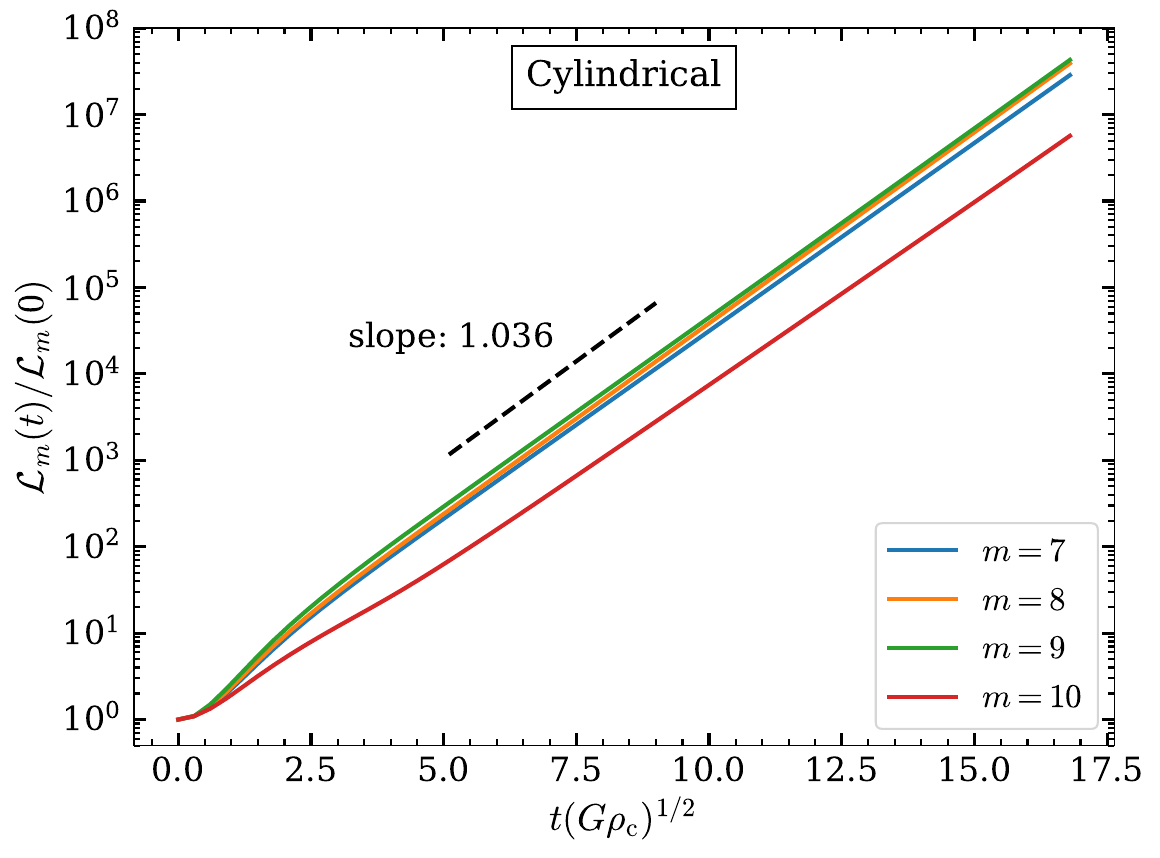}
\end{tabular}
  \caption{\emph{Top row}: Surface density ($\Sigma = \int\rho dz$) distribution for the ring fragmentation test at $t(G\rho_{\rm c})^{1/2} = 18.0$. The left panel shows the result from the spherical grid simulation, while the right panel corresponds to cylindrical grid.
  \emph{Bottom row}: Time evolution of the Fourier amplitude $\mathcal{L}_m$ of the integrated ring density for $m=7,8,9$ and $10$ in spherical (left) and cylindrical (right) grid. The amplitudes are normalized by their initial values. The black dashed line in both panel shows the slope of the growth rate of the fastest growing mode, closely following the predicted value by \citet{Kim_2016} from linear stability analysis.}
  \label{fig:ring_surface_growth}
\end{figure*}

As discussed by \citet{Truelove_1998}, in the purely isothermal case without additional physics such as heating or viscosity, continued collapse drives both the fragments and the connecting bar toward singular states \cite[see also][]{Miyama_1987,Inutsuka_1997}. 
The bar becomes narrower progressively and approaches filamentary collapse. 
In this regime, the Jeans length becomes unresolved unless adaptive refinement or additional physics is introduced, and the system is prone to artificial fragmentation if insufficient resolution is used. 
Nonetheless, the early nonlinear evolution up to this stage provides a stringent and widely accepted benchmark.

The bottom panel of Fig.~\ref{fig:rotatingP_proj} shows the surface density at $t = 1.26~\tff$. 
The two clumps continue to contract toward singular collapse, while the bar connecting them becomes progressively denser and thinner. 
The symmetry, perturbation growth, and overall structural evolution are consistent between the spherical and cylindrical grids and closely resemble the results reported in previous studies \cite[e.g.,][]{Burkert_1993, Truelove_1998}. 
This agreement demonstrates that the solver accurately captures the growth of non-axisymmetric perturbations and the nonlinear gravitational dynamics governing fragmentation.

\subsection{Ring fragmentation}
\label{sec:ring_fragmentation}

As a final validation of our Poisson solver, we examine the fragmentation of a self-gravitating isothermal ring, a problem that has been investigated extensively in previous work \cite[e.g.,][]{Kim_2016,Moon_2019,Ahn_2025}. 
The system consists of a rigidly rotating ring with angular velocity $\Omega_0$ and sound speed $\cs$, embedded in a hot, low density background medium. 
Initially, the ring is in steady equilibrium under the combined effects of self-gravity and external gravity. Assuming, the external gravitational acceleration, $\boldsymbol{g}_{\rm ext} = -\Omega_{\rm e}^2\sin\theta\boldsymbol{r}$ in spherical coordinate, or equivalently $-\Omega_{\rm e}^2\boldsymbol{R}$ in cylindrical coordinate, alone would produce rigid rotation at angular velocity $\Omega_{\rm e}$, the equilibrium condition is given by
\begin{equation}\label{eq:ring_equilibrium}
    \cs^2\boldsymbol{\nabla}\ln{\rho} + \boldsymbol{\nabla}\Phi = \Omega^2_{\rm s}\boldsymbol{R},
\end{equation}
where $\Omega_{\rm s} = \left(\Omega_0^2 - \Omega_{\rm e}^2\right)^{1/2}$ corresponds to the angular velocity associated with self-gravity alone \cite[e.g.,][]{Kim_2016}. For spherical coordinate, $\boldsymbol{R} = \sin\theta\boldsymbol{r}$. It can be shown that, the equilibrium structures are completely described in terms of two dimensionless parameters \citep{Kim_2016}: $\alpha \equiv \cs^2/(G R_{\rm A}\rho_{\rm c})$ and $\hat{\Omega}{\rm s}\equiv\Omega{\rm s}/\sqrt{G\rho_{\rm c}}$, where $\rho_{\rm c}$ and $R_{\rm A}$ denote the maximum density and maximum radial extent of the ring in equilibrium.

For the simulations, we use a radially uniform grid with radial extent $ [0.55,1.05]$ in both spherical and cylindrical coordinates. In spherical coordinates, the polar angle spans $\theta \in [0.4\pi,0.6\pi]$, while in cylindrical coordinates the vertical domain covers $z \in [-0.23,0.23]$. The full azimuthal domain is included, $\phi \in [0,2\pi]$. The grid resolution is $(N_r, N_\theta, N_\phi) = (128,128,1024)$ in spherical coordinates and $(N_R, N_\phi, N_z) = (128,1024,128)$ in cylindrical coordinates.

\begin{figure*}[t]
\centering
\setlength{\tabcolsep}{0pt}
\renewcommand{\arraystretch}{1.0}
\begin{tabular}{cc}
  \includegraphics[width=0.5\linewidth]{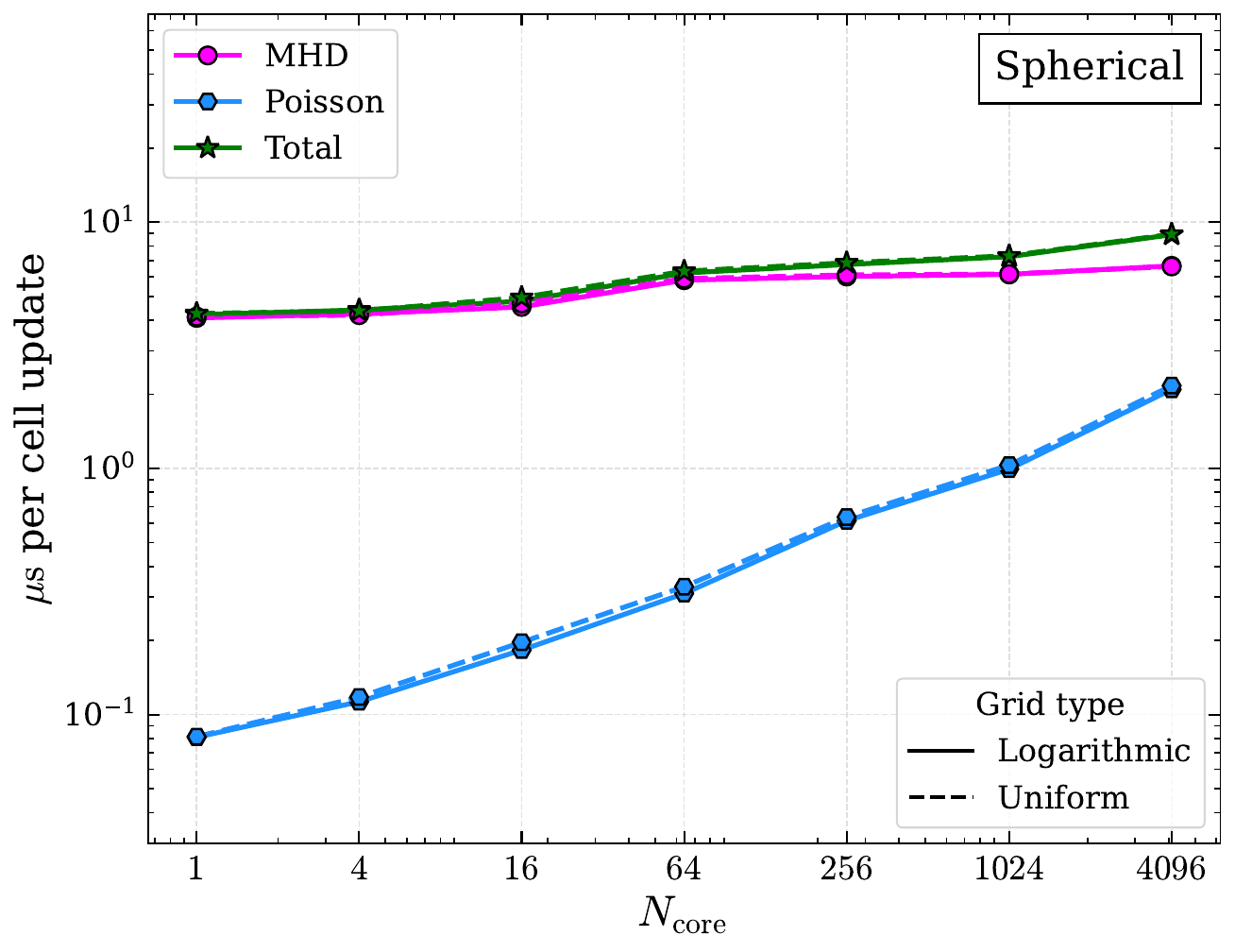} &
  \includegraphics[width=0.5\linewidth]{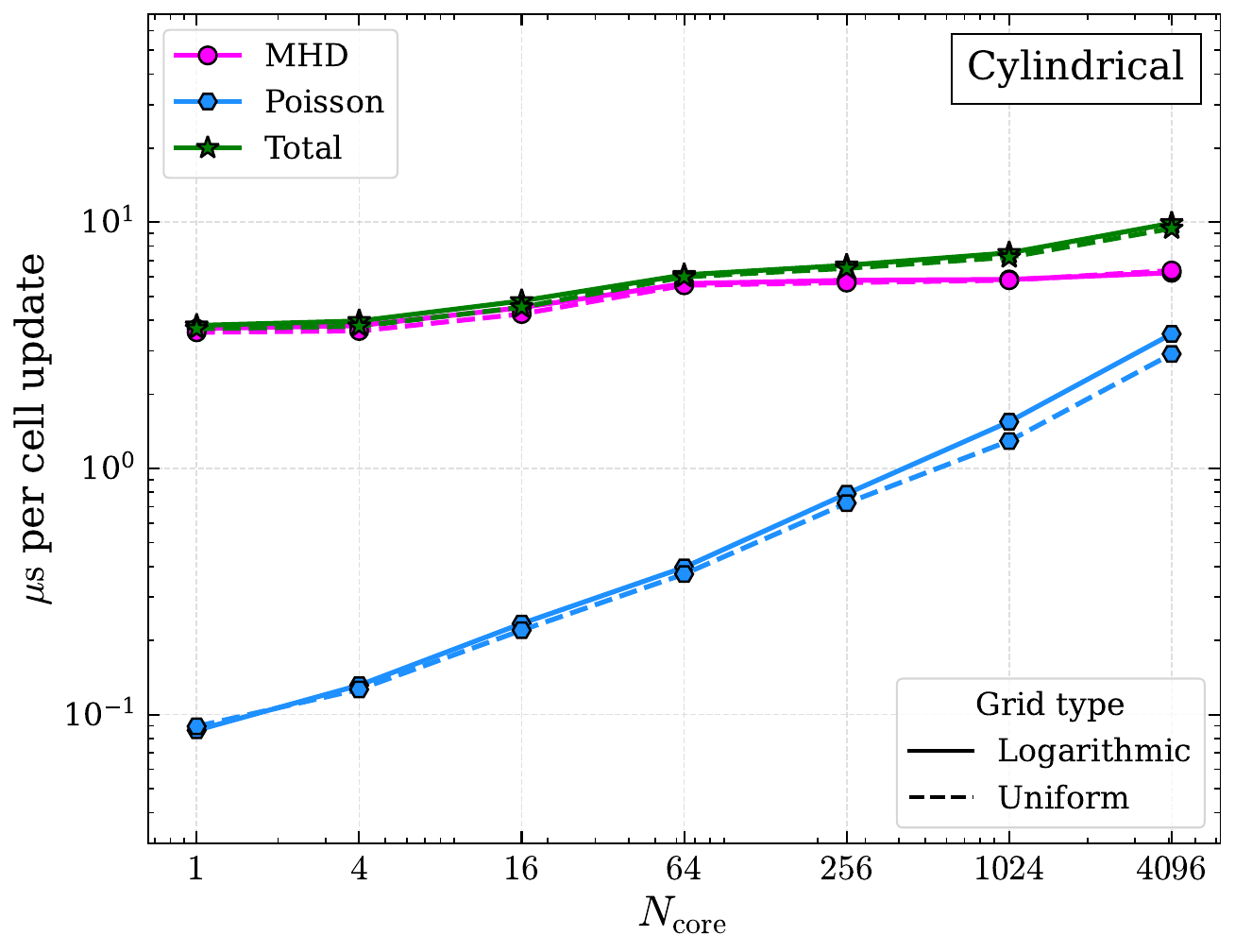}
\end{tabular}
  \caption{Result of weak scaling test. The average per-cell update time, $t_{\rm update}$, is plotted as a function of core count, $N_{\rm core}$, for spherical (left) and cylindrical (right) grids. Solid lines denote radially logarithmic grids, while dashed lines denote uniform radial grids. Blue hexagons show the total time taken by the Poisson solver, from density input to the computed potential. Magenta circles indicate the MHD update time per cell, and green stars show the total update time per cell.}
  \label{fig:weak_scaling}
\end{figure*}

Since no analytical solution exists for the equilibrium configuration described by Eq.~(\ref{eq:ring_equilibrium}), we construct the equilibrium state using the self-consistent field method of \citet{Hachisu_1986}, adopting $\alpha = 0.01$ and $\hat{\Omega}_{\rm s} = 0.27$. 
We then apply the external gravitational acceleration and increase the angular velocity to $\hat{\Omega}_0 = \Omega_0/(G\rho_{\rm c})^{1/2} = 0.4$. In the absence of perturbations, this configuration remains stable. 
However, \citet{Kim_2016} showed that the equilibrium becomes unstable to non-axisymmetric perturbations over a range of azimuthal mode numbers, depending on the values of $\alpha$ and $\hat{\Omega}_0$. 
For the parameters adopted here, the most unstable modes lie in the range $m \approx 7\mbox{-}10$ \cite[e.g., see Fig.~11 of][]{Kim_2016} and the growth rate of the most unstable mode (e.g., $m\approx8$) has a growth rate ${\rm Im}(\omega) \approx 1.01(G\rho_{\rm c})^{1/2}$. Note that this value was calculated by \citet{Kim_2016} for $\alpha=0.0115$; for our case with $\alpha=0.01$, the growth rate should be slightly higher.
We therefore examine whether our Poisson solver can correctly capture these unstable modes. 
To this end, we introduce random density perturbations with amplitude $10^{-5}$ to the equilibrium configuration and evolve the system.

The top row of Fig.~\ref{fig:ring_surface_growth} shows the surface density distribution of the simulations at $t(G\rho_{\rm c})^{1/2}=18.0$ on the spherical (left) and cylindrical (right) grids. 
In both cases, nine prominent clumps are visible along the ring, indicating the growth of unstable azimuthal modes. 
This suggests that the fastest-growing modes fall within the linearly predicted unstable azimuthal range of $7$–$10$.

This is further illustrated in the bottom row of Fig.~\ref{fig:ring_surface_growth}, which shows the evolution of the Fourier amplitude $\mathcal{L}_m$ of the integrated density, defined as $\bar{\rho}(\phi) = \int \rho r^2 \sin\theta dr d\theta$ for the spherical grid and $\int \rho dR dz$ for the cylindrical grid. 
The Fourier amplitude is given by $\mathcal{L}_m = e^{i m \phi_m} \int \bar{\rho}(\phi) e^{-i m \phi}  d\phi$, where $\phi_m$ is the phase angle of the $m^{\rm th}$ mode.

In both geometries, the $m=8$ and $9$ modes exhibit the fastest growth. The modes have a measured growth rate of ${\rm Im}(\omega)\approx1.04(G\rho_{\rm c})^{1/2}$ , which is very close to the analytical value predicted by the linear analysis of \citet{Kim_2016} (see discussion at the beginning of this section).
A small difference in the growth rates and the relative ordering of these dominant modes is observed between the spherical and cylindrical grids. This discrepancy likely arises from differences in grid geometry, cell aspect ratios, and resolution. 
Nevertheless, the overall agreement with theoretical expectations indicates that our Poisson solver accurately captures the development of the gravitational instability.

\section{Performance and scaling test}
\label{sec:scaling_test}

In this section, we evaluate the parallel scalability of our Poisson solver on the \texttt{Barnard} high-performance computing system at the NHR Center of TU Dresden\footnote{\url{https://compendium.hpc.tu-dresden.de/jobs_and_resources/barnard}}. 
Each compute node is equipped with two 54-core \texttt{Intel(R) Xeon(R) Platinum 8470} processors operating at a base frequency of 2.00~GHz, and nodes are interconnected via a \texttt{Mellanox ConnectX-6 InfiniBand} network. 
All MPI communications are handled using the \texttt{OpenMPI v4.1.6} library\footnote{\url{https://www.open-mpi.org}}. For the test, we, however, use 64 cores per nodes.

For this analysis, we consider a protoplanetary disk (PPD) in hydrostatic equilibrium under the gravitational potential of a central object \citep[e.g.,][]{Nelson_2013}. In constructing the initial radial and vertical profiles of density, pressure, and azimuthal velocity, self-gravity is neglected. The disk is permeated by a uniform vertical magnetic field, and a $1\%$ density perturbation is introduced. An ideal equation of state with $\gamma = 5/3$ is assumed.
The numerical setup employs an HLLD Riemann solver \citep{Miyoshi_2005} with piecewise parabolic reconstruction \citep{Colella_1984}, second-order Runge–Kutta time integration, and constrained transport \citep{Mignone_2021} to enforce the divergence-free condition of the magnetic field. These numerical methods are standard for accurately modeling PPDs. The specific values of the physical parameters are not important for the purposes of this section.

With this setup, we perform a weak scaling test to evaluate how the computational performance varies with the number of cores while keeping the workload per core constant. 
Since the azimuthal direction ($\phi$) is not partitioned, each core is assigned a domain of $16\times16\times512$ in spherical geometry and $16\times512\times16$ in cylindrical geometry. 
The total number of cores is varied from $N_{\rm core}=1, 4, 16, 64, 256, 1024$ up to $4096$, with processors arranged in a $\sqrt{N_{\rm core}}\times\sqrt{N_{\rm core}}$ decomposition over the $r$–$\theta$ (spherical) or $R$–$z$ (cylindrical) plane. 
Therefore, at $4096$ cores, we have a domain resolution of $1024\times1024\times512$ (spherical) or $1024\times512\times1024$ (cylindrical). 
The scaling tests are performed for all considered grid configurations, including both radially uniform and logarithmic grids in the spherical and cylindrical coordinate systems.

The simulation is run for 200 time steps, including both the MHD update and the Poisson solve. 
The wall time per step ($t_{\rm wall}$) is obtained by averaging over all time steps, and the corresponding time to update a single cell per step is computed as $t_{\rm update} = t_{\rm wall}/(16 \times 16 \times 512)$. 
The time required to compute the DGF is excluded, as this operation is performed only once during initialization.
To ensure robust measurements and minimize the impact of performance variability, the test is repeated five times and the results are averaged, reducing the influence of occasional slow runs caused by node-level fluctuations.

Unlike some previous studies \cite[e.g.,][]{Gressel_2024}, we do not fix the number of multigrid V-cycles. Instead, iterations proceed until a specified tolerance of $10^{-8}$ is reached. This criterion provides a more realistic and practically relevant assessment of performance under production conditions.

Fig.~\ref{fig:weak_scaling} shows the per-cell update time ($t_{\rm update}$) as a function of core count ($N_{\rm core}$) for spherical (left panel) and cylindrical (right panel) grids.
In both panels, solid lines correspond to radially logarithmic grids, while dashed lines correspond to uniform radial grids. 
Blue hexagons indicate the total time required for the Poisson solve, from the input density to the resulting potential, without separating individual solver components (see Sec.~\ref{sec:algorithm_overview}).
Remarkably, the total Poisson time remains below the MHD update time (magenta circles) up to 4096 cores for all grid configurations. 
For up to $N_{\rm core}=1024$, the Poisson steps require nearly an order of magnitude less time than the MHD updates. 
At larger core counts, its relative cost increases; nevertheless, even at $N_{\rm core} = 4096$, the Poisson solver remains approximately a factor of 2–3 faster than the MHD update.

We observe that the weak scaling behavior of the Poisson solver is not ideal, as the per-cell update time increases with core count. 
Specifically, $t_{\rm update}$ grows by nearly a factor of 20 between 1 and 4096 cores. To better understand the origin of this increase, we provide a component-wise breakdown of the Poisson solver (FFT, multigrid solvers, and DGF convolution) in Appendix~\ref{sec:scaling_grav}. The breakdown reveals that both the multigrid and DGF components increase with problem size, while the FFT remains flat and subdominant. At lower core counts, the multigrid step dominates; at higher core counts, the DGF convolution becomes comparable to or greater than the combined cost of the two multigrid steps.

Several factors contribute to the observed scaling.
While each core handles a fixed number of cells, the global resolution grows as more cores are added. 
Unlike a uniform Cartesian grid, the convergence rate of the multigrid solver for a variable-coefficient elliptic problem on a non-uniform grid is not grid-independent \cite[see, e.g.,][]{Trottenberg_2000}, which is the case for our Poisson solver in curvilinear coordinates.
Thus, the multigrid solver requires more V-cycles to reach a fixed tolerance at higher global resolution, leading to a mild increase in the multigrid cost per cell (see Appendix~\ref{sec:multigrid_convergence}).
The DGF convolution cost increases with problem size due to its $\mathcal{O}[(N_r+N_\theta)^2N_\phi/2]$ scaling.
In addition, because each Helmholtz equation is solved independently, boundary exchanges must be performed separately for each azimuthal mode, introducing additional global communication overhead that becomes more significant at large core counts.

Despite these effects, the Poisson solver remains highly efficient. 
Even at 4096 cores, it is 2-3 times faster than the MHD update, keeping its contribution to the total runtime small. 
These results demonstrate that our solver is both fast and practical for large-scale simulations, combining high computational efficiency with scalability across thousands of cores.

\section{Summary}
\label{sec:summary}
We have developed and implemented a Poisson solver for self-gravitating fluid dynamics in spherical and cylindrical coordinates that combines finite-volume discretization (Sec.~\ref{sec:discretization}), azimuthal Fourier decomposition (Sec.~\ref{sec:fourier_decompostion}), a geometrically consistent multigrid algorithm (Sec.~\ref{sec:multigrid_solver}), and the James method for vacuum boundary conditions (Sec.~\ref{sec:boundary_calculation}). The finite-volume formulation preserves conservation properties and achieves second-order accuracy in curvilinear geometries by consistently incorporating metric terms and volume-weighted cell centroids. This accuracy is maintained on both uniform and logarithmically stretched grids, enabling reliable treatment of problems spanning large dynamic ranges in spatial scale.

The azimuthal Fourier decomposition transforms the three-dimensional Poisson equation into a sequence of independent two-dimensional Helmholtz problems, one for each azimuthal mode. This formulation preserves the full non-axisymmetric structure of the gravitational field while enabling efficient solution via specialized multigrid solvers in two dimensions. A key conceptual advantage of the mode-by-mode approach is that different azimuthal modes converge at different rates: higher-$m$ modes, representing finer azimuthal structures, typically converge very rapidly—often within only one or two V-cycles, while lower-$m$ modes may require more iterations. Solving modes independently avoids unnecessary computation for rapidly converging modes, an efficiency that would not be available in a full 3D solver treating the entire grid simultaneously.

The multigrid algorithm described in Sec.~\ref{sec:multigrid_solver} employs area-weighted restriction and prolongation operators (Sec.~\ref{sec:transfer_operator}) to ensure consistency across grid levels and maintain second-order accuracy throughout the hierarchy. Convergence is further accelerated using the alternating line smoother introduced in Sec.~\ref{sec:line_smoother}, which is specifically designed to handle anisotropies arising from metric coefficients and non-uniform grid spacing. These anisotropies are particularly pronounced on logarithmically stretched grids, where line-based relaxation provides substantially more effective error damping than pointwise methods. Finally, the discrete Green’s function formulation used in the James method (Sec.~\ref{sec:boundary_calculation}) enables accurate and efficient treatment of vacuum boundary conditions, including configurations with inner cavities.

The solver has been validated through an extensive and diverse set of tests covering a wide range of physically relevant grid configurations and dynamical regimes. Static mesh-segment tests confirm that second-order convergence is achieved across all grid types, including both uniform and logarithmically stretched radial grids. Dynamical tests were designed to probe challenging scenarios, from logarithmic grids with large dynamic ranges ($R_{\max}/R_{\min} \sim 10^3$) typical of gravitational collapse and star formation simulations, to uniform or moderately extended grids relevant for protoplanetary disk and disk evolution studies. Across all configurations, in both spherical and cylindrical geometries, the solver maintains excellent accuracy. Collapse, disk formation, and fragmentation tests further demonstrate that the solver reproduces analytic expectations and previously published numerical results, confirming its robustness in fully time-dependent, self-gravitating flows.

Parallel performance tests on up to 4096 processors show that the solver scales efficiently and remains significantly faster than the MHD update across all tested core counts. Although weak scaling efficiency decreases at very large processor counts due to the intrinsic convergence properties of elliptic solvers and the communication overhead associated with the mode-by-mode solution strategy, the Poisson solve continues to represent a subdominant fraction of the total computational cost in production simulations. This performance, combined with its accuracy, geometric flexibility, and robustness across a wide range of grid configurations, makes the solver well suited for demanding astrophysical applications, including gravitational collapse, disk evolution, and fragmentation processes spanning many orders of magnitude in spatial scale.

\begin{acknowledgments}
This work was supported by the European Union (ERCCoG, Epoch-of-Taurus, No. 101043302). Views and opinions expressed are
however those of the author(s) only and do not necessarily reflect those of the European Union or the European Research Council. Neither the European
Union nor the granting authority can be held responsible for them. The authors gratefully acknowledge the computing time made available to them on the high-performance computer at the NHR Center of TU Dresden. This center is jointly supported by the Federal Ministry of Research, Technology and Space of Germany and the state governments participating in the NHR (\url{www.nhr-verein.de/unsere-partner}). A.~M. thanks Steven Rendon Restrepo for detailed feedback on the draft, and Moun Meenakshi and Marc van den Bossche for valuable discussions throughout the project.
\end{acknowledgments}

%

\software{\textsc{Pluto} \citep{Mignone_2007}, \textsc{FFTW3} \citep{FFTW}}


\appendix

\section{Approximate Optimal Over-Relaxation Parameter for Non-Uniform Curvilinear Grids}
\label{sec:SOR_omega}
We consider the variable–coefficient tridiagonal system
\begin{equation}\label{eq:tridiagonal_system}
    a_j u_{j-1} + b_j u_j + c_j u_{j+1} = f_j,  \quad j = 1, \dots, N,
\end{equation}
subject to Dirichlet boundary conditions
\begin{equation}
    u_0 = u_{N+1} = 0.
\end{equation}
Such systems arise naturally from finite–difference discretizations of elliptic operators on non-uniform grids, for example in cylindrical or spherical geometries where grid spacing and metric factors vary spatially.

The system matrix can be decomposed as
\begin{equation}
    \mathbf{A} = \mathbf{L} + \mathbf{D} + \mathbf{U},
\end{equation}
where
\begin{equation}
    \mathbf{D} = \mathrm{diag}(b_1, \dots, b_N)
\end{equation}
is the diagonal part, and $\mathbf{L}$ and $\mathbf{U}$ are the strictly lower and upper triangular parts, respectively.

The Jacobi iteration scheme is given by \citep{Press_1992}
\begin{equation}
    \mathbf{u}^{(k+1)} = \mathbf{M}_{\rm J}\mathbf{u}^{(k)} + \mathbf{D}^{-1}\mathbf{f},
\end{equation}
where the iteration matrix ($\mathbf{M_J}$) is
\begin{equation}
    \mathbf{M}_{\rm J} = -\mathbf{D}^{-1}(\mathbf{L}+\mathbf{U}).
\end{equation}
The convergence rate of the Jacobi method is governed by the spectral radius
\begin{equation}
    \varrho(\mathbf{M}_{\rm J}) = \max |\mu|,
\end{equation}
where $\mu$ are the eigenvalues of $\mathbf{M}_{\rm J}$.

\subsection{Generalized Eigenvalue Problem}
The eigenvalues satisfy
\begin{equation}
    \mathbf{M}_{\rm J} \mathbf{v} = \mu \mathbf{v},
\end{equation}
which is equivalent to the generalized eigenvalue problem
\begin{equation}
    -(\mathbf{L}+\mathbf{U})\mathbf{v} = \mu \mathbf{D} \mathbf{v}.
\label{eq:generalized_eigen}
\end{equation}

In component form, this becomes
\begin{equation}
    -(a_j v_{j-1} + c_j v_{j+1}) = \mu b_j v_j, \quad j = 1, \dots, N,
\label{eq:component_eigen}
\end{equation}
with boundary conditions
\begin{equation}
    v_0 = v_{N+1} = 0.
\end{equation}

Multiplying Eq.~(\ref{eq:component_eigen}) by $v_j$ and summing over all interior points yields the exact Rayleigh quotient \citep{Horn_Johnson_1985}
\begin{equation}
    \mu = -\frac{\DS\sum_{j=1}^N\left(a_j v_{j-1} v_j + c_j v_{j+1} v_j\right)}{\DS\sum_{j=1}^Nb_j v_j^2}.
\label{eq:rayleigh_exact}
\end{equation}

Thus, estimating eigenvalues reduces to constructing accurate approximations to the eigenvectors.

\subsection{Approximate Eigenvectors on a Non-Uniform Grid}
For uniform grids with constant coefficients, the eigenvectors are discrete sine modes,
\begin{equation}
    v_j^{(p)} = \sin\left(\frac{p \pi j}{N+1}\right), \quad p = 1, \dots, N.
\end{equation}
These modes correspond to standing waves satisfying the boundary conditions. On non-uniform grids, the eigenvectors remain oscillatory, but their phase evolves according to the physical coordinate rather than the grid index. This behavior can be understood using a WKB-type approximation \citep{Bender_1999} for second-order difference equations.

Let the physical grid points be denoted by $x_j$, and define the normalized coordinate
\begin{equation}
    \xi_j = \frac{x_j - x_0}{x_{N+1} - x_0} = \frac{x_j - x_0}{L},
\end{equation}
where $L$ is the domain length.

This coordinate increases monotonically from 0 to 1 across the domain. Motivated by the continuous eigenfunctions of the Laplacian, we approximate the eigenvectors as
\begin{equation}\label{eq:approx_eigenvector}
    v_j^{(p)} = \sin\left(p \pi \xi_j\right) = \sin\left(p \pi \frac{x_j - x_0}{L}\right).
\end{equation}

This approximation satisfies the boundary conditions exactly and reduces to the classical sine modes when the grid is uniform.

\subsection{Evaluation of the Rayleigh Quotient}
Define the phase
\begin{equation}
    \phi_j = p \pi \frac{x_j - x_0}{L}.
\end{equation}

Let the local grid spacing be
\begin{equation}
    \delta_j = x_j - x_{j-1}.
\end{equation}

Then the phase increment is
\begin{equation}
    \Delta \phi_j = p \pi \frac{\delta_j}{L}.
\end{equation}

Using trigonometric identities,
\begin{equation}
    \frac{v_{j-1}}{v_j} = \frac{\sin(\phi_j - \Delta \phi_j)}{\sin \phi_j} = \cos(\Delta \phi_j) - \cot(\phi_j)\sin(\Delta \phi_j).
\end{equation}

The second term oscillates rapidly with $j$ and averages to zero when summed over many grid points. Retaining only the slowly varying contribution gives
\begin{equation}
    \frac{v_{j-1}}{v_j} \approx \cos\left(p \pi \frac{\delta_j}{L}\right),
\end{equation}
and similarly,
\begin{equation}
    \frac{v_{j+1}}{v_j} \approx \cos\left(p \pi \frac{\delta_{j+1}}{L}\right).
\end{equation}

Substituting into the numerator of Eq.~\eqref{eq:rayleigh_exact} gives
\begin{equation}
    \begin{aligned}
        \mathcal{N}_p &= -\sum_{j=1}^N\left(a_j v_{j-1} v_j + c_j v_{j+1} v_j \right) \\
        &\approx \sum_{j=1}^N (a_j + c_j)
        \cos\left(p \pi \frac{\delta_j}{L}\right)\sin^2\left(p \pi \frac{x_j - x_0}{L}\right).
    \end{aligned}
\end{equation}

The denominator becomes
\begin{equation}
    \mathcal{D}_p = \sum_{j=1}^N b_j \sin^2\left(p \pi \frac{x_j - x_0}{L}\right).
\end{equation}

Thus, the eigenvalues are approximated by
\begin{equation}
    \mu_p \approx\frac{\DS\sum_{j=1}^N(a_j + c_j)\cos\left(p \pi \frac{\delta_j}{L}\right)w_j^{(p)}}
    {\DS\sum_{j=1}^N b_j w_j^{(p)}},
\end{equation}
where
\begin{equation}
    w_j^{(p)} = \sin^2\left( p \pi \frac{x_j - x_0}{L}\right).
\end{equation}
%

\subsection{Spectral Radius Approximation}
For elliptic operators on smoothly varying grids, the largest eigenvalue is generally associated with the $p=1$ mode. The spectral radius can therefore be well approximated by the lowest mode $p=1$,
\begin{equation}
    \varrho(\mathbf{M}_{\rm J}) \approx |\mu_1|.
\end{equation}
This approximation follows from the fact that the lowest mode corresponds to the smoothest spatial variation and therefore has the largest magnitude eigenvalue. Higher modes decay more rapidly and do not determine the asymptotic convergence rate.

Using the Rayleigh quotient estimate, this gives
\begin{equation}
    \varrho(\mathbf{M}_{\rm J}) \approx \left\lvert\frac{\DS\sum_{j=1}^N (a_j + c_j)\cos\left(\pi \frac{\delta_j}{L}\right)w_j^{(1)}}
    {\DS\sum_{j=1}^N b_j w_j^{(1)}}\right\rvert,
\end{equation}
where
\begin{equation}
    w_j^{(1)} = \sin^2\left(\pi \frac{x_j-x_0}{L}\right).
\end{equation}
This expression reduces exactly to the classical uniform-grid result when the coefficients and spacing are constant \citep{Strikwerda_2004}. For smoothly varying grids, it provides a close approximation consistent with WKB theory for difference equations \citep{Bender_1999}.

The preceding methodology extends straightforwardly to two dimensions. In particular, for a two-dimensional elliptic operator of the form
\begin{equation}
    a_{i,j} u_{i-1,j} + b_{i,j} u_{i,j} + c_{i,j} u_{i+1,j} + d_{i,j}u_{i,j-1} + e_{i,j}u_{i,j+1} = f_{i,j}
\end{equation}
the spectral radius of the associated Jacobi iteration matrix may be approximated by
\begin{equation}
    \begin{aligned}
    \varrho(\mathbf{M}_{\rm J}) \approx \max_{p,q}
    \left|
    \frac{\DS \sum_{ij} 
    \left[
    (a_{i,j}+c_{i,j})
    \cos\!\left(\pi\frac{\delta \xi_i}{L_\xi}\right)
    \right.}{\displaystyle \sum_{ij} b_{i,j} w^{p,q}_{i,j}}
    \right.
    \\
    \left.
    \qquad\qquad
    + \frac{\left.\DS \sum_{ij}
    (d_{i,j}+e_{i,j})
    \cos\!\left(\pi\frac{\delta \eta_j}{L_\eta}\right)
    \right] w^{p,q}_{i,j}}
    {\displaystyle \sum_{ij} b_{i,j} w^{p,q}_{i,j}}
    \right|,
    \end{aligned}
\end{equation}
with
\begin{equation}
    w^{p,q}_{i,j} \approx \sin^2\left(p\pi\xi_i\right)\sin^2\left(q\pi\eta_j\right).
\end{equation}

\subsection{Approximate Optimal Relaxation Parameter}

The optimal relaxation parameter is related to the spectral radius of the Jacobi iteration matrix by \cite{Young_1954,Strikwerda_2004}
\begin{equation}\label{eq:omega_opt_rho}
    \omega_{\mathrm{opt}} =\frac{2}{1 + \sqrt{1 - \varrho(\mathbf{M}_{\rm J})^2}}.
\end{equation}

Using the spectral radius estimate derived in the previous section,
\begin{equation}
    \varrho(\mathbf{M}_{\rm J}) \approx |\mu_1|,
\end{equation}
the optimal relaxation parameter can be approximated as
\begin{equation}\label{eq:omega_opt_mu}
    \omega_{\mathrm{opt}}\approx\frac{2}{1 +\sqrt{1 - \mu_1^2}}.
\end{equation}
Because this expression is based on an approximate estimate of the Jacobi spectral radius derived using a WKB approximation, the resulting value of $\omega_{\mathrm{opt}}$ is also an approximation. 
Nevertheless, it provides an accurate and practical estimate for selecting near-optimal relaxation parameters on non-uniform curvilinear grids.

\subsection{Tests}

In order to verify whether the above formalism predicts the near-optimal relaxation parameter $\omega$ in practical computations, we consider two one–dimensional model problems corresponding to the radial and polar operators.
To examine the dependence on grid resolution, all tests are performed using two different numbers of grid cells, $N=128$ and $N=256$. For the radial problem, we additionally consider both uniform and logarithmic grids spanning the domain $r\in[10^{-3},\,1]$.
The polar problem is solved on a uniform grid over $\theta\in[\pi/6, 5\pi/6]$.

In each case, the resulting linear system is solved using successive over-relaxation (SOR) for a sequence of $\omega$ values ranging from 1.6 to 2.0 in 70 equal steps. The empirically optimal relaxation parameter is taken to be the one that minimizes the iteration count required to reduce the residual below $10^{-8}$.

\begin{figure*}[t]
    \centering
    \includegraphics[width=\linewidth]{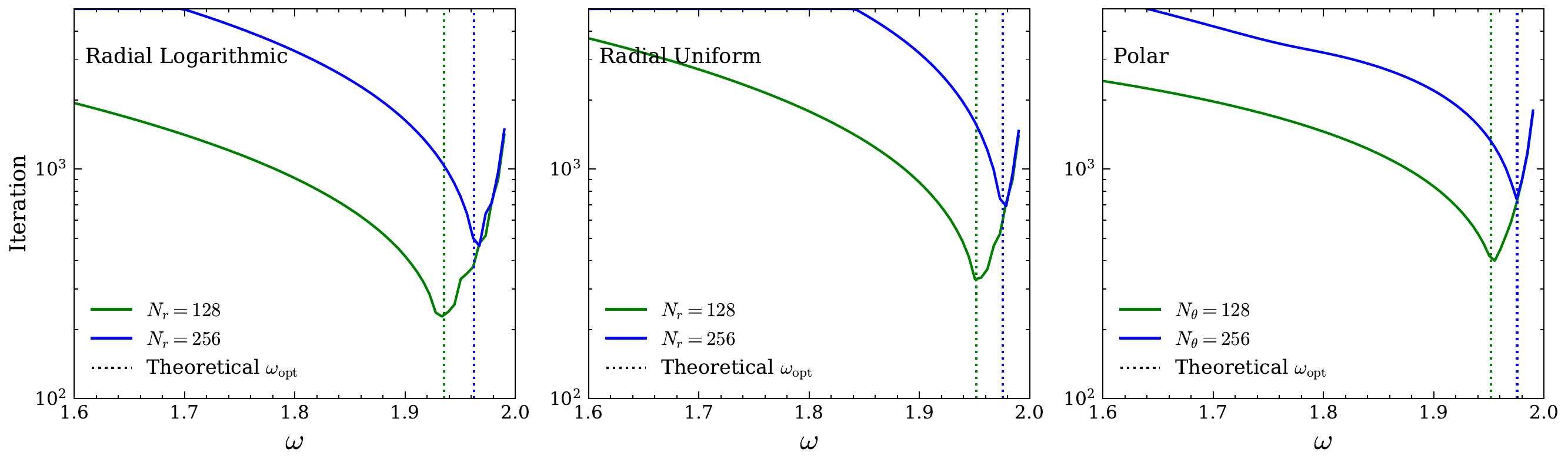}
    \caption{Number of SOR iterations required for convergence as a function of the relaxation parameter $\omega$ for the radial and polar problems. The left and middle panels correspond to logarithmic and uniform radial grids, respectively, while the right panel shows results for a uniform polar grid. In each case, two resolutions are considered: $N=128$ (green) and $N=256$ (blue). The theoretical predictions (vertical lines, color-coded to match each case) are in excellent agreement with the numerically determined optimal values.}
    \label{fig:omega_opt}
\end{figure*}

\subsubsection{Radial problem}
We first consider the spherically symmetric Poisson equation,
\begin{equation}\label{eq:source_3D}
    \frac{1}{r^2}\frac{\del}{\del r}\left(r^2\frac{\del\Phi}{\del r}\right) = 4\pi\rho(r),
\end{equation}
where the density is defined as
\begin{equation}
    \rho(r) =
    \begin{cases}
        1, & \text{if $R_1\le r\le R_2$},\\[4pt]
        0, & \text{otherwise}.
    \end{cases}
\end{equation}

The corresponding analytical solution is
\begin{equation}
    u(r) =
    \begin{cases}
        -2\pi\left(R_2^2-R_1^2\right), & r<R_1,\\[6pt]
        -\DS\frac{2\pi}{3r}\left(3R_2^2r - 2R_1^3 - r^3\right), & R_1\le r\le R_2,\\[6pt]
        -\DS\frac{4\pi}{3r}(R_2^3-R_1^3), & r>R_2.
    \end{cases}
\end{equation}

We discretize Eq.~(\ref{eq:source_3D}) using a finite–volume formulation, which is given by
\begin{equation}
    \frac{3}{\rphalf^3-\rmhalf^3}
    \left(
    \rphalf^2\frac{u_{i+1}-u_i}{r_{i+1}-r_i}
    -
    \rmhalf^2\frac{u_i-u_{i-1}}{r_i-r_{i-1}}
    \right)
    =
    4\pi\rho_i.
\end{equation}

This yields a tridiagonal linear system,
\begin{equation}
    a_i u_{i+1} + b_i u_i + c_i u_{i-1} = 4\pi\rho_i,
\end{equation}
with coefficients
\begin{equation}
    \begin{aligned}
        a_i &=
        \frac{3}{\rphalf^3-\rmhalf^3}
        \frac{\rphalf^2}{r_{i+1}-r_i},\\[6pt]
        c_i &=
        \frac{3}{\rphalf^3-\rmhalf^3}
        \frac{\rmhalf^2}{r_i-r_{i-1}},\\[6pt]
        b_i &= -(a_i+c_i).
    \end{aligned}
\end{equation}

We solve this system using SOR for a range of relaxation parameters $\omega$. The number of iterations required to reach convergence is shown in Fig.~\ref{fig:omega_opt} for both logarithmic (left panel) and uniform (middle panel) grids and for resolutions $N=128$ (green line) and $N=256$ (blue line). The corresponding colored dotted lines represent the calculated value of $\omega$ from Eq.~\eqref{eq:omega_opt_mu}.

We find that the measured optimal relaxation parameter agrees closely with the theoretically predicted value in all cases. As expected, the optimal value depends slightly on both the grid resolution and grid type. In particular, the logarithmic grid produces a small shift in the optimal value due to the nonuniform spacing, which alters the spectral properties of the discrete operator. Nevertheless, the theoretical prediction remains closely accurate, demonstrating that the analysis correctly captures the convergence behavior of the radial operator even for strongly nonuniform grids.

\subsubsection{Polar problem}
We next consider the angular operator,
\begin{equation}
    \frac{1}{\sin\theta}
    \frac{\del}{\del\theta}
    \left(
    \sin\theta\frac{\del u}{\del\theta}
    \right)
    =
    f(\theta),
\end{equation}
with source term
\begin{equation}
     f(\theta)
     =
     \frac{2\cos\theta\cos2\theta - 4\sin\theta\sin2\theta}{\sin\theta},
\end{equation}
which corresponds to the analytical solution
\begin{equation}
    u(\theta) = \sin2\theta.
\end{equation}

\begin{figure*}[t]
\centering
\setlength{\tabcolsep}{0pt}
\renewcommand{\arraystretch}{1.0}
\begin{tabular}{cc}
  \includegraphics[width=0.5\linewidth]{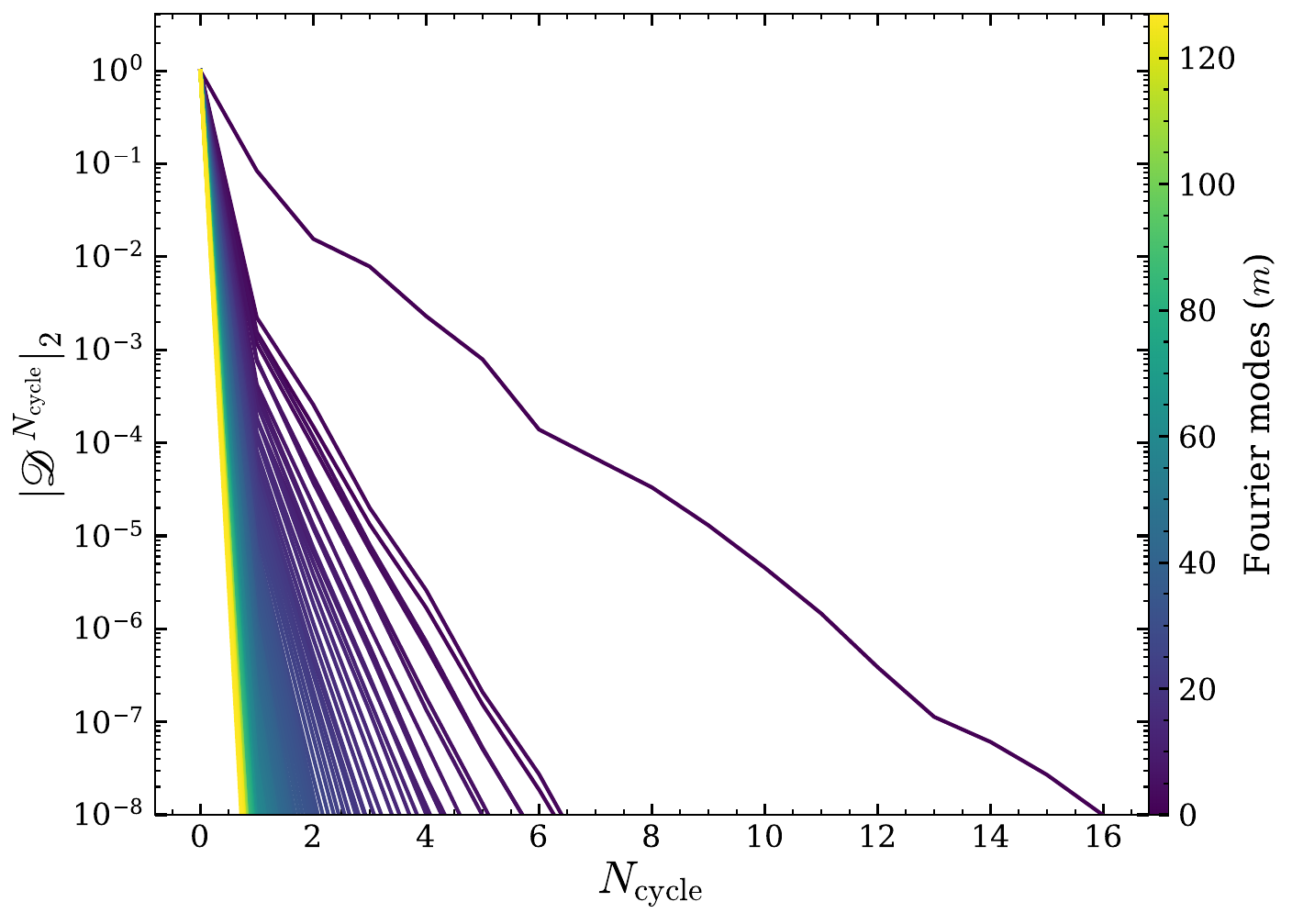} &
  \includegraphics[width=0.5\linewidth]{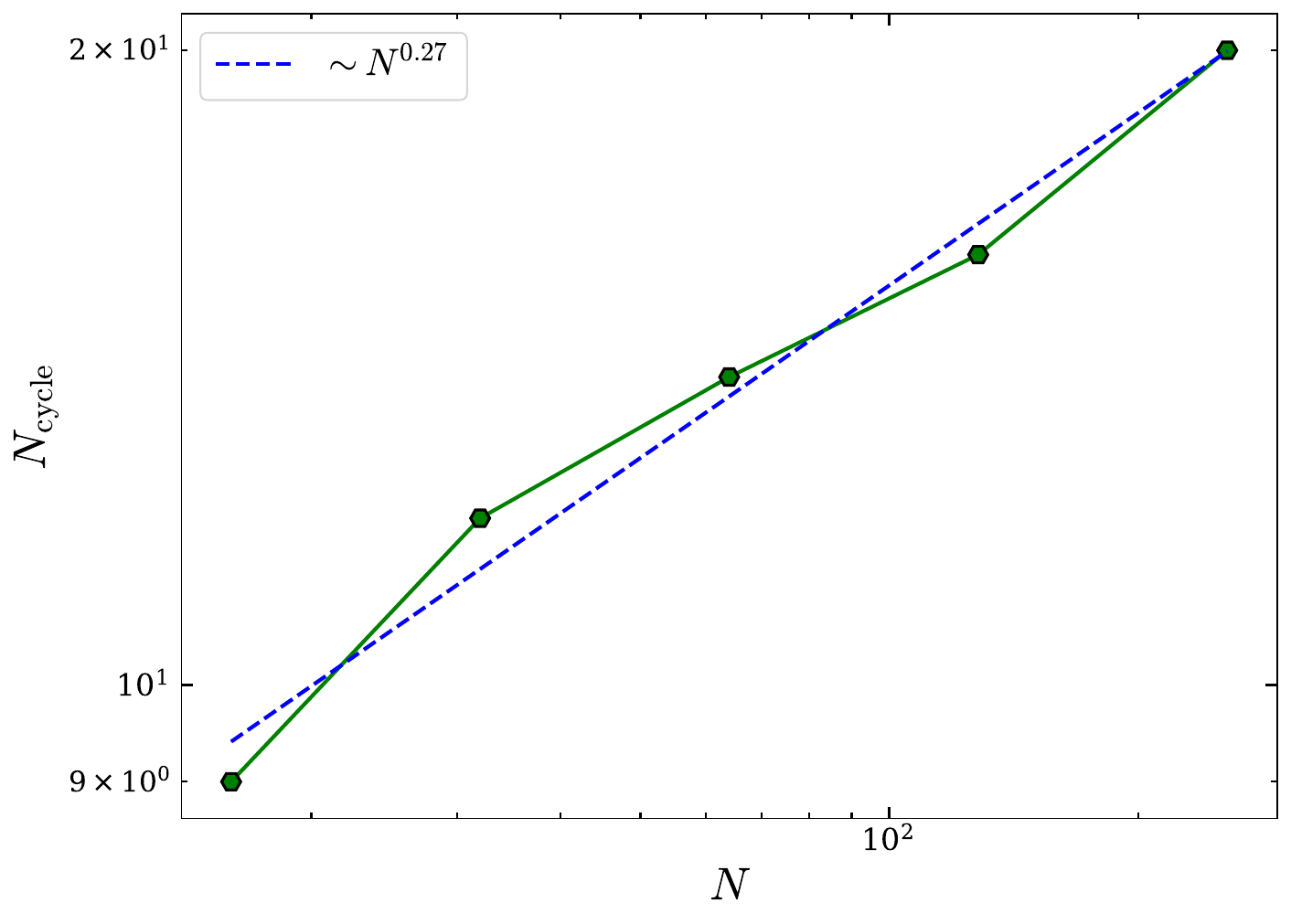}
\end{tabular}
  \caption{\emph{Left}: Residual reduction as a function of multigrid V-cycles for the mesh-segment test on a logarithmic spherical grid. Each curve corresponds to a Fourier mode $m$, color-coded from $m=0$ (dark blue) to $m=N_\phi/2$ (yellow). The residuals are normalized to their initial values. Lower-$m$ modes converge more slowly, while higher-$m$ modes converge within 1–2 V-cycles, consistent with the increasing diagonal dominance of the Helmholtz operator. \emph{Right}: Number of multigrid V-cycle ($N_{\rm cycle}$) needed to reach a tolerance of $10^{-8}$ as a function of grid resolution ($N$) for the slowest growing mode ($m=0$). The blue dashed line is a fit to the points, which shows a weak dependence of multigrid convergence on the grid resolution.}
  \label{fig:v-cycle_convergence}
\end{figure*}

Using a finite–volume discretization, the discrete equation becomes
\begin{equation}
    \frac{1}{\cos\tmhalf-\cos\tphalf}
    \left(
    \sin\tphalf\frac{u_{j+1}-u_j}{\theta_{j+1}-\theta_j}
    -
    \sin\tmhalf\frac{u_j-u_{j-1}}{\theta_j-\theta_{j-1}}
    \right)
    =
    f_j.
\end{equation}

This yields a tridiagonal system,
\begin{equation}
    a_j u_{j-1} + b_j u_j + c_j u_{j+1} = f_j,
\end{equation}
where
\begin{equation}
    \begin{aligned}
        a_j &=
        \frac{1}{\cos\tmhalf-\cos\tphalf}
        \frac{\sin\tmhalf}{\theta_j-\theta_{j-1}},\\[6pt]
        c_j &=
        \frac{1}{\cos\tmhalf-\cos\tphalf}
        \frac{\sin\tphalf}{\theta_{j+1}-\theta_j},\\[6pt]
        b_j &= -(a_j+c_j).
    \end{aligned}
\end{equation}

The system is solved using SOR for various values of $\omega$, and the number of iterations required for convergence is shown in the right panel of Fig.~\ref{fig:omega_opt} for resolutions $N=128$ (green) and $N=256$ (blue).

As in the radial case, the optimal relaxation parameter predicted by the theoretical analysis agrees very well with the measured optimum for both grid resolutions. 
The close agreement confirms that the theoretical formalism closely predicts the convergence properties of the polar operator.

\section{Convergence of the 2D Multigrid Solver}
\label{sec:multigrid_convergence}
In this appendix, we provide additional detail on the convergence behavior of the 2D multigrid solver for the Helmholtz equation described in Sec.~\ref{sec:multigrid_solver}. For this purpose, we consider the double sphere test on a radially logarithmic spherical grid, as described in Sec.~\ref{sec:double_sphere}.

After applying the Fourier transform to the source distribution, we obtain $N_\phi$ independent Helmholtz equations (one per Fourier mode), which are solved using the multigrid method described in Sec.~\ref{sec:multigrid_solver}. Each mode is solved independently, and its solution represents an intermediate quantity in Fourier space -- not the final solution of the full 3D Poisson equation in real space.

To characterize the convergence behavior of the multigrid solver for each mode, we compute the defect (residual) of the Helmholtz equation after $N_{\rm cycle}$ V-cycles as
\begin{equation}
    \mathscr{D}_{N_{\rm cycle}}^m = \nabla^2_m\Phi^m_{N_{\rm cycle}} - \rho^m
\end{equation}
and evaluate its $L_2$ norm using a volume-weighted average analogous to Eq.~\eqref{eq:L2_norm}

The left panel of Fig.~\ref{fig:v-cycle_convergence} shows the $L_2$ norm of the defect as a function of the number of multigrid V-cycles for a representative test problem. The norms are normalized to their initial values, computed using the initial guess $\Phi^m=0$. Each curve corresponds to a different Fourier mode $m$, with the mode number indicated by the color bar. The slowest converging mode is $m=0$, as expected since this mode is the least diagonally dominant. It requires approximately 16 V-cycles to reduce the initial defect by eight orders of magnitude. As $m$ increases, the convergence rate improves rapidly because the Helmholtz operator becomes increasingly diagonally dominant. For most higher modes, particularly $m\gtrsim 30$, a single V-cycle suffices to reduce the initial defect to near machine precision. Note that these numbers represent the worst-case scenario, where the initial guess is simply $\Phi^m=0$. In practical simulations, we use the solution from the previous timestep as the initial guess, which significantly reduces the number of V-cycles required.

\begin{figure*}[t]
\centering
\setlength{\tabcolsep}{0pt}
\renewcommand{\arraystretch}{1.0}
\begin{tabular}{cc}
  \includegraphics[width=0.5\linewidth]{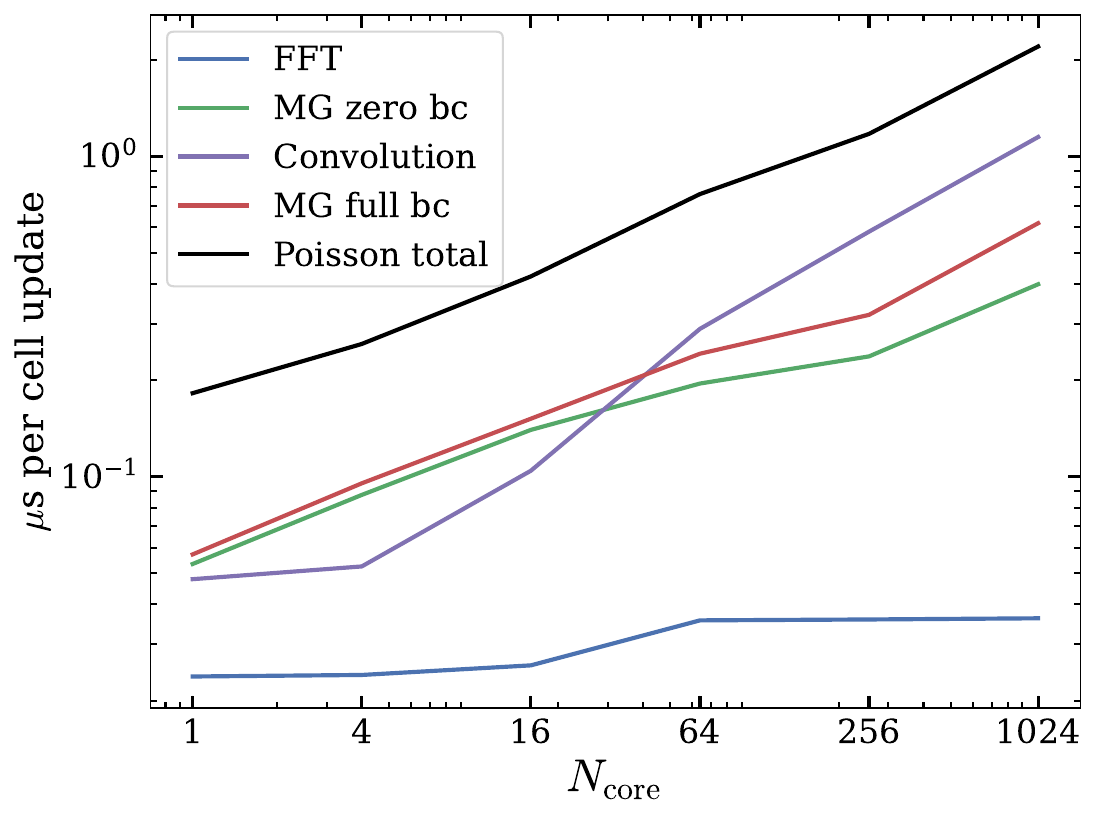} &
  \includegraphics[width=0.5\linewidth]{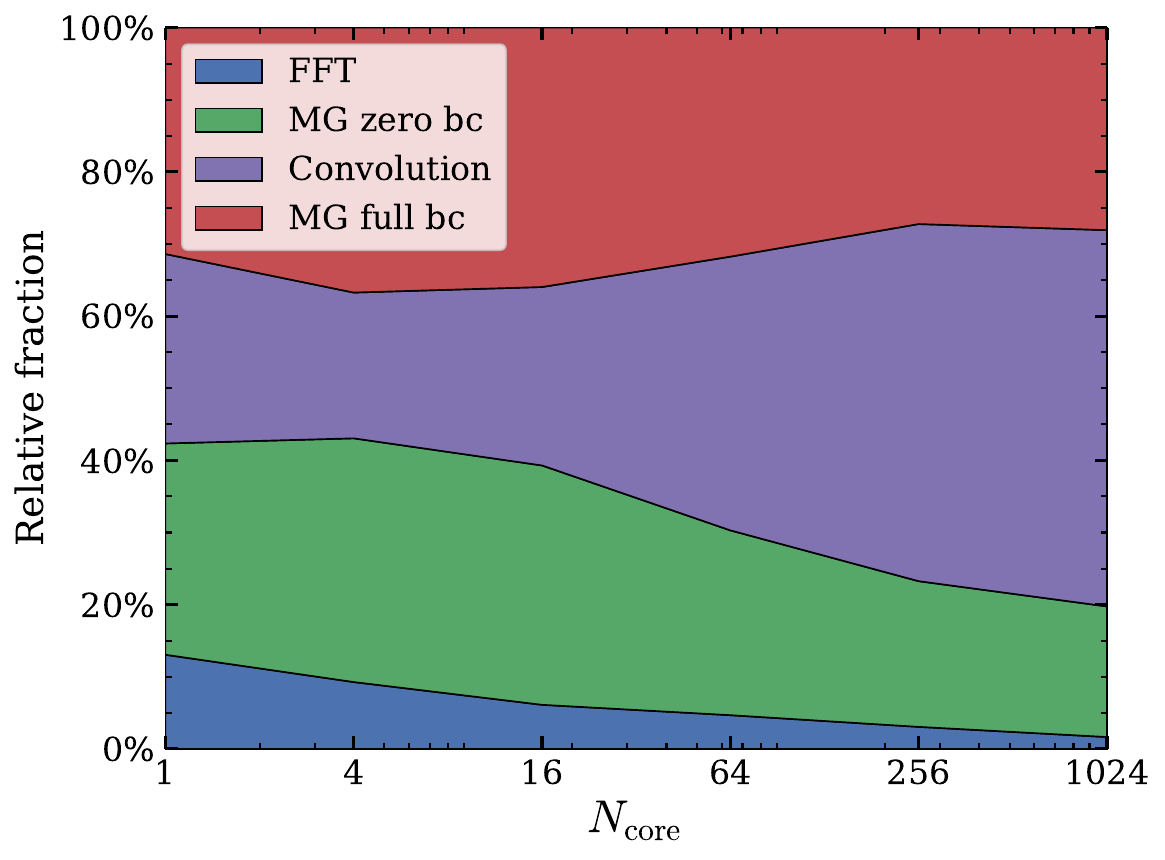}
\end{tabular}
  \caption{Weak scaling breakdown of the Poisson solver components: FFTs (blue), first multigrid solve (green), DGF convolution (purple), second multigrid solve (red), and total (black). Left: absolute time per cell per timestep. Right: relative fraction of total time.}
  \label{fig:weak_scaling_grav}
\end{figure*}

The right panel of Fig.~\ref{fig:v-cycle_convergence} examines the resolution dependence of the multigrid solver for the most demanding case, the $m=0$ mode. It shows the number of V-cycles required to reach a tolerance of $10^{-8}$ as a function of grid resolution $N$, where $N$ is the number of grid points along one direction (e.g., $N_r=N_\theta=N$). The measured scaling follows approximately $N^{0.27}$, indicating a very weak dependence on resolution. This is consistent with theoretical expectations for variable-coefficient anisotropic problems on non-uniform grids \citep[e.g., Chapter 5 of][]{Trottenberg_2000}. The overall efficiency of the solver remains high because the majority of modes converge in significantly fewer cycles.

This mode-dependent convergence behavior motivates our mode-by-mode solution strategy, in which each Fourier mode is iterated only until its individual tolerance is met, rather than iterating all modes simultaneously to a common convergence criterion. This approach avoids unnecessary computation for rapidly converging high-$m$ modes and improves overall efficiency.

We emphasize that the defect shown here is the residual of the Helmholtz equation for each Fourier mode, not the residual of the original three-dimensional Poisson equation. The final real-space gravitational potential is obtained only after solving all modes and performing an inverse Fourier transform (see Sec.~\ref{sec:algorithm_overview}).

\section{Component-wise weak-scaling of the Poisson solver}
\label{sec:scaling_grav}
In Sec.~\ref{sec:scaling_test}, we presented weak scaling results for the total Poisson solve time and the corresponding MHD update time. To better understand the origin of the increase in per-cell time with core count for the Poisson solver, Fig.~\ref{fig:weak_scaling_grav} breaks down the total gravitational update cost into its main components: the forward and inverse FFTs (blue), the first multigrid solve with zero boundary conditions (green), the DGF convolution for the James boundary treatment (purple), and the second multigrid solve with the full boundary conditions returned by the James method (red). The solid black line shows the total time taken by the Poisson solver. The left panel presents the absolute time for each component, while the right panel shows their relative fraction of the total Poisson solver time. We only present the analysis for the radially logarithmic spherical grid (results for the other grid configurations as considered in Sec.~\ref{sec:scaling_test} are similar).

The figure reveals that both the multigrid and DGF components increase with core count (i.e., with problem size) and have costs of similar order, while the FFT component remains almost flat and subdominant. As the core count increases, the DGF convolution cost becomes comparable or greater than the combined cost of the two multigrid steps, reflecting its $\mathcal{O}[(N_r+N_\theta)^2N_\phi/2]$ scaling. The mild increase in the multigrid cost with problem size is consistent with the expected weak resolution dependence of multigrid convergence for variable-coefficient anisotropic problems on non-uniform grids, as demonstrated in Appendix~\ref{sec:multigrid_convergence}. Therefore, the increase in total Poisson time seen in Fig.~\ref{fig:weak_scaling} is primarily due to the multigrid step at lower processor counts and to the DGF convolution at higher processor counts.


\bibliography{manuscript}{}
\bibliographystyle{aasjournalv7}



\end{document}